\documentclass{aa}  
\usepackage{graphicx}
\usepackage{txfonts}

\def\teff{$T_{\rm eff}$}

\def\cmm{cm$^{-1}$}
\def\cms{cm$^{2}$}
\def\cmq{cm$^{3}$}
\def\logg{log($g$)}

\def\taur{$\tau_{\rm ross}$}
\def\pg{P$_{\rm g}$}
\def\pe{P$_{\rm e}$}
\def\pr{P$_{\rm rad}$}

\def\Fr{F$_{\rm rad}$}
\def\Fc{F$_{\rm conv}$}
\def\vc{v$_{\rm conv}$}

\def\mnras{MNRAS}

\begin{document} 
	
	\title{A grid of self-consistent MSG (MARCS-StaticWeather-GGchem) cool stellar, sub-stellar, and exoplanetary model atmospheres} 
	
\titlerunning{MSG models of cool stellar and sub-stellar objects}
	
	\author{
   Uffe G. J{\o}rgensen
	\inst{1}
                 \and Flavia Amadio
	\inst{1,2}
                  \and  Beatriz Campos Estrada 
 	\inst{1,3}
                  \and   Kristian Holten M{\o}ller
 	\inst{1}     
                  \and   Aaron D. Schneider
 	\inst{1,2}
                 \and  Thorsten Balduin
 	\inst{1,3}
                  \and  Azzurra D'Alessandro
 	\inst{1}
                  \and Eftychia Symeonidou
 	\inst{1,4}
                \and	 Christiane Helling
	\inst{3}
                    \and  {\AA}ke Nordlund
	\inst{1,5}
                \and   Peter Woitke 
   \inst{3}
	}
	\institute{Centre for ExoLife Sciences, Niels Bohr Institute, University of Copenhagen,
               {\O}ster Voldgade 5,  1350 Copenhagen, Denmark\\
		\email{uffegj@nbi.dk}
     \and
     Institute for Astronomy, KU Leuven, Celestĳnenlaan 200D, 3001 Leuven, Belgium
     \and
     Space Research Institute, Austrian Academy of Sciences, Schmiedlstrasse 6, 8042 Graz, Austria
     \and
     Department of Biology, University of Copenhagen, Universitetsparken 15, 2100 Copenhagen, Denmark
     \and 
     Rosseland Centre for Solar Physics, University of Oslo, PO Box 1029 Blindern, 0315 Oslo, Norway
       }

	\date{Submitted to A{\&}A March 24, 2024; revised July 1, 2024; accepted xxx 2024}
	
	
	\abstract
	{Computation of self consistent 1D model atmospheres of cool stars, substellar objects and exoplanets in the effective temperature range 300K to 3000K, including cloud formation, chemical non-equilibrium effects, and stellar irradiation.}
	{To extend the classical MARCS model atmosphere grid from 2008 toward lower effective temperatures and a broader range of object types.
		}
	{The new model atmosphere computations, MSG, are based on a combination of three well tested codes, the classical MARCS 1D atmospheres, the StaticWeather cloud formation code, and the GGchem chemical equilibrium code. The combined code has been updated with new and more complete molecular and atomic opacities, cloud formation and advanced chemical equilibrium calculations, and addition of new numerical methods at low temperatures to allow a more robust convergence.}
	{The coupling between the MARCS radiative transfer and GGchem chemical equilibrium computations has made it possibly effectively to reach convergence based on electron pressure for the warmer models and gas pressure for the cooler models, enabling self-consistent modelling of stellar, sub-stellar and exoplanetary objects in a very wide 
range of effective temperatures.  The new cloud-free and non-irradiated models for solar metallicity and a selected variety of other chemical compositions are immediately available from our homepage. Illustrative examples of cloudy and irradiated models as well as models based on non-equilibrium chemistry are also presented, and will be described in more detail and made available on completion at the same place for a larger range of parameter space.
		}
   {For solar metallicity models, new additional molecular opacities only affect the structure of models cooler 
than \teff\ = 2500\,K, and becomes substantial for models below \teff\ $\sim$1500\,K. 
Atomic line opacities are important for models warmer than $\sim$3000\,K. 
The line profile of the molecular opacities may have larger effect on the model structure than previously anticipated,
in particular in the uppermost layers at low gas pressure. The qualitative changes in the relative abundances of
TiO, H$_2$O, CH$_4$, NH$_3$, and other molecules in our models follow the observationally defined M, L, T (and Y) sequences, but reveal more complex and depth dependent abundance changes, and 
therefore a spectral classification depending on more parameters. The self consistent coupling to Static-Weather cloud computations, allows detailed comparison between nucleation and observed relative dimming of different spectral bands, with advanced applications for new identification methods of potential exoplanetary biology.
}
	
	\keywords{
		stars: atmospheres - stars: late-type - exoplanets: atmospheres
	}
	
	\maketitle

	
	\section{Introduction}

We have developed a model atmosphere code consisting of an updated version of the classical MARCS (Model Atmospheres in Radiative and Convective Scheme) code (\citet{Gustafsson1975}, \citet{Gustafsson2008}) combined with the cloud formation code Static Weather  (\citet{Helling2006, Helling2008b}) and the chemical equilibrium code GGchem (\citet{Woitke2018}). We name the new code and the corresponding model atmospheres for MSG (MARCS-StaticWeather-GGchem) as an abbreviation of its 3 major components. For warmer models, the atmospheric temperature versus gas pressure structures of the MSG models are quite similar to the corresponding models of the MARCS grid from 2008 (\citet{Gustafsson2008}), which we in the following will refer to as "the classical MARCS models". Minor differences are found when one approaches the lower end of effective temperatures of \teff\ = 2500\,K in the 2008 classical MARCS grid. There are no published classical MARCS models below \teff\ = 2500\,K, but we will show here what they would have looked like, and illustrate what was missing in the classical models to describe these lower temperatures and what the differences to the new grid would approximately have been. The new grid is at present computed down to effective temperatures of \teff\ = 300\,K, which will allow studies of the Earth-like exoplanetary temperature regime and in models still to be computed also the atmospheric effect of rocky surface development and biological evolution.

A basic philosophy behind the MARCS models, as well as the Static Weather and GGchem computations, has been to keep all computations fully self consistent with no free or adjustable parameters to improve fits to observed spectra, and this philosophy is inherited into the MSG grid. We believe that this approach will maximize the visibility of unknown (or non-included) physics and missing data, with a long-term goal of revealing for example a potential variety of known or unknown biological effects on exoplanetary atmospheric structures. The existing grid of classical models available at the MARCS web site (http://marcs.astro.uu.se) contains a few tens of thousand models, covering a broad range of dwarf and giant stars of many
different metallicities in the effective temperature range from 8,000\,K to 2,500\,K, as described in detail in 
\citep{Gustafsson2008}. This huge number of easily available models has made it possibly to test the model predictions against a large variety of stellar objects, and the more than 3,000 citations in the scientific literature to the two basic model grid papers  (\citet{Gustafsson1975, Gustafsson2008}) witness of a broadly applied and well tested use of the classical MARCS models.

The MARCS code was originally written \citep{Gustafsson1975}
with "solar-type" (i.e.\ F, G, K type) stars in mind. Such stars are well represented by plane parallel, LTE models in chemical equilibrium, with relatively few opacity sources, and they converge well in a scheme where variations in the opacity is calculated as function of temperature and electron pressure, and the adiabatic index is computed based on only few molecular and other opacities. During the almost 5 decades that have passed since the initial 1975-grid, the models have been modified, updated, expanded, and augmented several times by introduction of new physics and basic input data, suitable for a much larger parameter space, including introduction of more complete atomic and molecular opacities (e.g., \citet{Jorgensen1985}), and updating the opacity distribution function (ODF) method to the opacity sampling (OS) method in handling of the absorption coefficients (\citet{Saxner1984}, \citet{Jorgensen1992a}), which all made it possible to compute stars of a wider range of effective temperatures (for an overview see \citet{Gustafsson1994}).
Inclusion of spherical geometry has made it possible to reach larger stars of lower surface gravity (\citet{Plez1992}, \citet{Jorgensen1992}), and the first attempts of a description of cloud formation in the models was introduced in \citet{Juncher2017}.

To assure the reliability of the classical models, it was decided to limit the 2008-grid to temperatures above \teff\ = 2500\,K.  On top of this, models usually had convergence problems at the lowest temperatures, which could further support the suspicion that something was physically unrealistic at the lowest effective temperatures. In the present paper we will show that this suspicion to a large extent was correct, and we will introduce new data and methods that are consistent with the classical models for \teff\ = 2500\,K and above, and at the same time extends the grid from its present lower limit of \teff\ = 2500 K down to T$_{\rm eff}$ = 300\,K. This extension completes the grid to the coolest known stars and brown dwarfs, at temperatures overlapping with the bulk of the hot-jupiters and Earth-like exoplanets too, including all types of M, L, and T-dwarfs and probably 
reaching the bulk of the disputed (see e.g.\ \citet{Burningham2008}, \citet{Cushing2021} and our discussion of NH$_3$ in section 6) Y-types too. 
A corresponding extension of the grid to warmer models than the present upper limit of \teff\ = 8,000\,K is in progress by Edwardson et al.\ 2024, in preparation).

The extension of cloud-free models down to \teff\ = 300\,K has been made possible by inclusion of a considerably more complete set of molecular opacities assembled from the ExoMol database (\citet{Tennyson2012}, \citet{Tennyson2016}, \citet{Tennyson2020},  \citet{Chubb2021}),
by inclusion of the extensive new chemical equilibrium routines of GGchem (\citet{Woitke2018}) that allow computation of relevant species down to T = 100\,K, and by including a new way of handling the electron pressure in the iterative convergence scheme, and other numerical improvements. The corresponding cloudy models have been made possible by the integration with the self consistent cloud formation program Static Weather (\citet{Helling2006, Helling2008b}). 
The MSG computations are able to compute, in a self consistent and homogenous way, both the host stars and their irradiated exoplanets, based on an improvement of the more simple radiation schemes that began with the series of papers by Nordlund and Vaz for the description of binary stars already in the 1980'es (\citet{Vaz1985}, \citet{Nordlund1990}).

The present updates as well as the planed extensions have been made possible due to two major grants (mentioned in the acknowledgments) forming the backbone of a new Centre for ExoLife Sciences (www.cels.nbi.ku.dk). Under the auspices of this project, effects of irradiation (Amadio et al., in preparation), the application of the new results to the atmosphere of rocky and cloudy exoplanets (Campos Estrada et al., in preparation), the effects of cloud formation  and its consequences for the effects of biological evolution on the atmospheric structures (D'Alessandro et al, in preparation), are part of this ongoing extensions of the grid. 
 
In the present paper we will describe the overall results, the basic methods, and the grid of cloud-free models and corresponding synthetic spectra, and introduce the ongoing extensions to cloudy, irradiated, and non-equilibrium exoplanet models. The paper is structured with section 2 giving a short summary of the basic assumptions and methods of the three merged codes, section 3 outlining the degree of consistency with the 2008 classical MARCS-grid, section 4 focussing on some challenges,  
describing the addition of the new molecular line lists, partition functions, and chemical equilibrium routines, 
section 5 discussing the importance of the correct treatment of the line profiles, section 6 focusing on aspects of the total opacities, section 7 describing effects of irradiation, section 8 effects of cloud formation, section 9 considerations about chemical non-equilibrium, section 10 describing the boundary between M-type and L-type as well as the transition from L-type to T-type, section 11 the difference between oxygen-rich and carbon-rich models, and finally conclusions and outlook is outlined in section 12.

\section{Summary of the three merged model codes}
\label{sec:3codes}

	\subsection{MARCS}

The classical MARCS models are cloud-free, 1D, stratified, flux-constant, LTE models, computed in radiative-convective equilibrium in an optical depth scale, and solved iteratively by the Newton-Raphson method. With some modifications, they form the basic structure for the solution of the energy balance in the MSG models too. 
The models are characterized by the effective temperature, \teff, the surface gravity, $g$, and the relative abundances of the elements. Typically, the gravity is given as log$_{10}(g)$ in $cgs$-units, such that log($g$) of the Sun, Jupiter, and the Earth are 4.44, 3.39, and 2.99, respectively, and therefore log($g$) of respectively cool stellar and brown dwarfs, exo-jupiters, and exo-Earths can be considered as being $\sim$ 4.5, 3.5, and 3.0. 
The code can compute the models in plane parallel as well as in spherical geometry. In plane parallel approximation planetary/stellar radius is an irrelevant parameter, and only gravity needs to be specified. If sphericity is chosen, mass is specified in the input too, such that radius can be computed from mass and gravity. The full spherical atmospheric structure is then solved by use of the numerical method developed by \citet{Nordlund1984} and further described in for example \citet{Jorgensen1992}, \citet{HeiterEriksson2006}. It is a computationally fast, iterative method giving better than 1$\%$ accuracy in the source function after only 2 to 3 iterations, and therefore requiring typically only a factor of 5 to 10 times more CPU time than a corresponding plane parallel computation. The atmospheric structure of extended stars of low gravity, such as red giants, are substantially affected by sphericity, but for the type of objects discussed here sphericity effects are in general negligible.

The input variables for the calculation of the thermodynamic quantities and the opacities are temperature, T$(\tau)$, and electron pressure, \pe($\tau$). The use of \pe\ as input variable in the MARCS-based part of the MSG scheme is somewhat modified to compensate for the extremely low abundance of free electrons in the coolest MSG models. 

The other dependent variables are the gas pressure, \pg, and radiative pressure, \pr, as well as the integrated radiative and convective fluxes, \Fr\ and \Fc, and the convective velocity \vc. Together these variables can be expressed in a highly non-linear closed system of equations that includes the hydrostatic equilibrium, the radiative-convective flux-constancy, expressions for the gas pressure and radiative pressure, calculation of the convective flux from mixing-length theory, and calculation of the radiative flux in LTE, based on the Feautrier-method (\citet{Feautrier1964}), with boundary conditions as described in \citet{Nordlund1984}. The adopted formalism of the mixing length follows \citet{Henyey1965} who besides the mixing length parameter $\alpha$ itself (in units of the scale height) also includes a parameter $y$ which relates to the temperature distribution within the convective elements, and $\nu$ which relates to dissipation by the turbulent viscosity, and which in all our models are fixed to ($\alpha, y, \nu$) = (1.5, 0.076, 8), following the recommendation in \citet{Henyey1965}, but easily can be varied in the input file of the models. Additionally, also turbulent pressure $P_t$ can be included as a dependent variable in the system of equations, but following the arguments in \citet{Gustafsson2008} all the models in the present MSG grid are computed with $P_t$=0. More details about the specific equations, the choices of methods, and the introduced approximations can be found in \citet{Gustafsson2008}.

The code solves for the radiative energy balance (weighted sum over heating and cooling for each wavelength point) 
above a height where the convective flux is negligble, while below it solves for constancy of total flux, 
in both cases using linearization of the relevant equations. The point of transition between the two methods is not critical, except it should occur where the convective flux is zero or insignificant. With the second order Feautrier method, constant radiative flux and radiative balance are numerically equivalent, given suitable definitions and centering of optical depth increments, but numerical precision is better for radiative equilibrium at small optical depths, and for constant flux at large optical depths. Due to the linearization of the coupled equations (which while not including everything does include the most important dependencies), the temperature corrections decrease rapidly towards the end of iterations.  We typically stop the iterations when the temperature corrections from one iteration to the next are smaller than 1\,K in all layers, and tests have shown that this corresponds closely to remaining temperature errors. When clouds form in the model structure, it is the sum of the increased cloud opacities plus the reduced gas opacities that defines the energy 
balance (as described below in section \ref{sec:StaticWeather}), and this usually requires additional iterations compared to cloud free models and often demands a stepwise introduction of the cloud particles in order to avoid too abrupt changes in the opacities from one iteration to the next. 

 In the overlap regions between the classical MARCS grid and the corresponding parts of the PHOENIX grid (\citet{Hauschildt1999}, and in particular the following later versions) and the grids by Kurucz and collaborators (\citet{Kurucz1979}, \citet{Castelli2003}), there is a good general agreement between models from those three grids. This is obviously very reassuring given the large degree of independence between the three codes, concerning adopted methods, approximations, and (to some extent) adopted input data. We will comment in detail below about the level of agreement between the classical MARCS models and our MSG models in the overlap region between these two grids.

The extent of the atmosphere is defined by the input choice of the Rosseland optical depth, \taur, in the top (where \pg $=$ 0) and bottom of the atmosphere. It should be noted that although \taur\ is a very poorly defined physical quantity in gasses where the opacity is dominated by line absorption, the adaption of \taur\ is only used as a mean for the discretisation of the equation system, and does not have any effect on the physical solution. As such it can be thought of mainly as a convenient labelling of the individual discrete layers of the stratified atmosphere. The step length between the layers in any physical meaningful variable (e.g. gas pressure, temperature, height, etc) can be adjusted in the input by adjusting the step length in units of \taur\ between any of the individual layers. In gaseous objects of our grid we will typically define the bottom of the atmosphere as the layer where log(\taur) = 2.5, while for solid objects (e.g.\ Earth-like exoplanets) we will obviously define the bottom of the atmosphere as the planetary surface. In the bulk of the models presented in the following, we define the "top"  of the atmosphere as the layer where log(\taur) = $-$5, and use a steplength of $\Delta$log(\taur) ranging between 0.1 and 0.15 in various regions of the atmosphere, resulting in typically 53 layers in our "standard" models. For stellar atmospheres it is customary practice to call the top of the atmosphere for the "surface" of the star. In order to avoid confusion toward the obvious use in planetary practice with the bottom of the atmosphere being also the surface of the (solid) planets, we recommend, and try to be consistent here, to use only the terms top and bottom of the atmosphere, avoiding the term surface in the meaning of top of the atmosphere.

	\subsection{GGchem}
\label{sec:GGchem}
We use the publicly available thermo-chemical equilibrium code GGchem
of \citet{Woitke2018} to calculate the concentrations of all neutral
and single ionized atoms, electrons, molecules and molecular ions.  In
all models of the presented MSG grid, higher than single ionized atoms
are irrelevant, but in order to be consistent with the classical MARCS
grid, and in order to compute exoplanets and their irradiating host
star self-consistently, there is an input option to include higher
ionization from the classical MARCS code (subroutine JON) if desired.
GGchem is based on the principle of minimisation of the total Gibbs
free energy, including condensates, and applicable in a wide
temperature range between 100\,K and 6000\,K. However, in this paper
we use GGchem solely for the purpose to calculate the gas phase
particle densities based on temperature, pressure and the gas phase
elemental abundances. In MSG the solid state condensates are obtained 
from subsequent calls to the more advanced Static Weather computations 
of the dust grains (see next sub-section).
Hence the gas phase elemental abundances changes layer-by-layer though 
the MSG iterations and convergence, keeping everything self consistent.

The thermo-chemical data has been collected from different sources, in
particular from the JANAF thermochemical tables
\citep{Chase1996,Chase1998} and from \citep{Barklem2016}. The data have
been carefully compared to other data sources by \citet{Worters2017}.
We consider 22 elements (H, He, Li, Be, C, N, O, F, Na, Mg, Al, Si, P,
S, Cl, K, Ca, Ti, V, Cr, Fe, Zr).  Each element can be present in the form
of neutral atoms, or singly charged positive or negative ions.
According to this selection of elements, GGchem finds additional 543
molecules, molecular ions and cations in its database. GGchem has been
benchmarked down to 400\,K for a pure neutral gas phase chemistry
against TEA \citep{Blecic2016}, and down to 500\,K for chemistry,
ionisation and condensation against \citet{Sharp1990}. Below these
temperatures, the other codes start to have problems to converge.
GGchem uses a hierarchical approach of pre-iterations to stepwise
increase the number of elements, combined with quadruple-precision
arithmetics, to solve the thermo-chemical equilibrium down to
100\,K. GGchem is written in ultra-fast Fortran-90 and needs about
0.003 CPU-sec per call in this setup.

	\subsection{Static Weather}

\label{sec:StaticWeather}
The kinetic cloud formation model StaticWeather was developed by
\cite{Woitke2003,Woitke2004}.  The model calculates the rate of
nucleation of seed particles from the gas phase by modified classical
nucleation theory \citep{Gail1984}, and considers an explicit set of
surface chemical reactions to calculate the growth rates of all cloud
particles by molecule-surface collisions, and the reverse rates, the
sublimation rates.  Each pair of growth and sublimation rates is
related to each other via the supersaturation ratio $S$ of the
considered solid material in the atmosphere.  In order to calculate
the molecular densities required for the effective growth rates and
the supersaturation ratios, the GGchem code is used internally (see
Sect.\ref{sec:GGchem}).  The StaticWeather code uses the dust moment
method of \cite{Gail1988} to consider a cloud particle size
distribution function that changes with height in the atmosphere.  The
size-dependent gravitational settling of the cloud particles is part
of the modelling, which leads to a downward transport of condensable
elements.

The element abundances in the gas phase are hence consumed by
nucleation and growth, and replenished by sublimation and convective
mixing. For the replenishment by convective mixing, the StaticWeather
code uses a simple approach based on the mass exchange timescales
derived from the 3D hydrodynamical models for M-star atmospheres by
\cite{Ludwig2002}.  At each atmospheric layer, the gas with cloud
particles is exchanged by the cloud-free gas from the deep atmosphere,
on a timescale that depends on atmospheric height.  

In praxis the inclusion of clouds into the self-consistent modelling of the MSG atmospheric structure is done by starting with computing (or reading in) a cloud-free model structure of given elemental abundances (that are the same in all layers) as specified in the input file. Then Static Weather will compute the amount of gas that will condense into solids for each layer of this structure. In next iteration the opacities from the computed abundances of the solids and from the corresponding (reduced) gas species are computed and added up, and the radiative transfer is calculated. A new structure is then computed, and so on until a self-consistent convergence is reached. After full convergence, the input relative elemental abundances of the gas phase are only conserved in the regions outside the cloud formation area. Inside the cloudy region, the gas phase elemental abundances of those elements that participate in the mineral cloud particles are reduced, sometimes by large factors. The total elemental abundances present in the solid and the gaseous states in each cloudy layer do not in general sum up to the input elemental abundances, but depends on the local condensation, sublimation, and mixing time scales. The emerging spectrum of such fully self-consistent models can be substantially different from non-self-consistent modelling.


\cite{Helling2006} have generalised the StaticWeather model to mixed
condensates, assuming that all stable solid materials grow on the same cloud 
particle surfaces. The model is anchored in experimental laboratory data, all the way 
from the chemical reactions it uses to the optical properties of the condensates it considers. 
A hand-picked set of about 20 condensates is
selected, including several magnesium-silicates, solid iron, a few
high-temperature Al-Ca-Ti oxides, and some simple metal oxides and
sulphides.  In atmospheres that are cool at the top and hot at the
bottom, the nucleation kick-starts the formation of ``dirty'' cloud
particles materials high in the atmospheres, but as these particles
settle down into warmer regions, they stepwise ``purify'', until only
the most stable condensates remain -- mineral clouds -- before even
these materials sublimate at the cloud base. \cite{Helling2008a} have
compared the StaticWeather model to other cloud formation
models. \citet{Helling2008b} and collaborators have included the StaticWeather into
PHOENIX stellar atmosphere models of \cite{Hauschildt1999} for brown
dwarfs, as well as into gas-giant atmospheres (e.g.\, \citet{Helling2021}, \citet{Lee2015, Lee2016}, \citet{Juncher2017}, \citet{Samra2022}), including 3D GCM models
\citep{Helling2016}.

\subsection{Computational speed}

The computational speed of the code scales to a first approximation with the number of adopted frequencies, $n_\lambda$, in the radiative transfer computation, times the number of layers in the atmosphere, $n_\tau$, times the number $n_\mu$ of individual directions in the radiation field adopted in the computation of the mean intensity. In order to self consistently include clouds into the model computation, the MSG code allows for varying relative elemental abundances in the gas phase chemistry from layer to layer and from iteration to iteration, which obviously adds considerable time to the computation of the opacities (as does the fact that the MSG models include many more individual opacity sources than the classical MARCS models), but a combination of increased core memory and computer capacity and improved coding has allowed the total time per iteration to remain approximately the same as 20 years ago (for a corresponding classical MARCS model). This is very reassuring, because it allows using the MSG models not only for comparison of models over an extensive range of parameter space, but also as a test base for self consistent inclusion of advanced physics and chemistry into dynamically advanced models such as modern 3D GCM models. With typically 100,000 $\lambda$-points, 50 depth-layers and 6 directions (in the plane parallel approximation, and about 5 times more in the spherical case), and 10 iterations for full convergence (depending strongly on the applied input model for the first iteration), a typical cloud-free model convergence time when executing at a single core of our present hpc computing system (Intel(R) Xenon(R) CPU E5-2680 v4 @2.40GHz) is of the order of 15 minutes.

	\section{Consistency between non-irradiated, cloud-free MSG models and the classical MARCS models}

As described above, the classical MARCS grid was limited to models of \teff\ = 2500\,K and warmer, among other things due to the lack of sufficient input data for cooler models. Such data have now become available due to the huge work on molecular line data as computed and collected in the ExoMol data base (\citet{Tennyson2020}), due to the extensive development in cloud formation data and theory (\citet{Helling2006}), and the improved chemical equilibrium computations (\citet{Woitke2018}), as well as due to extensively updated atomic data bases (\citet{Grimm2021}) and updated data in general. In this section we will compare the cloud-free, non-irradiated, plane parallel, chemical equilibrium, solar metallicity MSG models with the corresponding classical MARCS models available at the Uppsala-MARCS-webpage ({\it https://marcs.astro.uu.se/}) and described in \citet{Gustafsson2008} and the DRIFT-MARCS models described in \citet{Juncher2017}.

The classical MARCS models include the line opacity from 16 molecules (HCN, H$_2$O C$_2$, CH, CN, CO, CaH, FeH, MgH, NH, OH, SiH, SiO, TiO, VO, ZrO) from various sources that were available at that time, as listed in the 2008 grid paper, plus atomic lines from the VALD data base. The DRIFT-MARCS models from \citet{Juncher2017} include the same 16 molecules (but not in all cases from the same sources) plus 6 additional molecules (CO$_2$, TiH, CrH, NO, LiH, H$_2$). Cloud formation and cloud opacities were included in the \citet{Juncher2017} models, but in general atomic line data were not included. The standard version of the present MSG model grid includes the same 22 molecules as in \citet{Juncher2017} plus additionally 28 molecules 
(SO$_2$, AlCl, PO, NaCl, SiS, LiF, LiCl, PH$_3$, KCl, HCl, CP, NaH, AlF, BeH, MgF, CaF, CH$_4$, H$_2$CO, NH$_3$, AlO, PN, CS, KF, PS, SN, AlH, NaF, HS). All included molecules are now from the ExoMol database. References to the original work behind the individual line lists adopted from the ExoMol database are 
given in Tab.\,1 below
together with the original sources for the individual atomic line list date adopted from the DACE database. The data for continuum and collision induced opacities 
are the same as used in the classical MARCS models, and listed in Table 1 of \citet{Gustafsson2008}.
Some of the classical MARCS and the DRIFT-MARCS models included two more carbon rich molecules (C$_2$H$_2$ and C$_3$), that are not generally included here. They only affect carbon rich atmospheres and not the solar metallicity models that are the focus of the present paper. They and other molecules will be included in forthcoming versions of the MSG model grid where relevant.
 
\begin{figure}
              \vspace{-1.0cm}
              \hspace{-1.8cm}
             \includegraphics[width=11.6cm,angle=180.]{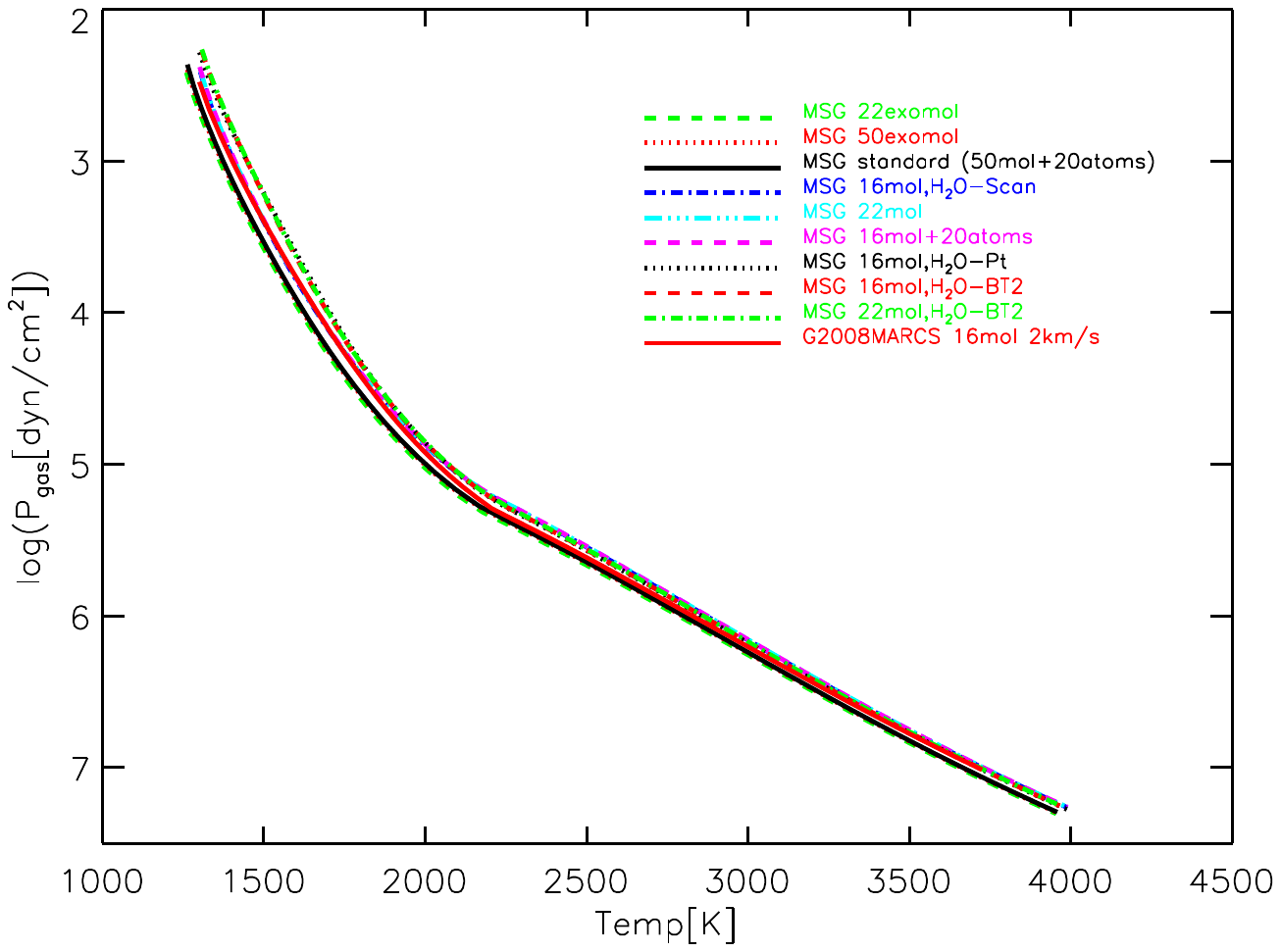}

              \vspace{-1.0cm}
              \hspace{-1.8cm}
             \includegraphics[width=11.6cm,angle=180.]{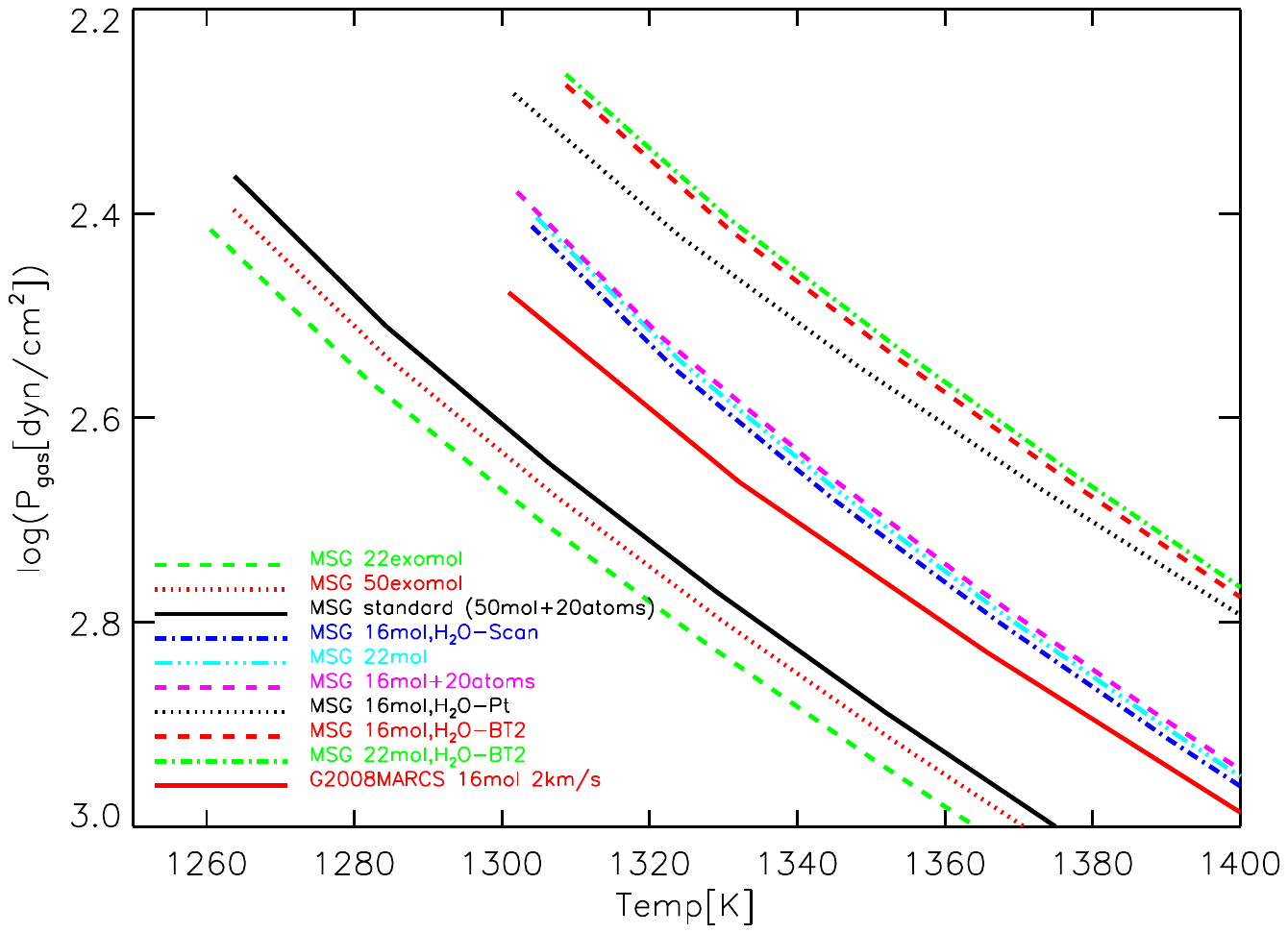}

                \vspace{-1.0cm}
		\caption{Comparison of the classical 2008-MARCS model at the lower boundary of the 2008-grid (labeled 'G2008-MARCS') with models computed with the MSG code and based on various levels of completeness and choices of input data as described in the text. Models labeled 'MSG 22...mol..' are based on molecular opacities from the 22 molecules included in the DRIFT-MARCS computations, while models labeled 'MSG 16mol...' are based on opacities from the 16 molecules included in the classical MARCS grid.
Upper panel show the T-P$_g$ structure of the full models from log($\tau_{ross}$) = --5.0 to +2.5 while the lower panel zooms in on the upper part of the atmosphere. }
	\label{FigTPg2500K}
	\end{figure}

The upper panel of Fig.\ref{FigTPg2500K} shows that for models of \teff\ = 2500\,K there is a good general agreement between the MSG-, the classical MARCS-, and the 2017-models, computed based on their adopted input as mentioned above. For higher values of \teff\ the agreement can be expected to be even closer, since the additional data added in the MSG grid computations will have a smaller importance than for the \teff\ = 2500\,K models. This is very reassuring, because it shows that the new grid of cloud-free and non-irradiated MSG models and the classical MARCS models are consistent and in general agreement with one another at the overlap region between the two grids. 

The lower panel of Fig.\ref{FigTPg2500K} zooms in on the upper most layers of the same models as in the upper panel of the figure, to quantify clearly the region where the various models differ the most. It shows that the models  in the most extreme cases differ from one another with up to 70\,K. The classical MARCS models fall close to the middle of the range. 

We consider the presently best MSG models to be those that are based on line data  from the 50 molecules of the ExoMol data base mentioned above plus from 20 different neutral atoms available in the DACE database (\citet{Grimm2021}) as listed in Tab.\,1, 
plus of course updated chemical equilibrium computations from GGchem and other updates as well. For \teff\ = 2500\,K, Fig.\ref{FigTPg2500K} shows that this model is only $\sim$20\,K cooler in the upper layers than the shown classical MARCS model (for a given value of P$_{gas}$). We note that most of the classical MARCS grid is computed for models with microturbulence of $\xi$=2\,km/s as well as $\xi$=5\,km/s, while the \teff\ = 2500\,K model was computed for $\xi$=2\,km/s only, and the present MSG grid is computed for a fixed value of $\xi$=3\,km/s. Typically, cool models get $\sim$3\,K hotter per increase of 1 km/s in $\xi$, so we conclude that the preferred \teff\ =2500\,K MSG model is approximately 25\,K cooler than the corresponding classical MARCS model (in the upper layers and for given value of P$_{gas}$). The choice of microturbulence determines the broadness of the Gaussian line profiles used in the computation of the molecular opacities, and it is this effect that accounts for the difference between the $\xi$=2\,km/s and $\xi$=5\,km/s models. 
We will discuss in detail in a separate section below the effect of the important choice of line profile in general for the resulting structure of cool-star and exoplanetary models.

In order to understand what effect various updates have played for the models on the boundary to the classical MARCS grid, we show in Fig.\ref{FigTPg2500K} also a number of \teff\ =2500\,K MSG models computed with a variety of chosen molecular and atomic input data. This will hopefully add some hints about where future improvements in exoplanetary modelling in general could be most beneficial. 

The model labeled 'G2008MARCS 2km/s' is the classical MARCS model (full drawn red line). The group of 3 models in Fig.\,\ref{FigTPg2500K} that are a few degrees cooler than the classical MARCS model are based on line opacities from respectively the set of 22 molecules that were also adopted in the 2017-grid (but here all take from the ExoMol data base), all the 50 molecules taken from the ExoMol data base, and all these 50 plus the 20 atoms from the DACE data base (full drawn black line). This latter is our preferred combination, used as "standard" for the MSG grid.
All these 3 models differ less than $\sim$10\,K from one another in the uppermost layers of the atmosphere, and it is therefore concluded that the 28 newest molecules added from the ExoMol data base only have a marginal effect (of slightly more than 5\,K) on the model structure (of a \teff\ =2500\,K model) and the addition of atomic lines has an even smaller effect. This latter conclusion (about the atomic lines) has always been implicitly assumed in our older cool MARCS models (although not in the 2008-grid), but has never been thoroughly and quantitatively tested before the results shown in Fig.\,\ref{FigTPg2500K}. We will return to it in a separate section below.

Of the group of 3 models in Fig.\,\ref{FigTPg2500K} that are all approximately 10 to 15\,K warmer than the classical model for a given P$_g$, the two coolest are computed with respectively the 16 and 22 molecules from (approximately) the line lists adopted in the 2008-grid and in the 2017-grid, while the warmest of the three is computed based on the 16 molecules from the classical grid augmented with the 20 atoms from the DACE database that we use in the MSG grid (labeled 'MSG 16mol+20atoms'). The difference between these 3 models is very small, and we conclude that the effect of the difference in the opacities included in the 2008- and 2017-grids were very small for these values of \teff. 

Finally, the group of the 3 warmest models shown in Fig.\ref{FigTPg2500K} are almost 50\,K warmer than the published classical MARCS model. The two (almost) identical of these models are based on the 16 molecules in the classical MARCS grid or the 22 from the 2017 grid, respectively, and both with the water line opacity from the BT2 ExoMol list (\citet{Barber2006}), 
while the third of these models are with the Prokazatel H$_2$O linelist from ExoMol (\citet{H2O_Polyansky}). We conclude from these 3 models that H$_2$O is likely to be the most important individual molecule for the opacity at these temperatures (and compositions), and that the difference between choosing the BT2 or the Prokazatel ExoMol water line list give rise to only minor changes.


\begin{table}[hbt!]            
\label{tab:mol_opac}      

\centering
\caption{Molecular and atomic line lists input data sources}
\begin{tabular}{l l}    
\noalign{\smallskip}
\hline\hline
\noalign{\smallskip}
Molecule/atom & Reference \\    
\hline
\noalign{\smallskip}
AlCl & \citet{AlCl_Yousefi}  \\
AlF & \cite{AlCl_Yousefi} \\
AlH & \citet{AlH_Yurchenko} \\
AlO & \citet{AlO_Patrascu} \\
BeH & \citet{BeH_Darby} \\
$\mathrm{C_2}$ & \citet{C2_Yurchenko} \\
CH & \citet{CH_Masseron}  \\
CH$_4$ & \citet{CH4_Yurchenko2017}  \\
CN & \citet{CN_Brooke}  \\
CO & \citet{CO_Li} \\
CO$_2$ & \citet{CO2_Yurchenko} \\
CP & \citet{CP_Ram}\\
CS & \citet{CS_Paulose}\\
CaF & \citet{CaF_Hou} \\
CaH  & \citet{CaH_MgH_Owens} \\
CrH & \cite{CrH_Burrows} \\
FeH & \citet{FeH_Wende} \\
H$_2$CO & \citet{H2CO_AlRefaie}  \\
H$_2$O & \citet{H2O_Polyansky}  \\
HCN & \citet{HCN_Barber} \\
KCl & \citet{NaCl_KCl_Barton} \\
KF & \cite{NaF_KF_Frohman} \\
LiCl & \citet{LiCl_Bittner} \\
LiF & \cite{LiCl_Bittner} \\
LiH & \citet{LiH_Coppola}\\
MgH & \citet{MgH_Gharib} \\
NaCl & \citet{NaCl_KCl_Barton} \\
NaF & \cite{NaF_KF_Frohman} \\
NaH & \citet{NaH_Rivlin} \\
NH$_3$ & \citet{NH3_Coles} \\
NS & \citet{SN_Yurchenko} \\
PH$_3$ & \citet{PH3_SousaSilva} \\
PN & \citet{PN_Yorke} \\
PO & \citet{PO_PS_Prajapat} \\
PS & \citet{PO_PS_Prajapat} \\
SiH & \citet{SiH_Yurchenko} \\
SiO & \citet{SiO_Yurchenko}\\
SiS & \citet{SiS_Upadhyay}\\
SH & \citet{SH_Gorman}\\
SO$_2$ & \citet{SO2_Underwood}\\
TiH & \citet{TiH_Burrows}\\
TiO & \citet{TiO_McKemmish}\\
VO & \citet{VO_McKemmish}\\
Li, Be, N, O, F, Al, Si, P, S  & \citet{VALD3Ryabchikova2015}\\
Cl, Ca, Ti, V, Cr, Fe, Zr  & \citet{VALD3Ryabchikova2015}\\
C, Mg & \citet{NISTRalchenko2010}\\  
Na & \citet{Allard2019}\\
K & \citet{Allard2019}\\   
\hline                              
\end{tabular}
\end{table}

We summarize from the comparisons in Fig.\ref{FigTPg2500K} described above that the coolest end of the classical 2008-MARCS-grid models and the cloud-free, non-irradiated MSG models in the present grid are in very good agreement with one another. Our \teff\ =2500\,K, log(g)=4.5, solar composition, cloud-free, non-irradiated MSG models with our "standard" input choices are 
$\sim$25\,K cooler in the upper layers than the corresponding classical MARCS models at a given value of the gas pressure, and has a $\sim$0.1dex lower gas pressure at the formal top layer of log($\tau_{ross})=-$5.0. Further we conclude that the effect of the additional 28 molecules that were added in the MSG computations, but not included in the DRIFT-MARCS models, is marginal (at \teff\ = 2500\,K and solar composition), and that the effect of atomic line opacities only have a negligible effect on the model structure at \teff\ = 2500\,K (and cooler). We therefore conclude, to our great satisfaction, that the conclusions of the many tests and analyses of observational data that have been done based on the classical MARCS grid will remain unchanged by the introduction of the MSG grid; even for the coolest models of the published MARCS grid. It is only for the models of lower effective temperatures there are big differences between the old (unpublished) MARCS models and the new MSG models.  We do, however, caution that the large amount of different molecular line opacity sources individually gives rise to heating as well as to cooling of the atmospheric layers, and that the effect of different combinations of the available line computations demonstrated in Fig.\,\ref{FigTPg2500K} combined give rise to what might or might not be seen as an uncertainty in the results of 
$\Delta$T $\sim$70\,K and $\Delta$log($g$)$\sim$0.2\,dex in the upper layers of the models, and corresponding spectral interpretations. We have chosen to hopefully lower this uncertainty by consistently choosing all molecular line data from the ExoMol data base and all atomic line data from the DACE data base
as listed in Tab.\,1. The imposed uncertainty introduced by the choice of line profile applied to these data will be discussed in a separate section below. 


\section{The cooler MSG models} 

To a first order approximation, the physical temperature at a given atmospheric optical depth decreases almost linearly with the effective temperature. With decreasing physical temperature, an increasing number of atoms will combine into molecules. Since molecules can have electronic transitions (as atoms) as well as a cascade of associated vibrational and rotational transitions, the absorption coefficient per atom will generally increase substantially when neutral atoms combine into diatomic and polyatomic molecules. Cooling the atmosphere will therefore usually result in an increasing opacity per gram of stellar material. As a response to the increased opacity, the atmosphere will usually expand, and the gas pressure at any given optical depth will decrease. Note that $\tau_{ross}$ only has a strict physical meaning when the gas opacity is dominated by continuum sources, and therefore the "surface" or "top" of the atmosphere, defined here as where 
$\tau_{ross}$ = 10$^{-5}$, is mainly to be considered as a computational label of the atmospheric top, and the more meaningful comparison of individual models is for example what the physical temperature is at a place of a given gas pressure.

Since the input data to the classical  MARCS models were known to lack opacity sources from species that are observed in stars cooler than \teff\ = 2500\,K, it could be suspected that the models would be too compact at the lowest effective temperatures, due to the missing opacities, and the result would be not only that the computed spectra would be absent of the missing opacity sources, but the computed spectra would also be correspondingly too strong in the other spectral features because of too high gas pressure (and hence partial pressures) in the models. This would translate into the model spectra predicting lower than real abundances when used for interpreting observed spectra (at least to a first approximation), or alternatively giving erroneous estimates of the effective temperatures. A now classical and extreme example of this effect was the analysis of the hydrogen abundance in giant stars (\citet{Jorgensen1989}), and it would have been in effect for the classical MARCS models too, if they had been squeezed below \teff \,$\sim$\,2500\,K. 

\begin{figure}
             \vspace{-0.5cm}
               \hspace{-0.5cm}
               \includegraphics[width=10.0cm,angle=180.]{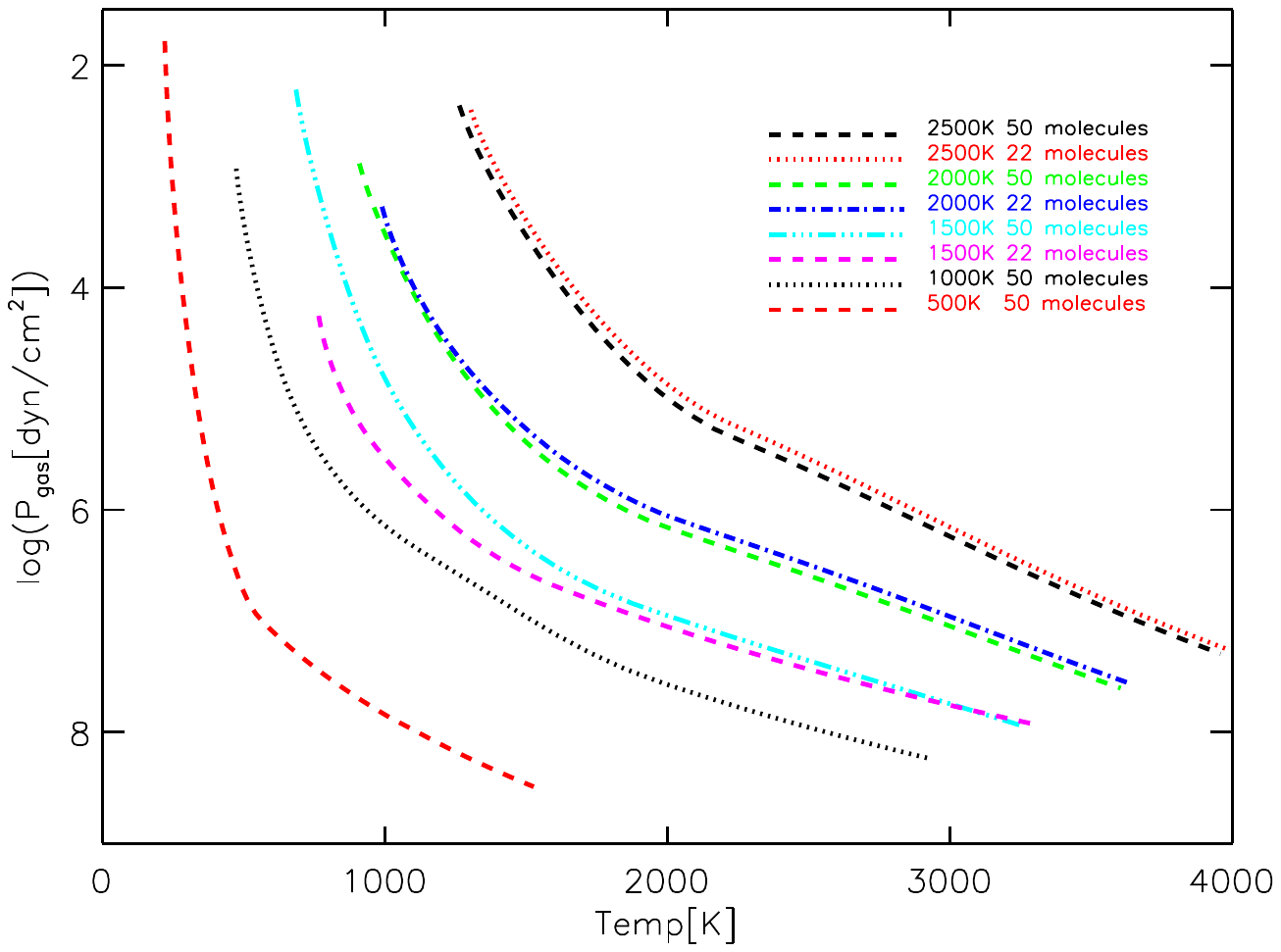}

                \vspace{-0.5cm}


               \hspace{-0.5cm}
               \vspace{-1.5cm}
               \includegraphics[width=10.0cm,angle=180.]{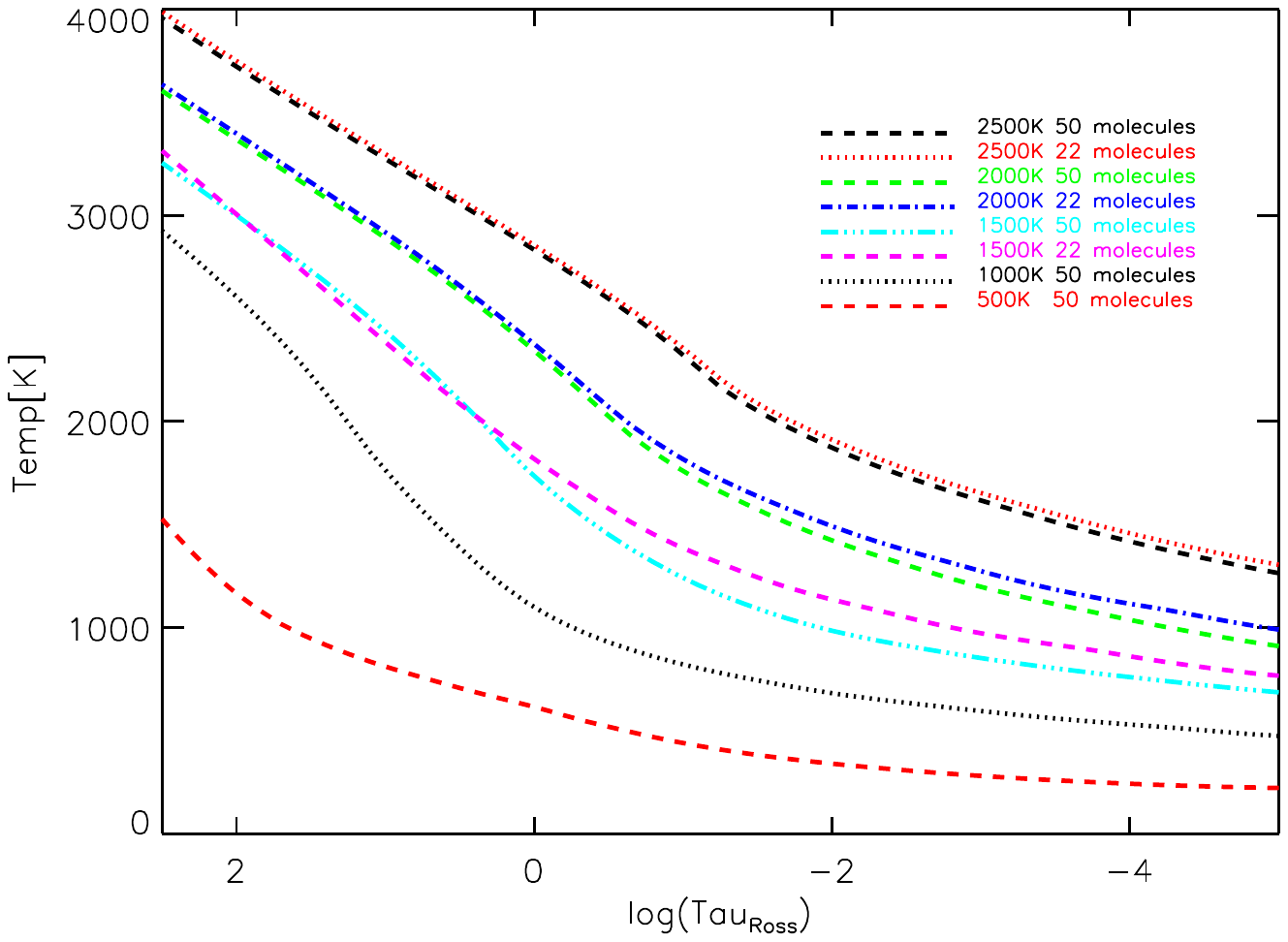}

                \vspace{1.5cm}
		\caption{The T-P$_g$ (upper panel) and T-$\tau$ (lower panel) structure from log($\tau_{\rm ross}$) = +2.5 to --5.0 of models with T$_{\rm eff}$ from
                   2500\,K to 500\,K. For each \teff\ is shown a classical MARCS model 
                    with 22 molecules in the opacity calculation as well as the models from the present grid 
                   (i.e., 50 molecules plus atoms included  in the opacity), except for the \teff=1000\,K and \teff =500K models where it has not been possibly to converge 
                   the models with only 22 molecules.
		}
		\label{FigBelowMarcs}
	\end{figure}

Fig.\ref{FigBelowMarcs} illustrates both how the physical temperature scales with the effective temperature as well as quantify the effect of insufficient inclusion of molecular opacities in the classical grid for temperatures below \teff=2500\,K. For \teff=2500\,K the $T-P_g$ and $T-\tau$ structures of the classical MARCS models (with 22 molecules in the opacity) 
and the present models (with 50 molecules in the opacity) are seen to be virtually identical. This is of course 
reassuring, since it shows that the classical grid as published (i.e., down to \teff=2500\,K) is not affected by lacking (now known) molecular opacities. For \teff=2000\,K the 
difference between the models with respectively 22 and 50 molecules in the opacity are noticeable, and for \teff=1500\,K the difference amounts to more than 100\,K all the way from $\tau_{\rm ross}$=1 and upward in the atmosphere, i.e.\ basically throughout the whole spectrum forming region.

Also derived structures such as the physical size of the atmosphere are for \teff\ = 2500\,K almost identical in the classical MARCS and the MSG grid. For \teff\ = 2000\,K, however, although the $T-\tau$ structure of the old classical MARCS and the new MSG models are quite similar, the gas pressure at the top of the atmosphere (defined as $\tau_{\rm ross}=10^{-5}$) of the MSG models are $\sim$30\% lower than in the ones simulating the classical MARCS models (for the reasons described above, due to the larger molecular opacities), giving rise to a quite different abundance determination when compared to observed data. At \teff\ = 1500\,K the MSG models are completely different from the classical MARCS ones, with the gas pressure in the top layers ($\tau_{ross} \sim 10^{-5}$) reduced by two orders of magnitude in the MSG models compared to models simulating the classical MARCS ones, and the physical size of the atmosphere (from log($\tau_{\rm ross}$) = 2.5 to --5) increased with 40\% (from 140 to 190 km for log($g$)=4.5 models). Below \teff\ = 1500\,K it was no longer possibly (nor meaningful) to attempt to converge the models with input data and computational methods corresponding to what was used in the classical MARCS models.

In general, one sees from Fig.\,\ref{FigBelowMarcs} that the temperature (and from Fig.\,\ref{FigPe-tau} the electron pressure as well) of the MSG models decrease with a larger and larger amount compared to the classical MARCS models with decreasing effective temperature,
resulting in a slight increase in the physical temperature for deeper atmospheric layers with values of log$(\tau_{\rm ross})$ above 0 (best understandable from Fig.\ref{FigBelowMarcs} lower
panel for the \teff\,=1500\,K models).
Taken isolated, these changes would result in a decreased continuum opacity and an increased molecular line (and pseudo-continuum) opacity, which isolated seen would result in an increased intensity in the predicted spectral lines and hence a lower 
abundance estimated when compared to observed spectra. 
As usual in stellar atmosphere and self-consistent planetary atmosphere theory, it is, however, not possible to give a general scaling, because the problem is non-linear.

	\section{Three "standard challenges" in self consistent cool star and planet atmosphere modelling}

We will discuss here three "standard challenges" in deriving the final gas-pressure versus temperature structure in atmospheric computations for a sufficiently wide range of input parameters to make a general-purpose grid. 
The first is somewhat 
specific to the way classical MARCS models were computed, while the two others are of more general character. 

	\subsection{The choice of independent variables}

The first challenge is to chose "the right" independent variables in the convergence scheme to make the models converge efficient and accurately for a large temperature range. 
The original MARCS code (\citet{Gustafsson1975}) introduced temperature and electron pressure as the independent variables in the iteration scheme toward  flux-constant atmospheric model structures. 
The same choice is done in for example Gray's text book on stellar atmospheres \citep{Gray2022}.
This choice is very convenient and valuable for solar type and warmer stars and even for giant stars in globular clusters, which was an early focus of the application of the MARCS models. When the stars become considerably colder, or we enter the substellar and planetary regime, however, electrons are sparse and the iteration on electron pressure becomes a challenge.

	\begin{figure}[h!]     
   \vspace{-0.5cm}
   \hspace{-1.2cm}
               \includegraphics[width=10.5cm,angle=180.]{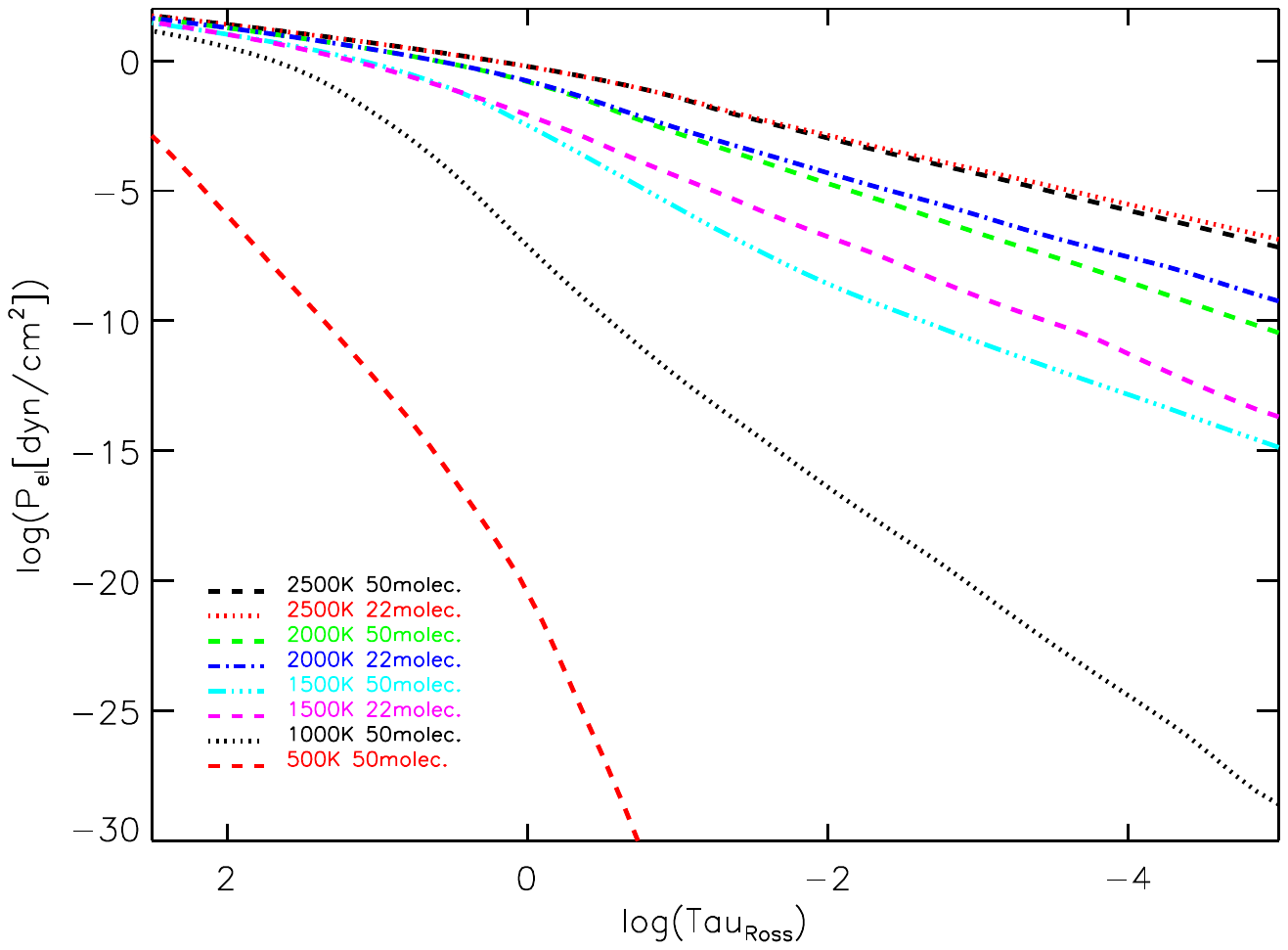}
                \vspace{-0.9cm}
		\caption{Electron pressure versus optical depth for models with T$_{\rm eff}$ from 
                    2500\,K (upper curves) to 500\,K (lower curve), computed 
                    from log($\tau_{\rm ross}$) = +2.5 to --5.0. For the warmer of the models are shown both the results 
                    based on including only the original 22 molecular opacities in the equilibrium
                    computation as well as 
                    the full set of the 50 molecules, as discussed in the text. For the coolest models  
                   only those with opacities from all the 50 molecular opacities adopted in the MSG models could converge. 
		}
		\label{FigPe-tau}
	\end{figure}

At high temperatures the opacity is dominated by continuum sources, including electron scattering and H$^-$ b-f and f-f transitions. At somewhat lower temperatures, more and more ions recombine with electrons and become neutral atoms, and the electron pressure (and therefore also H$^-$) decreases. At even lower temperatures the atoms combine to form molecules, and in chemical equilibrium the electron pressure rapidly approaches zero. Fig.\ref{FigPe-tau} shows how the electron pressure in the uppermost layers drop 20 orders of magnitude from 10$^{-7}$ dyn/cm$^2$ to 10$^{-27}$ dyn/cm$^2$ when lowering the effective temperature from \teff\ = 2500\,K to \teff\ = 1000\,K (for models of \logg\ = 4.5 and solar metallicity). While this is not a physical problem, it creates a numerical challenge in the computation of the derivative of the opacity as function of electron pressure, $\delta O /\delta Pe$, and in derived thermodynamical quantities. Since the convergence scheme in MARCS builds on $\delta O /\delta T$ and  $\delta O /\delta Pe$, the MARCS code traditionally has problems converging at low temperatures.

However, for the low temperatures, one can take advantage of the fact that almost all non-noble gas elements are bound in molecules and the small number of atoms that are not bound in molecules are basically all neutral. Hence the gas pressure is to a very good approximation the sum of the partial pressures of molecules and noble-gas atoms only (ie no electrons and no ions). The small number of free electrons can therefore be computed from a simplified Saha equation that sums over just a few atoms. Adopting the nomenclature from \citet{Gray2022} one can write the electron pressure as
\begin{equation}
P_e = \sum_j kT \cdot N_{ej}  = \sum kT\left(N_j {\Phi_j(T)/P_e \over 1 + \Phi_j(T)/P_e}\right)
\label{eqpe}
\end{equation} 
with 
\begin{equation}
\Phi_j(T) = 1.20\,{\cdot}10^9 (u_{1j}/u_{0j})\theta^{-5/2}10^{-\theta I_j}
\label{eqpe2}
\end{equation} 
\noindent
where $N_{j}$ is the number density of element $j$,
 $N_{ej}$ is the number density of electrons coming from element $j$,
 $u_{0j}$ and $u_{1j}$ are the partition functions of respectively the neutral and first ionized state of element $j$,
$I_j$ is the ionization potential from the neutral to the first ionized state of element $j$ (in units of eV). $\theta = 5040/T$ is the "temperature"  in 
units of per eV when $T$ is measured in $K$.

For temperatures where the electron pressure is very small compared to the gas pressure, the correct temperature versus gas pressure model structure is not dependent on the actual value of $P_e$ used for the convergence, as long as the correct electron pressure is used for the opacity computations. In order of convenience and to make the MSG code efficient and accurate for high as well as low temperatures, we therefore kept the convergence method from the classical MARCS models, but allowed the MSG convergence scheme at low temperatures to be a function of $P_e$, $P_{ae}$ = $f(P_e)$, instead of the actual electron pressure itself, with $f( P_e)$ computed from Eq.\,\ref{eqpe} as described above. 

Once the gas pressure has been related to the artificial electron pressure, $P_{ae}$, we use GGchem to calculate the true electron pressure, $P_e$, which is then subsequently used for all calculations of physical quantities (such as the continuum opacities, the molecular opacities etc) to estimate $\delta O(P_e) /\delta P_{ae}$. Since 
$\Phi_j(T)$ (and hence also $P_{ae}$) is a linear function of $10^{-\theta I_j}$ (Eq.\,\ref{eqpe2}) we have essentially changed the code to iterate over gas pressure instead of electron pressure, where the real electron pressure $P_e$ has now become a dependent variable instead of the artificial electron pressure $P_{ae}$. 

At the very lowest temperatures this same reasoning can be applied to allow further efficiency in the convergence by artificially lowering the ionization potential, $I_j$,  of the relevant atoms (which for solar composition is potasium only), and in this way secure efficient convergence of the model structure for all models where no physical layer reach a temperature close to the convergence criterium of GGchem at around 100\,K. For non-irradiated models this corresponds to \teff$\approx$\,300\,K, and for irradiated models it corresponds to a lower value of \teff. Once the low-temperature models have converged, the thermodynamic quantities (such as heat capacities) are now self consistently calculated using the NASA polynomials from data/nasa9.dat (\citet{McBride2002}). This assures that 
the correct adiabatic index (and hence boundary between layers in radiative and convective equilibrium), and the 
correct gas pressure, etc, are computed at low temperatures, while still allowing high accuracy by include both
single and double ionized atoms (and only few molecules) at
the highest temperatures, as in the original MARCS code. In this way the flexibility, computational speed and high
accuracy in the convergence to a final T-P$_g$ structure is assured for a sufficiently wide range of \teff\ to compute 
the exoplanetary system and its host star together with the same code and input data.

	\begin{figure}[h!]  
   \vspace{-0.4cm}
   \hspace{-1.5cm}
   \vspace{-1.1cm}
               \includegraphics[width=9.2cm,angle=180.]{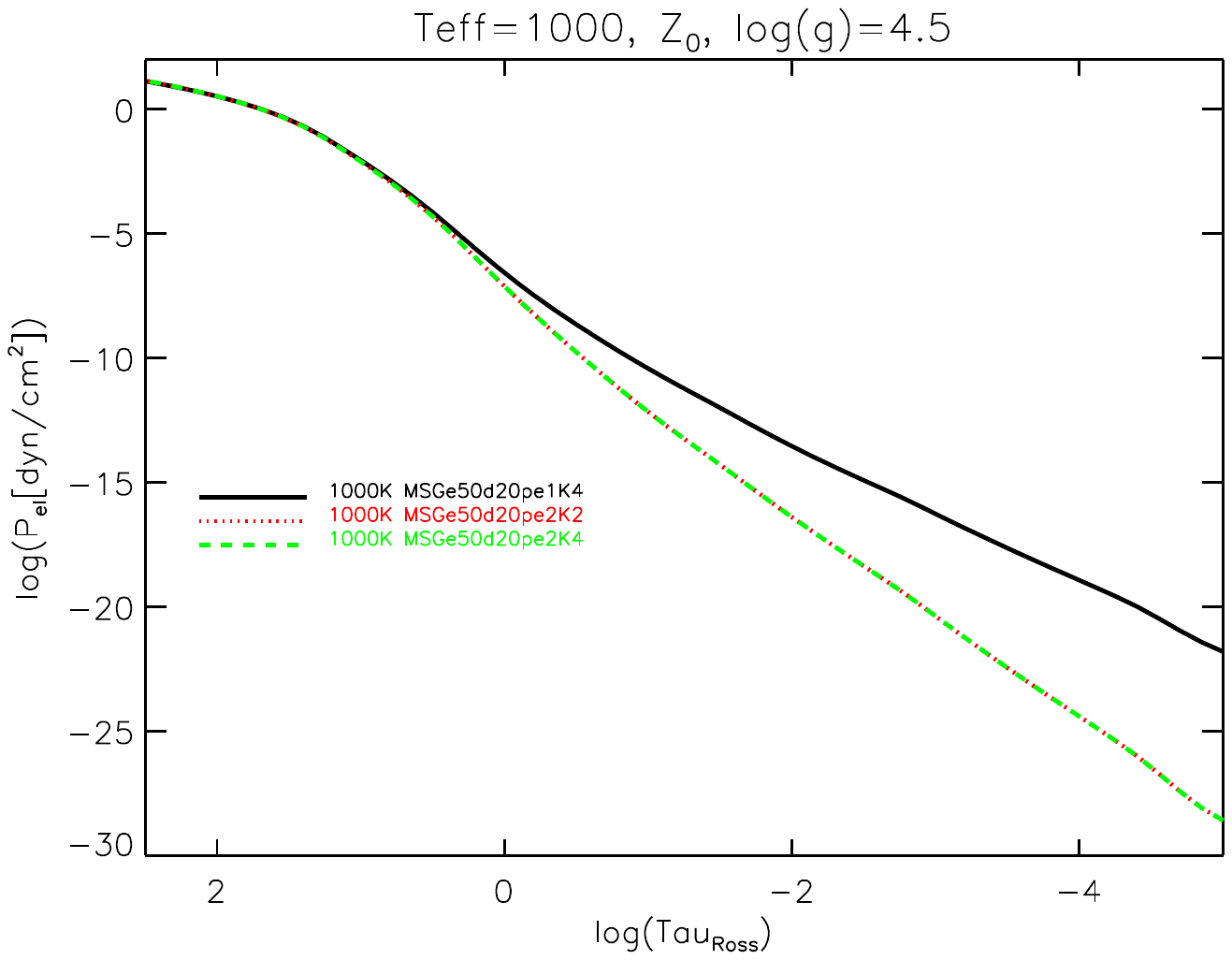}
   \vspace{-0.8cm}
   \hspace{-0.2cm}
               \includegraphics[width=9.2cm,angle=180.]{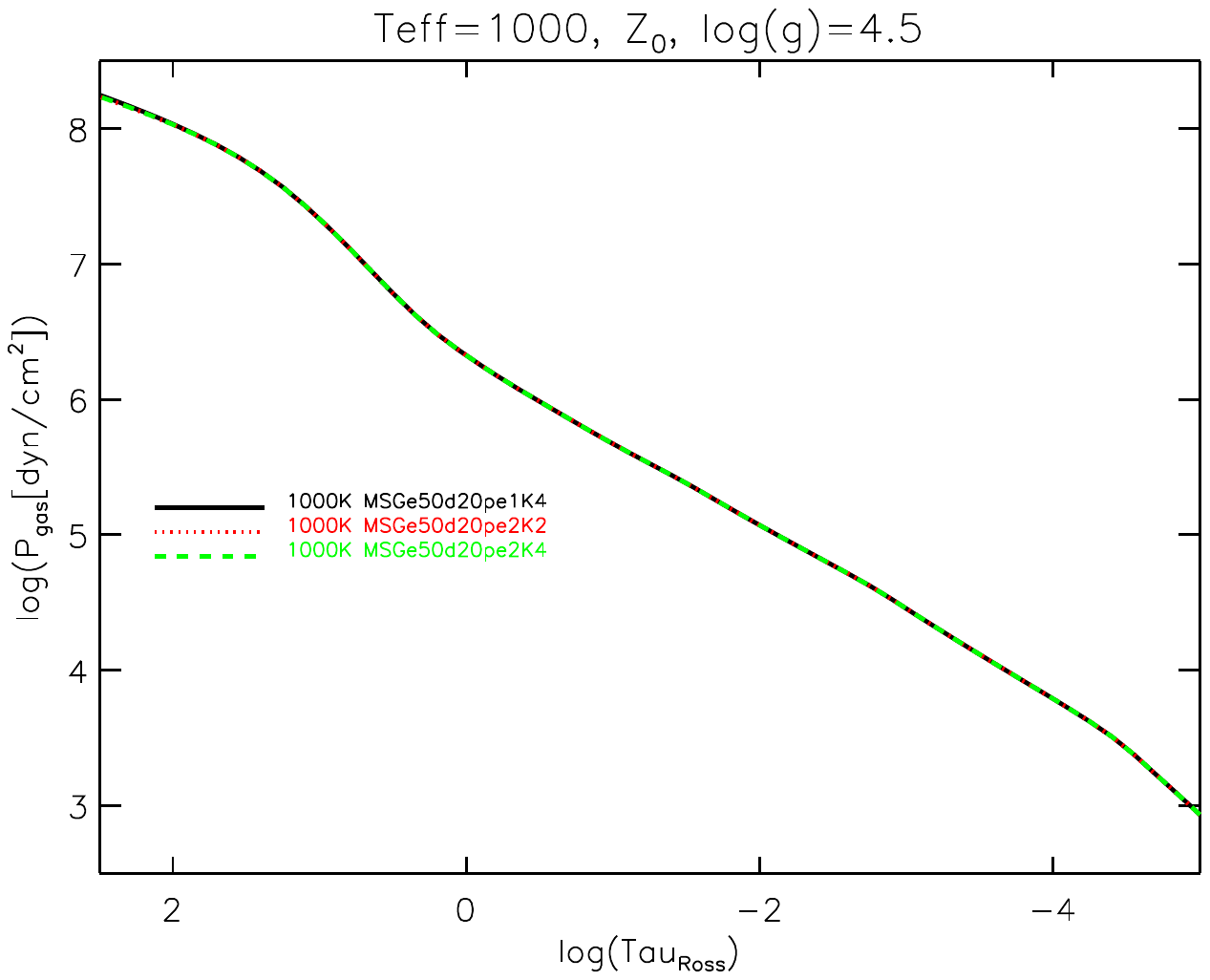}
               \vspace{0.2cm}
		\caption{Models with T$_{\rm eff}$ = 1000\,K computed based on  the "old" way (labeled "...Pe1K4") of estimating 
                      the electron pressure,  as well as the new way with $I_j$ equal to 2.339\,eV (labeled "...Pe2K2) 
                    and 4.339\,eV (labeled "...Pe2K4), respectively.  The upper and lower panels show, as function of 
                   log($\tau_{ross}$), respectively P$_e$ and P$_g$.
		}
		\label{FigPae}
	\end{figure}

 Several tests have assured us that for the models that could converge with both options in handling of $P_e$ the resulting structures were identical, 
as is illustrated in Fig.\,\ref{FigPae} where the computed electron pressure as function of $\tau_{Ross}$ is seen to be 7 orders of magnitude higher for 
$P_{ae}$ than $P_e$ (upper panel) in the "old convergence scheme" (models labeled "...p1...") than the new (which is further seen to be independent of the assumed value of the ionization potential of potasium, as it should), but yet, as expected and explained in the text above, leading to exactly the same $T-P_g$ model structure (as is seen in the lower panel).

	\subsection{The number of transitions to include in the opacity computation}

A second "standard assumption" one needs to take when computing a self consistent atmospheric model, is 
how many transitions to include in the opacity computation (unless one just adopt an opacity table accessibly on the web, but then somebody else have take the decision instead). Choosing too few lines will make the outgoing radiation escape between the lines, with the effect that the atmospheric structure becomes too compact. A too compact model will not have too much effect on the observed spectrum of the dominating 
opacity source(s) that is (are) computed wrong. This is because the computed atmospheric structure will balance 
itself such that the hydrostatic equilibrium, flux constancy, and energy balance is preserved, and hence also the emerging flux from the top of the atmosphere (the spectrum). However, the opacity from the sources (mainly molecules because they for cool objects carry most of the opacity) that are not dominant will be wrong, often by a substantial factor, and the 
sensitivity to for example abundance analysis will therefore often be sensitive to the choice of number of lines included in the opacity
computation of the model structure used for the analysis. On the other hand choosing too many lines can have the opposite effect. The problem is the same concerning the choice of line profiles that we will return to below.
One could be tempted to think that one can never choose too many lines, because one could just choose "all of the lines" and be on the safe side. The problem here is that there is no such concept as "all the lines". 
The following will illuminate why such a concept does not exist.

In modern quantum mechanics, transition moments ("intensities of electronic, vibrational or rotational transitions")
are computed as integrals over the corresponding wavefunctions, but the factors that define how many lines a 
given electronic or vibrational transition ends up producing relies on the quantum numbers associated to the 
eigenstates as well as the definition of the energy of the highest energy level in the energy potential,
both of which relies on quantum numbers from classical mechanics derived from the harmonic oscillator 
theory or some more or less empirical modifications of it. The same is true if one attempts to evaluate the completeness of an observed linelist.The highest eigenstates are most affected by the incorrectness of the harmonic oscillator theory, and one will often argue that since the highest eigenstates are least populated in a Boltzmann distribution, then it is probably not too important to compute them accurately, or one could just disregard them.
In the pure harmonic oscillator theory, there are infinitely many lines originating from the uppermost states in the potential, each approaching infinitely small intensity. The "unimportance-argument" is probably true concerning the total integrated absorption coefficient, but not necessarily concerning how much energy escapes between the included lines, which affect the model structure in a complex and non-linear way which is not straight forward to predict. The only realistic way forward may be a rather cumbersome comparison between high-resolution observed spectra (which has to include transitions from high and low excitation energies in opacity-dominant as well as opacity un-important sources) of high-temperature objects (stars and planets) of various types compared to theoretical results based on different attempts to do something "reasonable". 
We tested and described in detail in \citet{Jorgensen2001} the numerical effect on the model structure of successively including from a few million water lines to a few billion water lines, and we described in detail in \citet{Popovas2016} various choices one can make for including or excluding lines from excitation energies above a certain level, and calculated as an example what could be a reasonable choice for H$_2$, which was later tested and found in good agreement with laboratory experiments for H$_2$ (\citet{Retter2020}).

	\subsection{The choice of line profiles in the opacity computation}

A third "standard assumption" one needs to take when computing a self consistent atmospheric model,
is the selection of line profiles in the opacity (and spectrum) computation. This challenge is quite analogue to the 
"second standard assumption" described above, since it also concerns how much flux one chose to allow the 
object to let escape between the lines (in the opacity).

In all computed MARCS atmospheric models so far, the molecular line profiles were treated as Gaussians, ie. that the wings of the full Voigt profile were ignored.
If one assumes that a line list from a given molecule (or the sum of line lists of all the molecules included in the computation of the model structure) includes for example a billion lines from somewhere in the visual region to $\lambda$ equal to say 10 microns, then we have 10,000 lines per angstrom.
With a microturbulence dominated Gaussian line width of say $\xi$ = 3 km/s, we then have of the order of 1000 lines per Gaussian halfwidth. Intuitively, one could guess that the sum of the opacity of such a huge number of even relatively weak line cores would always dominate over the wings of any even relatively strong single line. This would intuitively justify the exclusion of the line wings, but it has never really been possible to test this assumption. 

One reason for excluding the wings is that they are temperature and pressure dependent, while 
the Gauss profile is temperature dependent only. One therefore save substantial computing time (and core memory) by limiting the computation to Gaussian profiles. Another reason for doing it was of more practical character, namely that the necessary data for computing the full Voigt profile for the huge number of lines were unavailable. This problem was overcomed with the creation of the ExoMol data base, and the test of the "correct" line profile is therefore now in principle possible. Opacity computations based on (almost) the full list
of transitions in the ExoMol data base  has been done by us for the present work assuming Gaussian line profiles 
with $\xi$ = 3km/s, and by \citet{Chubb2021} for Voigt profiles with $\xi$ = 0km/s. The use of the two set of 
line lists in the opacity computation leads to differences in the model structures somewhat larger than the differences in the model structures based on Gaussian profiles with respectively 2km/s and 5km/s that 
in the classical MARCS models were implicitly assumed to encompass the uncertainty due to line profiles.
It is not obvious which of the two existing approaches (excluding the Lorenz broadening or excluding the microturbulence) are most severe (most likely none of them are really "correct") and which values of the microturbulence  are most correct for the sub-stellar objects, only that the difference between the two approaches seems to give rise to larger structural differences than the choice of $\xi$=2 and 5km/s do.   
A  proper solution may require an observational estimate of the line widths as function of
sub-stellar parameters, such as it was fruitfully done through decades of observational studies for stars.

\begin{figure}
              \vspace{-0.3cm}
              \hspace{0.8cm}
              \includegraphics[width=12.6cm,angle=-90.]{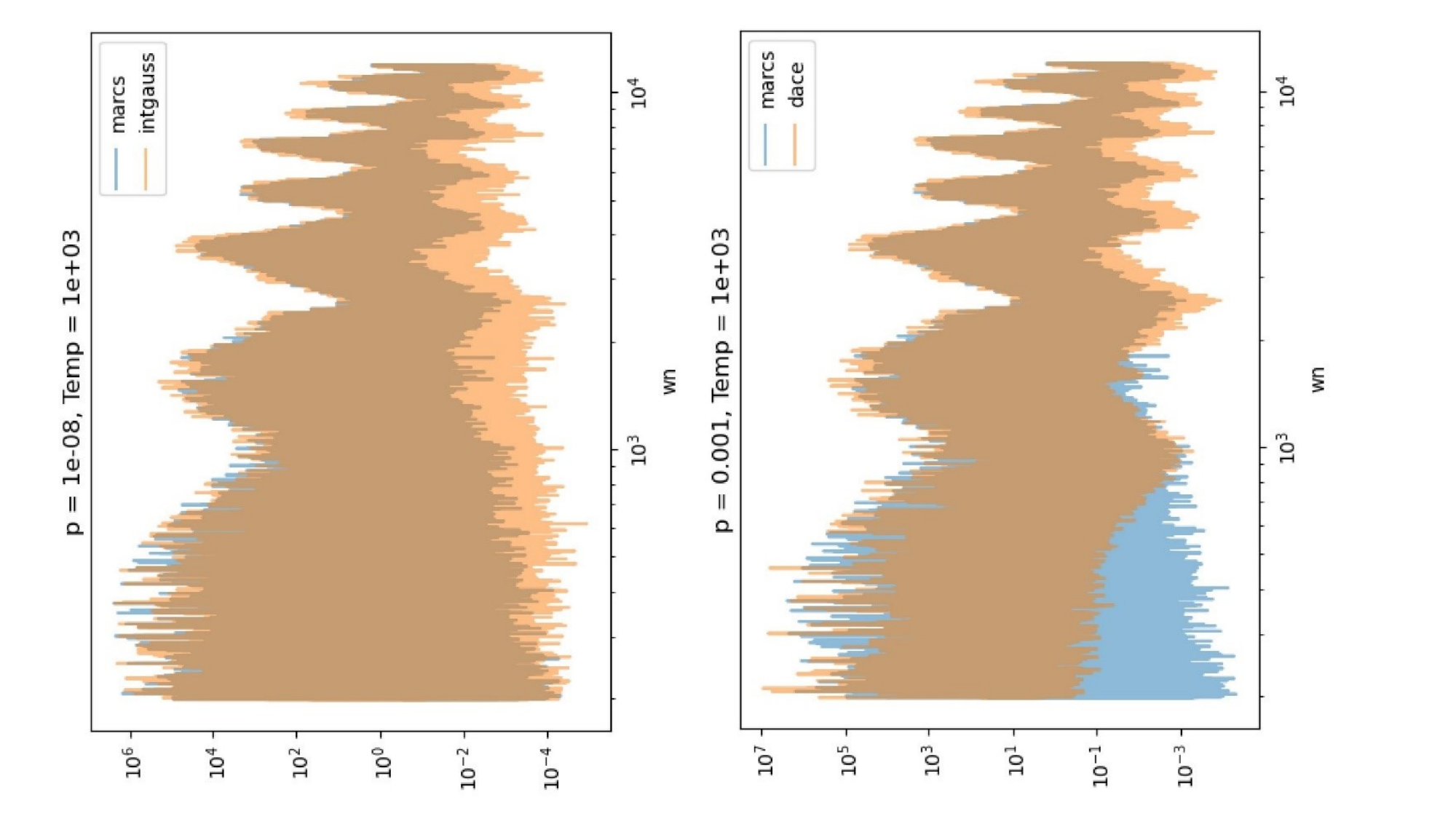}

              \vspace{-1.2cm}
              \hspace{0.8cm}
               \includegraphics[width=13.2cm,angle=-90.]{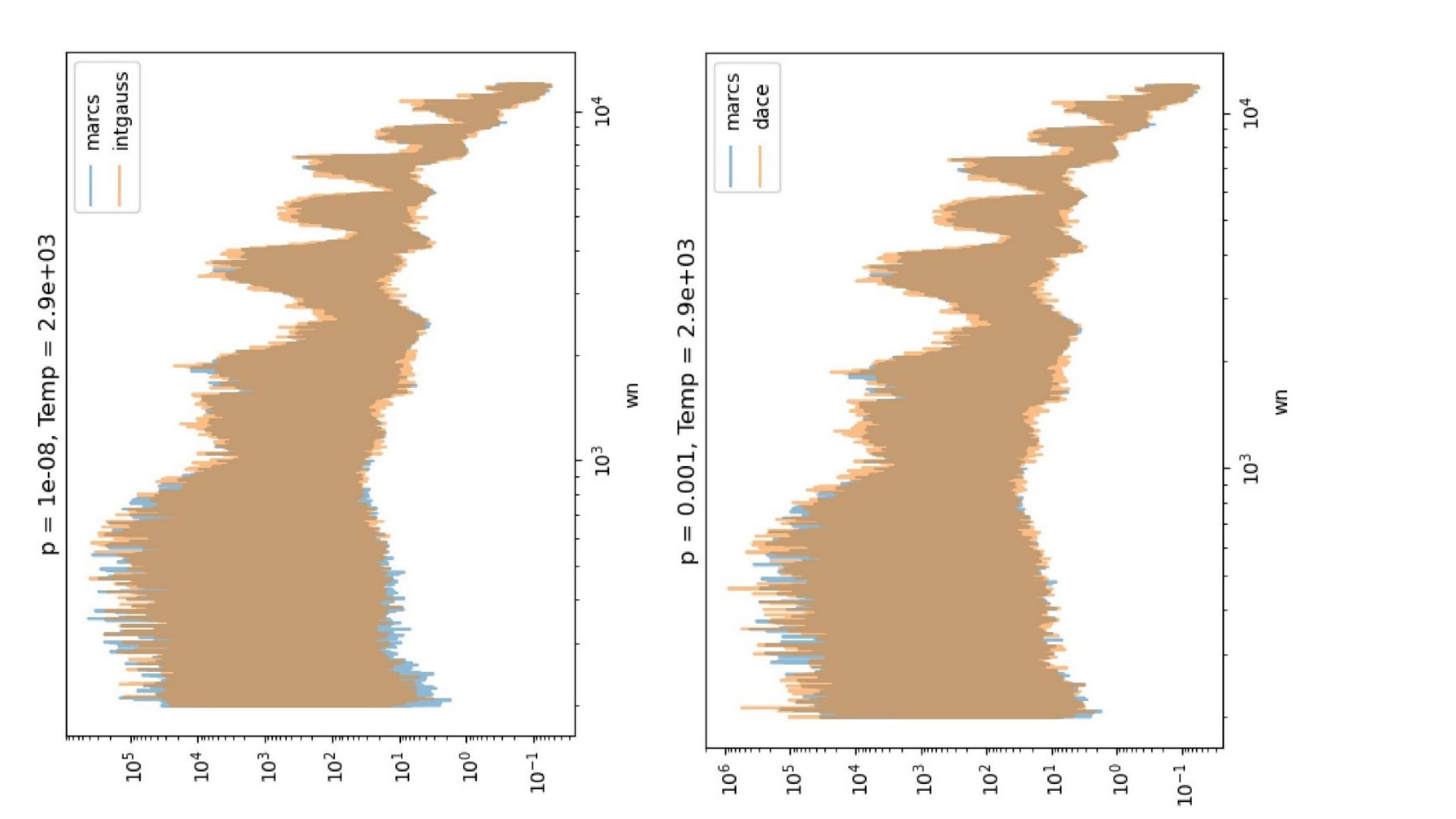}

                \vspace{-1.4cm}
		\caption{The OS line absorption coefficient as function of wavenumber, computed from the ExoMol linelist 
                  under the assumption of Gaussian molecular line
                  profiles (blue) and Voigt profiles (light brown), respectively, and for two different values of temperature 
                  (T = 1000\,K and 2900\,K), and two different values of P$_g$ (10$^{-8}$ and 
                  10$^{-3}$ dyn/cm$^2$). }
	\label{FigProfileOS}
	\end{figure}

Fig.\,\ref{FigProfileOS} illuminate the complexity of the problem. For low temperatures (here 1000K) and low gas pressures (here 10$^{-8}$ bar), the Lorenz broadening gives rise to a larger blanketing at all wavelengths, whereas for high temperature and high pressure (here 10$^{-3}$ bar) the effect of the microturbulence broadening of the Gauss profile on the blanketing strongly exceeds the effect of the Voigt wings at wavelengths below 1$\mu$m. For higher temperatures (here 2900K) the effect is opposite, but also much smaller. Qualitatively this shows us that we will have blanketing from the microturbulence line broadening mainly in the deeper layers (short-wavelength Gaussian blanketing at high pressures) and therefore cooling of most of the atmospheric structure, while the effect of the Lorentz wings in the Voigt profile will 
dominate at lower pressures and therefore block more light in the upper layers and backwarm the bulk of the lower atmosphere. Generally we therefore expect that excluding the microturbulence cools the models a bit (the lines or
line cores get more narrow, and more radiation therefore escapes between the lines), while substituting the Gauss profiles with Voigt profiles heats the models a bit (there are fewer regions with ultra-low opacities, so more radiation is blocked). Experimenting with the two sets of opacities showed us that the effect is larger for the cooler models than for the warmer models. One has, however, to be cautious to generalize this conclusion because the effect is complex and non-linear, but it could be because fewer levels of the individual molecules are populated at lower temperatures, but on the other hand there are also more different molecules present. We also noted that the bulk of the effect generally moves downward in the models when one goes from warmer to colder models and that for the cooler models introducing Voigt profiles heats the inner structure and removing microturbulence cools the inner structure, which most likely reflects a temperature dependence of the degree of backwarming due to the pseudo-continuum of the veil of weaker lines. It could therefore be that the effect of lacking modelling of the Lorenz wings might have had a 
somewhat larger effect in the upper layers of existing models than we have usually assumed.

	\section{The sampling method and the changing partial pressures}
    \label{sec:PartialPressures}

In order to transform the line lists into molecular line absorption coefficients and further into a total monochromatic molecular absorption coefficient  (e.g.\ in units of \cms/molecule = cm/molecule/\cmm) that can be used for the radiative transfer calculations, some kind of statistical approach to the resulting opacity  (e.g.\ in units of \cms/g$_*$, where g$_*$ is grams of stellar material) has to be adopted, which here (as in all MARCS models computed after $\sim$1990) is chosen to be the Opacity Sampling scheme (as described for example in \citet{Jorgensen1992a}). We typically use a sampling density of $R = \lambda/\Delta\lambda$ = 15,000 which is sufficient for medium resolution spectrum calculations, and then sample each 10$^{th}$ point for the radiative transfer calculation. In the OS approximation the line frequencies for the transfer computation has to be chosen randomly, but this is well satisfied with a fixed step length as long as the line positions can be regarded as random relative to the chosen OS frequencies. 

The chosen sampling is a good compromise between speed of computation and accuracy of the final model structure, and it corresponds to solving the radiative transfer problem in $\sim$10,000 frequency points. The computing time scales approximately linearly with the number of frequency points. Such sampling gives essentially the same result as the ODF method (which has some similarity to the often used correlated K-method; \citet{Goody1989}) that was used and described in the oldest versions of the MARCS code (\citet{Gustafsson1975}). The OS method does not have the problem of lacking frequency correlation 
between individual species that the ODF method has (\citet{Saxner1984}), and is therefore more flexible and considerably faster when many opacity species are included, and it is relatively easy to estimate how many frequency points are needed to reach a correct solution (\citet{Helling1998}). It further has the advantage that the same OS's can be used consistently in the model iteration and the spectrum computation.

	\begin{figure}
               \hspace{-1.0cm}
               \includegraphics[width=10.7cm,angle=180.]{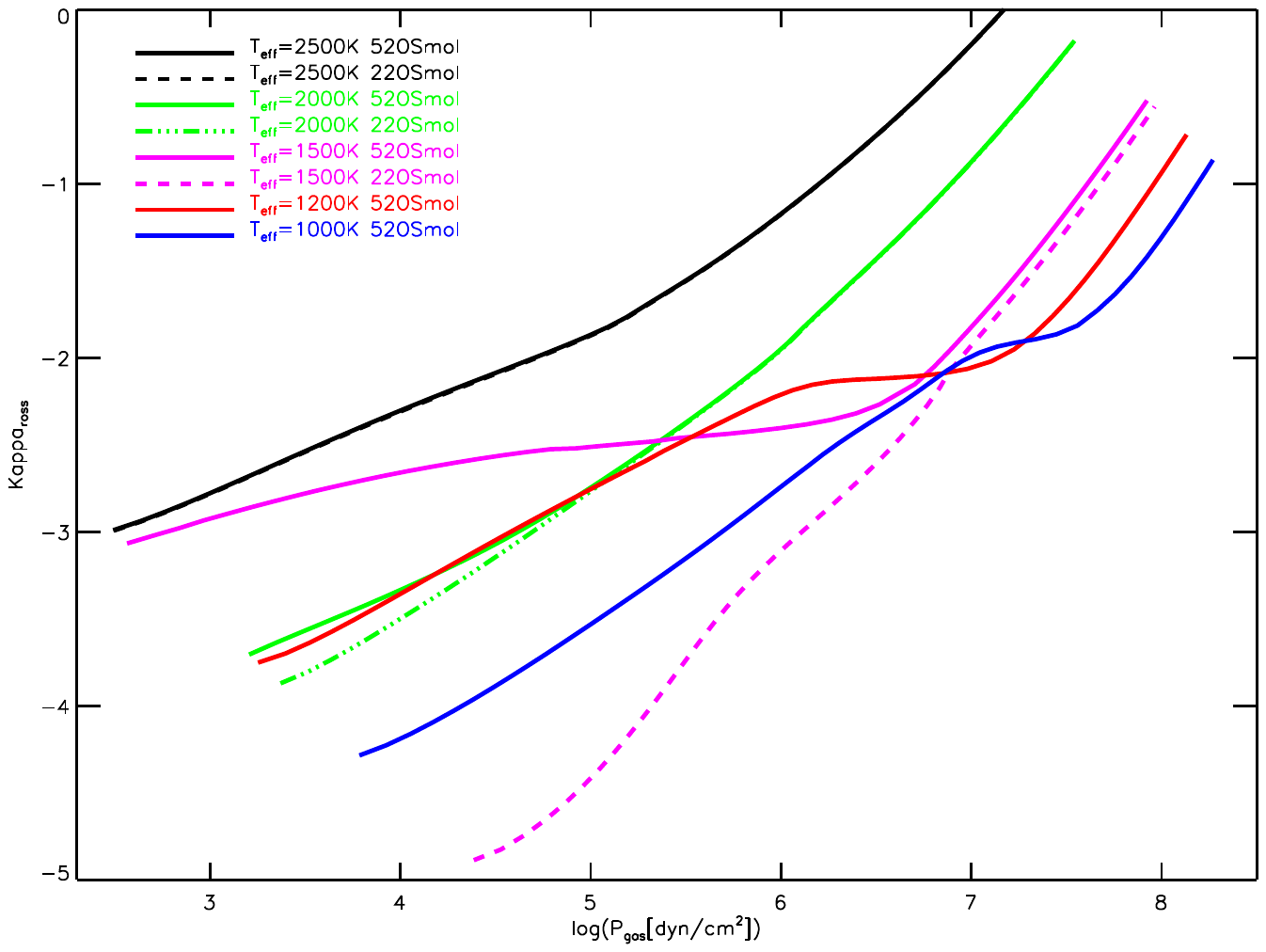}
                \vspace{-0.5cm}
		\caption{The Rosseland mean opacity as function of gas pressure, for models with T$_{\rm eff}$ from 
                 2500\,K (upper curves)  to 1000\,K (lower curves), based on the full set of 52 molecular opacities (full drawn lines)
                and the original 22 molecular opacities, respectively, in the radiative transfer. 
                 Colour coding correspond to the effective temperature of the models, as shown in the figure legends. 
		}
		\label{FigKapPg}
	\end{figure}


Once the individual monochromatic absorption coefficients are calculated and tabulated, it is then simple to compute the individual molecular opacities in the sampling points by multiplying with the individual partial pressures (for example in units of number density of species per number density unit of stellar mixture, $\rho_{\rm molecule}/\rho_*$, which when applying the ideal gas equation as the equation of state is identical to the molecular partial pressures divided with the stellar gas pressure) such that the opacities of the individual species becomes in units of per stellar mixture (for example cm$^2$ per gram of stellar material). These opacities can then be added linearly, since they are all in units of per gram stellar material, and the computing time of the stellar opacity therefore scales only approximately linearly with the number of opacity sources included. 

Fig.\,\ref{FigKapPg} shows the total opacity as function of gas pressure in units of \cms\ per gram of stellar material. The figure illustrates well the reason behind the changes in the $P_g-\tau$ structure described above and the changing $T-\tau$ structure shown in Fig.\,\ref{FigBelowMarcs}. It is seen that for the lower limit (\teff\ = 2500\,K) of the classical 2008-grid, the inclusion of the many new molecules from ExoMol compared to the only 16 molecules included in the classical grid, does not change the stellar opacity at any depth of the atmosphere, and hence the model structure does not change either. For \teff\ = 2000\,K the opacity changes with approximately a factor of 2 (0.3 dex), while for \teff\ = 1500\,K the new opacities are 2 orders of magnitude larger than the old ones. For \teff\ = 1000\,K we must expect the differences to be even larger, and it is somehow reassuring that it was not possible, as described in connection with Fig.\,\ref{FigBelowMarcs}, to converge models with such unrealistic input data.

It should be stressed that the opacities shown in Fig.\,\ref{FigKapPg} are not physical entities that as such could be measured at any place  in the atmosphere. They are the Rosseland-mean opacities, arising from the sum of the individual monochromatic absorption coefficients multiplied with the respective partial pressures for the relevant molecules, which only has a physical meaning for non-monochromatic absorption coefficients. They are therefore only illustrations of the approximate reason for the changing model structure as function of effective temperature. While the real opacity that goes into the model structure computation is monochromatic (in the OS statistical sense), the Rosseland opacity (sometimes a bit misleading called the Rosseland mean absorption coefficient, 
$\kappa_{\rm ross}$), is one over the integral of one over the monochromatic opacity times the planck function.  It is a bit like the way we talk about thin and thick cloud cover on Earth, meaning how the opacity is qualitatively sensed with our eyes, through the sensitivity function of our eyes multiplied with the intensity distribution of the sun light. It is therefore also to be understood that the two orders of magnitude decrease in the gas pressure in the top of the atmosphere mentioned above, because of the 100 times increased value of the opacity seen in  
Fig.\,\ref{FigKapPg} for the new linelists, is influenced by the definition of "top of the atmosphere", but the figure serves well its illustrative purpose.

	\begin{figure}[h!]
               \hspace{-1.0cm}
               \includegraphics[width=10.7cm,angle=180.]{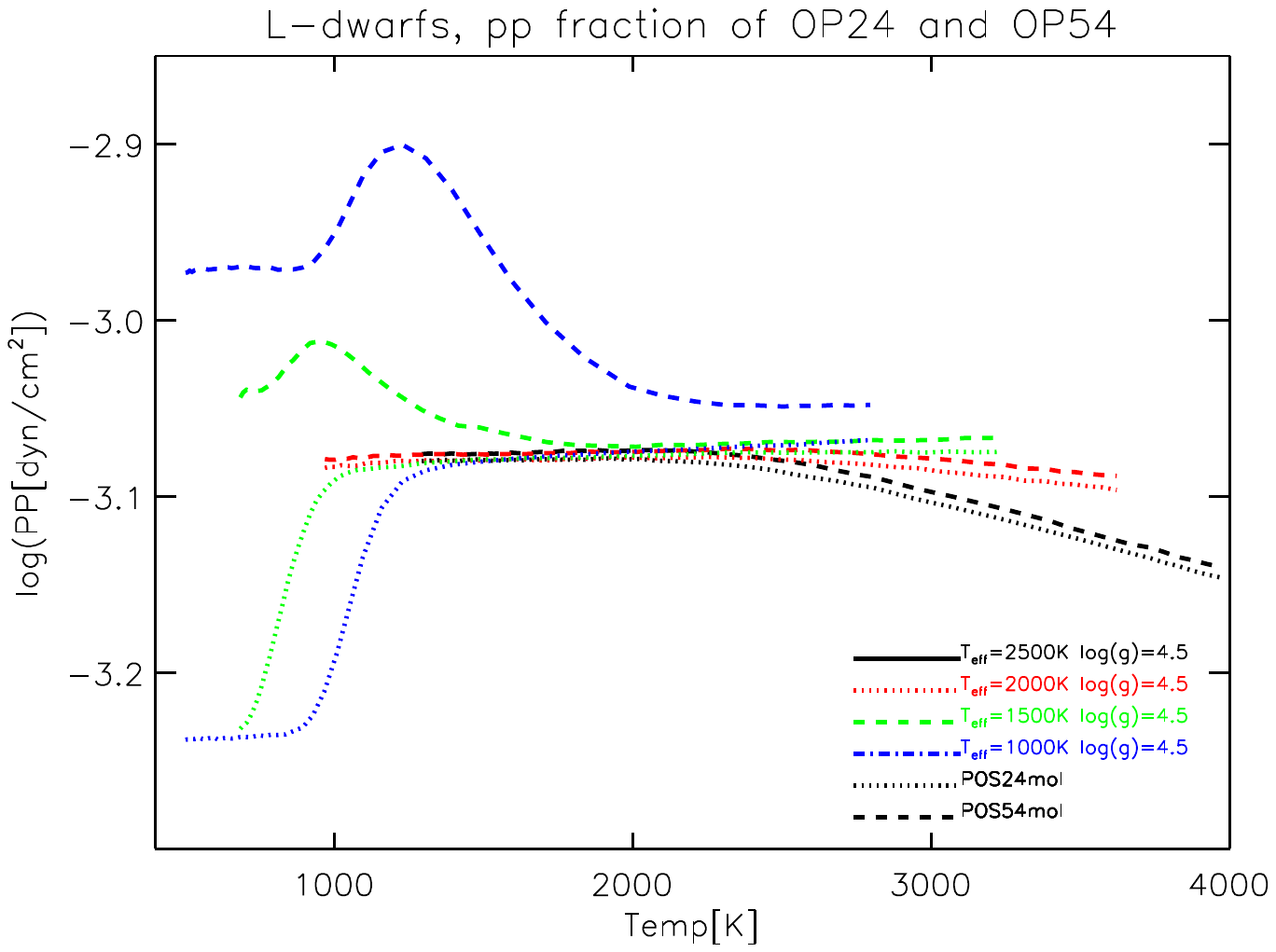}
                \vspace{-0.5cm}
		\caption{The logarithm of the sum of the partial pressures (minus the partial pressure of H$_2$, 
        and normalized to the gas pressure)
        of respectively the original 22 opacity bearing molecules (dotted lines) and 
        the 22 plus the new 30 ones (dashed lines), for the new models presented here of T$_{\rm eff}$ = 2500\,K
        (black lines), 2000\,K (red lines), 1500\,K (green lines), and 1000\,K (blue lines). 
		}
		\label{Figp24p54}
	\end{figure}

As described above, the opacity consist of a product between the absorption coefficients and the corresponding partial pressures. Fig.\,\ref{Figp24p54} shows the sum of the partial pressures of the 22 molecules that went into the computation of our classical DRIFT-MARCS models, compared to the sum of the partial pressures of the full set of 50 molecules introduced in the present grid, for the new models of \teff\,=2500\,K, 2000\,K, 1500\,K and 1000\,K, respectively. The effect on the final atmospheric structure depends both on how large the partial pressures are and how strong the absorption coefficients are, and sensitively on how the absorption coefficient is distributed as function of frequency, but the relative change in the partial pressures between the models gives a good feeling for the reasons behind the changing model structures. 

H$_2$ is the dominant contributor to the gas pressure at the temperatures and compositions discussed here, but H$_2$ in itself  has a very small absorption coefficient concentrated on a few narrow frequency intervals. In order to illustrate most clearly the effects of the 28 new molecules added to the present grid, the sum of the partial pressures of the 22 opacity bearing molecules in the old grid and the 50 in the new grid are therefore shown as the sum of these minus the partial pressure of H$_2$ (normalized to the gas pressure of the respective models). 
Figure \ref{Figp24p54} indicates, as expected, that the sum of the 22 partial pressures (black dotted line) and 50 partial pressures
(black dashed line), defined as described above, are almost identical for the (new) model of \teff\ = 2500\,K, while for the (new) 
models of 1500\,K (green lines) and 1000\,K (blue lines) the sum of the partial pressures of the 50 opacity bearing species (dashed lines) is substantially higher than the corresponding sum of the 22 molecules (dotted lines) in the classical models, in particular for the upper and coolest atmospheric layers at temperatures around and below 1000\,K.

	\begin{figure}[h!]
               \vspace{-0.8cm}
               \hspace{-1.0cm}
               \includegraphics[width=10.7cm,angle=180.]{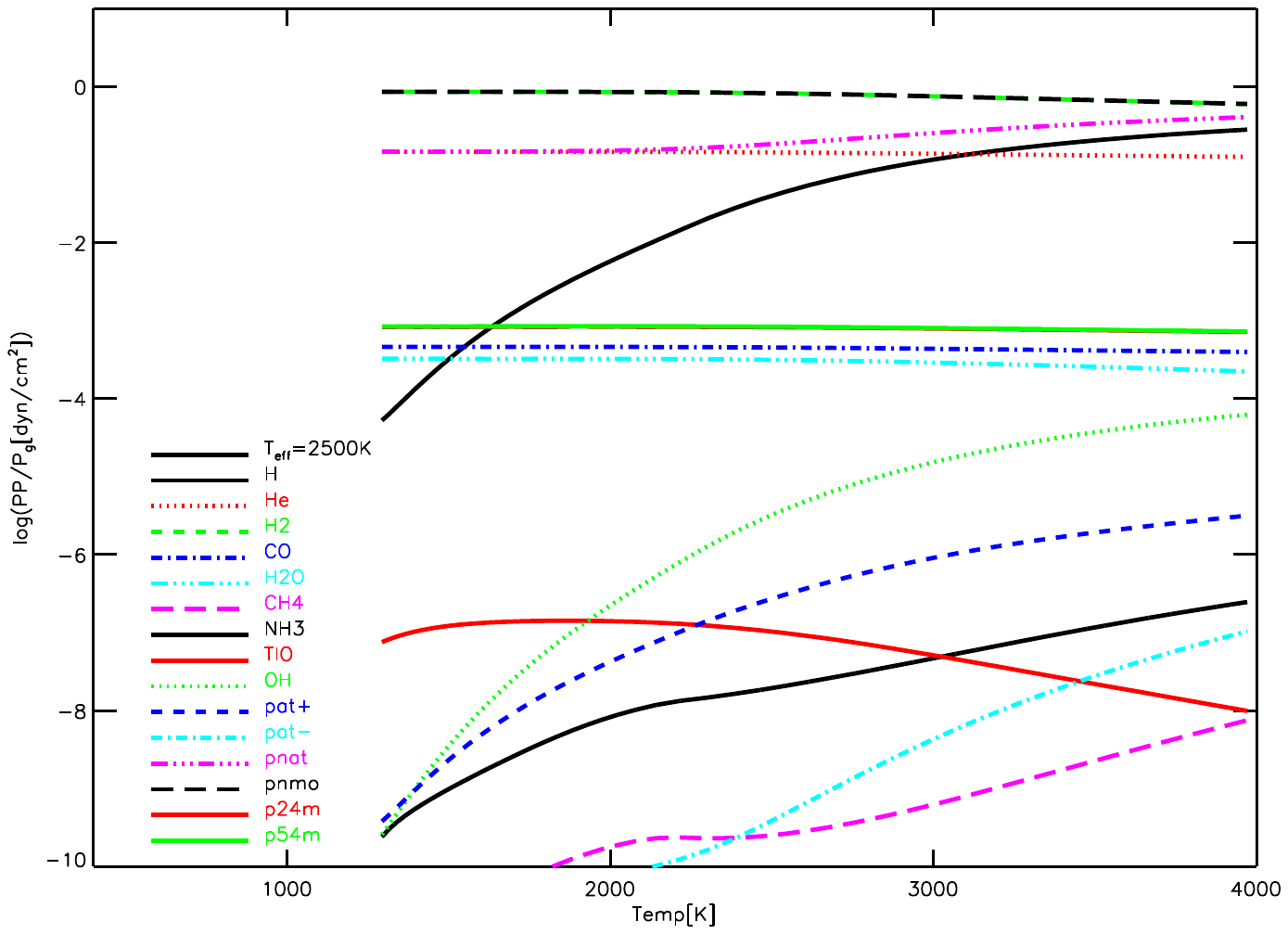}
                \vspace{-0.8cm}
		\caption{Distribution of the total gas pressure onto its various components, for a
          model of \teff\,=2500\,K. As indicated with the legend, the curves show the partial 
          pressures of the molecules H$_2$, CO, H$_2$O, CH$_4$, NH$_3$, TiO, and OH,
          the neutral H and He atoms, the sum of all positively charges species ("pel+"), 
          all negatively charged species ("pel-"), sum of the partial pressures of all neutral 
          atoms together ("pnat"), all neutral molecules ("pnmo"), and finally the sum of all 
          the opacity bearing molecules in the classical MARCS models ("p24m") as well as in the 
          new models presented here ("p54m"). Note that the curves for H$_2$ and all neutral molecules (pnmo) 
          as expected plot on top of one another, as do the curves p24m and p54m (therefore p24m/p54m show 
          as a single full drawn green curve, and the only red curve is for the partial pressure of TiO). 
         The upper black curve is for H and the lower for NH$_3$
		}
		\label{FigPP2500}
	\end{figure}

	\begin{figure}[h!]
               \vspace{-0.5cm}
               \hspace{-1.0cm}
               \includegraphics[width=10.7cm,angle=180.]{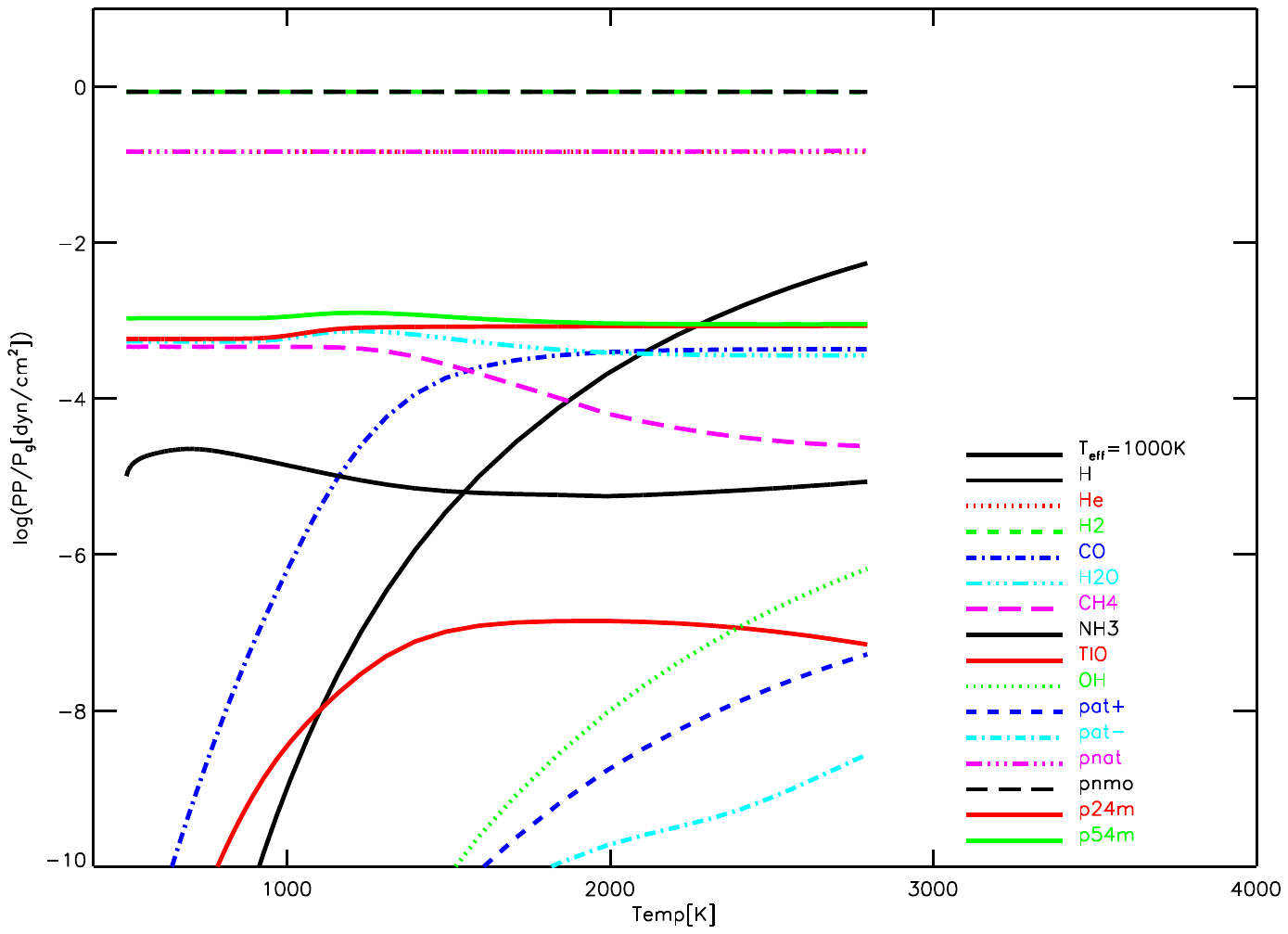}
                \vspace{-0.7cm}
		\caption{Same as Fig.\,\ref{FigPP2500}, but for at model of \teff =1000\,K.
         Note that for this model the curves p24m and p54m do deviate, in particular for the upper layers, illustrating 
          the importance of of the additional molecular opacities in the MSG models compared to the classical MARCS models.
		}
		\label{FigPP1000}
	\end{figure}

Figure \ref{FigPP2500} and Fig.\,\ref{FigPP1000} show how the distribution of the gas pressure onto various individual species develops from \teff\,=2500\,K (Fig.\,\ref{FigPP2500}) to \teff\,=1000\,K (Fig.\,\ref{FigPP1000}). From top to bottom in the two figures, 
the long-dashed black line and the short-dashed green line show the sum of the partial pressures of all the neutral molecules together and the partial pressure of H$_2$ alone, respectively. For the \teff\ = 1000\,K model these two lines are identical and practically equal to the total gas pressure (log($pp/p_{\rm gas})\approx$0), while for the \teff\ = 2500\,K model they are also identical but deviate a bit from the total gas pressure in the deeper layers of the model because of the relative increase in the sum of all neutral atoms
(dash-triple-dot purple line) which mainly consist of hydrogen (full drawn black line) and helium (yellow dotted line) in solar metallicity models. 

With 3 orders of magnitude lower partial pressures we find (in Figure \ref{FigPP2500} and \ref{FigPP1000})       
the sum of the partial pressures of all the 22
molecules that contribute to the opacity of the classical models (full drawn red line) and the corresponding 50 in the new models (full drawn green line). In the \teff\ = 1500\,K (not shown) and \teff\ = 1000\,K (Fig.\,\ref{FigPP1000}) models the partial pressures of the 50 molecules (minus H$_2$) are of the order of twice that of the original 22 opacity carrying molecules in the old models. It is worth noticing how small a fraction of the total number of molecules per \cmq\ of the stellar gas that actually carries all of the opacity and hence determines the structure. For solar metallicity objects it is approximately 0.1\%\ of the gas pressure that carries 100\%\ of the opacity and for metal poor stars this percentage is correspondingly lower. This is in fact approximately the same in the Earth's (cloud-free) atmosphere where N$_2$ and O$_2$ carries approximately 99\%\ of the gas pressure but almost no opacity (because of their homonuclear structure). It is the reason for why so relatively small changes (in temperature or chemical composition) have so dramatic consequences for the structure (for stars as well as for Earth).

Right below the two lines with the sum of the 22 or 50 opacity bearing molecules (which are identical to one another in the \teff\ = 2500\,K case) we find the partial pressures of CO and H$_2$O. In the \teff =2500\,K model the relative partial pressures of CO and H$_2$O are an almost constant fraction of the gas pressure at all atmospheric depths, with CO dominating, while the ratio of the two form a more complex pattern in the cooler models, as is well illustrated at \teff\,=1000\,K. 

It is seen by comparing Fig.\,\ref{FigPP2500} and \ref{FigPP1000}
that while the partial pressure of CH$_4$ (methane) is orders of magnitudes below those of CO and H$_2$O in the \teff\ = 2500\,K model, even at such low temperatures as 1500\,K, the partial pressure of methane approaches that of CO and water already at 2000\,K in the \teff\ = 1000\,K model, and supersedes the one of CO at T$\sim$1500\,K. The reason for this difference has to do with the gas pressure, which is 5 orders of magnitude higher at 1500\,K in the \teff\ = 1000\,K model than in the \teff\ = 2500\,K model, and the general tendency of the chemical equilibrium to shift toward larger molecules at higher pressures. This illustrates in an extreme form why the transition between
L-type spectral class and T-type spectral class (defined as the visibility of methane in the spectrum) must depend not only on 
\teff, but also on log($g$). If the effective temperature is well determined, 
the ratio between the intensity of CH$_4$ and CO bands can therefore in principle be used to determine log($g$) if the object is a star or a brown dwarf, and the mass of the object if it is a transiting exoplanet.

It is the same phenomenon that explains the huge difference between the H$_2$O (light blue dashed-triple-dot line) to OH (dotted green line) ratio at similar temperatures in the \teff\,=2500\,K and \teff\,=1000\,K models.
We will further see (in a later section) that TiO will be much stronger in the spectrum of the warmer of the models in the 
present grid than in the colder ones, and we see here that it is not because the partial pressure of TiO is larger in the 
warmer models (in fact it is very low in both of the models presented in Fig\,\ref{FigPP2500} and \ref{FigPP1000}), but because the partial pressure of TiO increases outward in the warmer 
models and inward in the cooler models, while the opposite effect is present for NH$_3$ (ammonia). This difference in the behaviour in the partial pressures of the two molecules makes them not only a good indicator of \teff, but the difference in the reduced observed band intensities of the two molecules compared to what can be computed from cloud free models, as those presented here, becomes a direct indicator of the height in the atmosphere of the presence of clouds.
 
TiO has a relatively low partial pressure in both models, although several orders of magnitude lower in the uppermost layers of the \teff\ = 1000\,K model than in the \teff\ = 2500\,K model. However, its high absorption coefficient per molecule and its many electronic transitions in the visual part of the spectrum, makes it an important contributor to the structure and spectrum for the \teff\ = 2500\,K models, and it will also play a role even in the \teff\ = 1000\,K models. In addition, it plays a crucial role in the temperature inversion for irradiated models, as discussed below, and therefore is also a good diagnostics in this. 
OH is the most important species for initiating atmospheric oxidation on Earth, and (in chemical equilibrium) it is considerable more abundant in high temperature models than in low temperature models.

Finally, the blue dashed lines show
the partial pressure of the sum of all positively charged atoms and the 
dash-dot light blue lines show the sum of the partial pressures of all negatively charged atoms. The ratio between the two curves is a direct illustration of the origin of the electron pressure, and is another way of quantifying the strongly decreasing electron pressure in the cooler models already discussed in connection with Fig.\,\ref{FigPe-tau}. 
The partial pressures of the negative ions as well as the positive have become very small in the \teff\ = 1000\,K model, and the partial pressures of the positive and negative ions are almost identical in the upper atmospheric layers, hence the electron pressure (not shown) goes toward zero as already discussed above.

	\section{The effect of irradiation}

The influence of irradiation of one component onto the other in eclipsing binary stellar systems was already included into the MARCS code by Vaz\,\&\,Nordlund in the 1980's (\citet{Vaz1985}, \citet{Nordlund1990}) in a plane parallel treatment. We use here these early initiatives as a starting point to develop an irradiation scheme in the MSG code that can realistically describe the effects of host star irradiation on an exoplanet (as well as being able to treat the atmospheric structure and spectrum formation in binary stars in a more up-to-date radiative transfer and geometry description). The MSG scheme includes angle dependent irradiation from the host star, and can in principle be computed in plane parallel as well as full spherical geometry of the host star and/or the irradiated exoplanet atmospheres.

\begin{figure}[h!]
               \vspace{-0.4cm}
               \hspace{1.0cm}
               \includegraphics[width=11.7cm,angle=-90.]{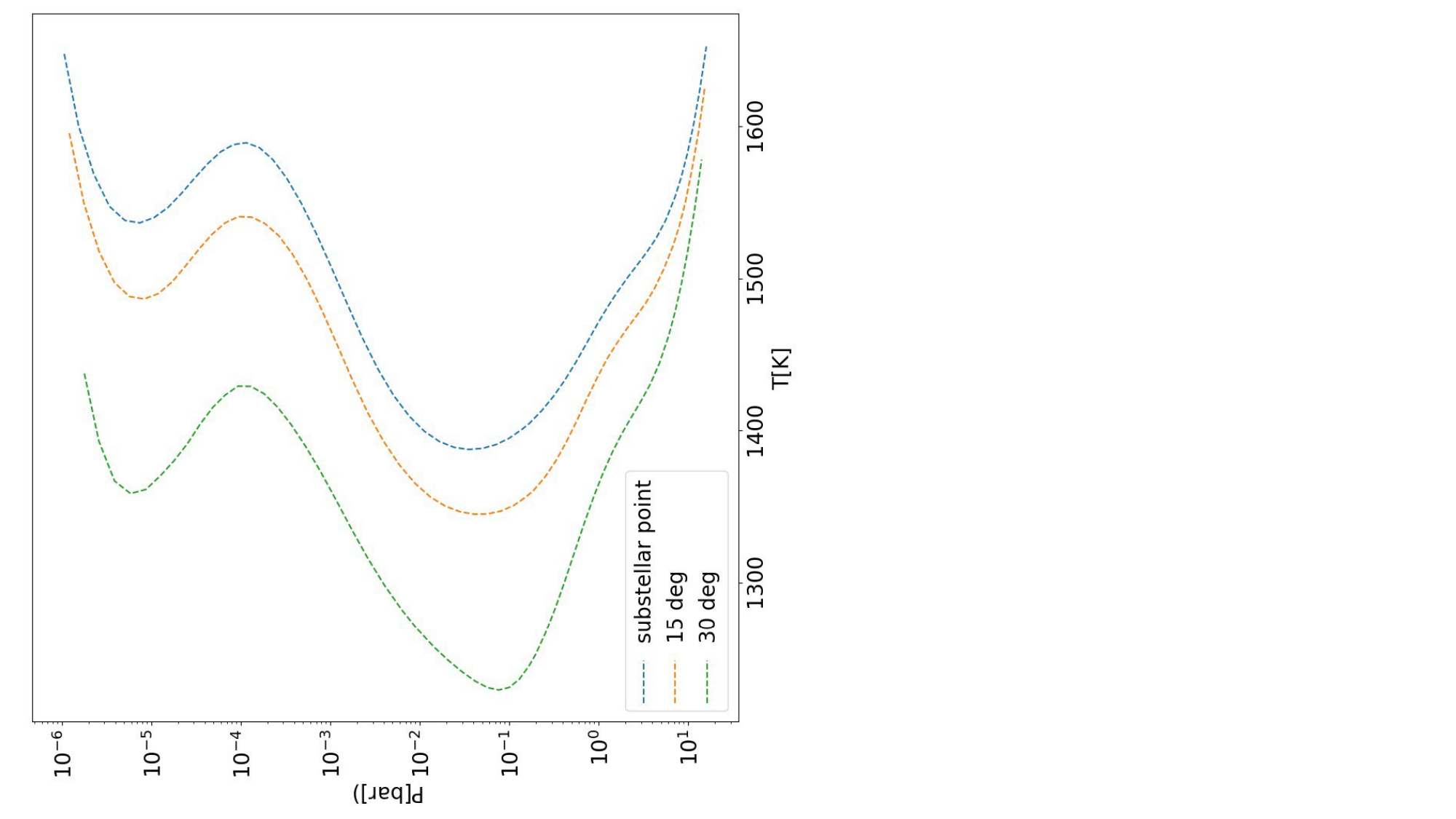}
                \vspace{-5.6cm}
		\caption{The $T-P_g$ structure of our irradiated 1.5D model simulation of an exoplanet with the parameters of WASP-39b at the substellar point (right-most curve), at $\Theta$ = 15$^\circ$ and at  $\Theta$ = 30$^\circ$ 
(left-most curve).  
		}
		\label{FigIrAngle}

	\end{figure}

\begin{figure}[h!]
               \vspace{-0.0cm}
               \hspace{1.0cm}
               \includegraphics[width=11.7cm,angle=-90.]{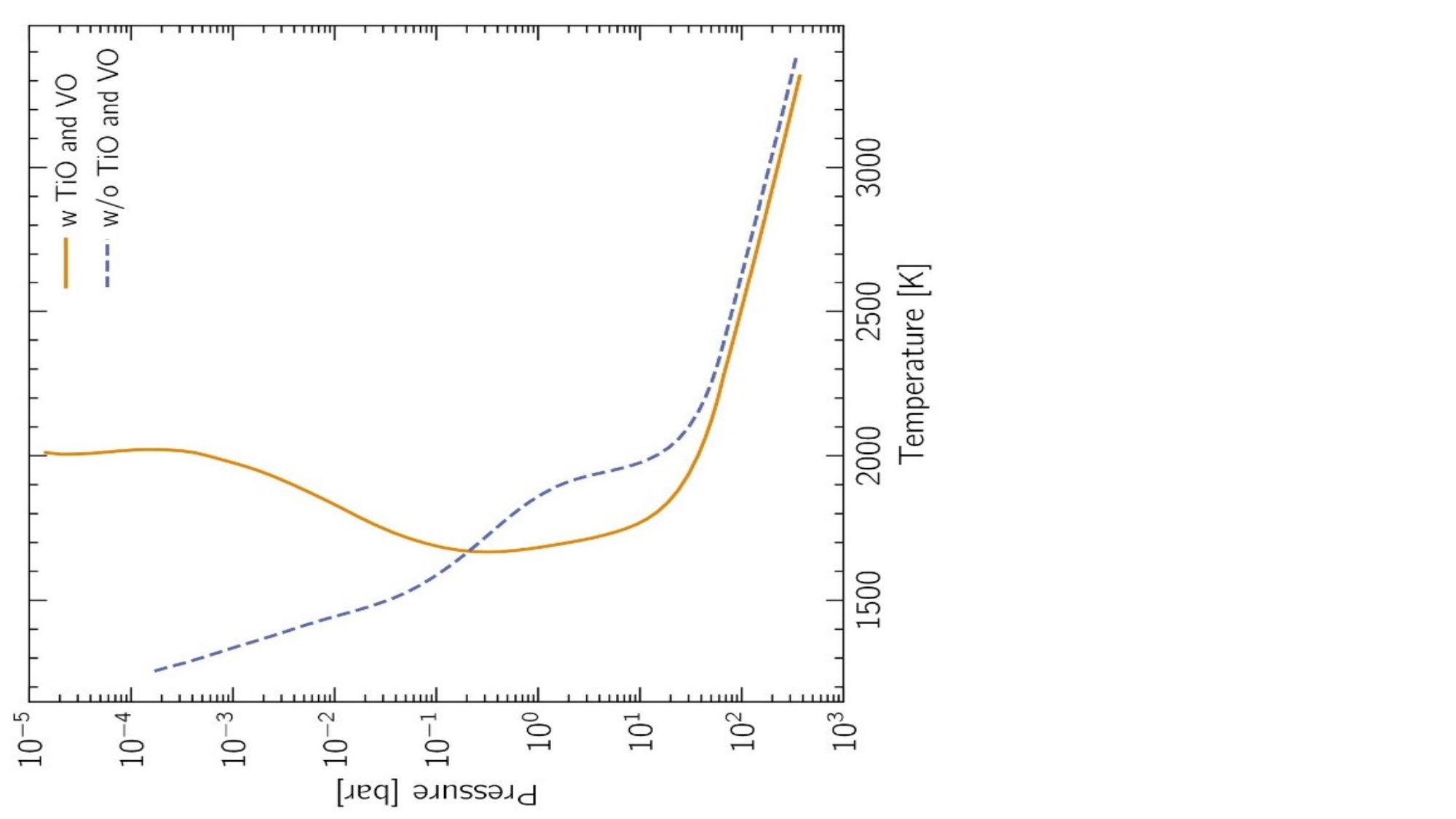}
                \vspace{-4.0cm}
		\caption{The TiO and VO caused temperature inversion at the substellar point in irradiated MSG 1.5D model simulations of WASP-43b..
		}
		\label{FigIrInversion}

	\end{figure}

In plane parallel geometry of the planet, the optical path of the incident irradiation converges toward 
infinity for 90 degrees angle toward zenith (ie. at the "morning and evening terminator",
Eq.\,\ref{pp_path}\,below). To compensate this,
we followed in our plane parallel computations the approximations introduced by {\citet{Malik2019} (see in
particular their equation 27 and associated figure 2). 
Since the models are static, and therefore do not include rotation and dynamical
effects as in a GCM 3D modelling, we term the method 1.5D
atmospheric modelling, in the sense that there will be no difference in the angle dependence of the 
atmospheric structure between planetrary longitude and latitude.
In our future irradiated models we plan to implement a parametrization
of computationally extensive GCM results (which was presented within our team in e.g.\,\citet{Schneider2022}), 
which will allow keeping the high flexibility
and speed that 1D self-consistent studies offer for a large grid of exoplanets
with complex physical, chemical and biological effects.
A route forward for this coupling of 3D modelling results into the 1D computations was developed in detail in \citet{Hubeny2017}, in particular for irradiated models in his section 2.2. More details of our own experiments with hybrid coupling of the MSG modelling with 3D results is described in Kiefer et al. (2024). Here we describe the most basic features of our irradiated models as they are included in the present grid, with illustration of a few examples in Fig.\,\ref{FigIrAngle} and \ref{FigIrInversion}.}

To a first approximation, the irradiation intensity J$_{\nu ,star}$ reaching layer $k$ of an exoplanet from a host-star shining onto the planet from and angle $\Theta_{star}$ with the zenith direction at a given longitude and latitude of the planetary surface, 
can be expressed as
\begin{equation}
J_{star}(\nu, \mu, k)=B_\nu(T_{star})\,\Delta\Omega\,exp(-\tau_{star})
\end{equation}
where B$_\nu$(T$_{star}$) is the black body flux of a host star of effective temperature of T$_{star}$.
$\Delta$$\Omega$ is the solid angle subintended by the star onto the planet,
\begin{equation}
\Delta\Omega = \pi \left( {R_{star}\over a_{au}} \right)^2 {1\over 4\pi},
\end{equation}
where $a_{au}$ is the semimajor axis of the exoplanet's orbit and $R_{star}$ is the radius of the host star.
In the plane parallel approximation, $\mu_{star}$ is a simple function of $\Theta_{star}$,  
\begin{equation}
\mu_{star} = 1/cos(\Theta_{star}).
\label{pp_path}
\end{equation}
$\tau_{star}$ is the optical depth scale that the stellar light travels inside the planetary atmosphere to reach layer $k$ of the zenith direction at the local position of the planet, $\tau_{star}$ = $\tau_k \mu_{star}$. 

Since the MSG grid can be used to compute the structure (and spectrum) of the exoplanets as well as their host stars, B$_\nu$(T$_{star}$) can be substituted with the real stellar spectrum to compute a correct and fully frequency dependent planetary 
transit spectrum, which can eventually be folded with a filter function to give a transit light curve for any arbitrarily chosen filter, without any assumed relation between scale height, radius and frequency.

The full description of the effect of irradiation onto the models will be given in a separate paper 
(Amadio et al 2024 in preparation). Here we give two examples of preliminary results from this part of the grid.
Figure \ref{FigIrAngle} illustrates how the model structure changes as function of irradiation angle of the host star relative the planetary surface. Quantitatively the figure is a simulation of a cloud-free exoplanet with parameters as those of WASP-39b (i.e.\ a 0.28 M$_{jup}$ mass planet orbiting a \teff\ = 5400\,K G8 type star at a semimajor axis of 0.0486 au), with the internal model temperature estimated from the relations between equilibrium temperature and internal temperature given by the formulas in \citet{Thorngren2019}. Within the limitations of 1.5D modelling, the 3 curves show the gas-pressure temperature structure of the atmosphere "at noon" (the substellar irradiation corresponding to $\Theta$ = 0$^\circ$; the right-most curve),  $\Theta$ = 15$^\circ$ 
(middle curve) and  $\Theta$ = 30$^\circ$ (left most curve). 
Figure \ref{FigIrInversion} illustrates the temperature inversion caused by the strong absorption of the irradiated star light by TiO and VO in our model of WASP-43b (a 2 jupiter mass planet orbiting a K7V star of \teff\ = 4500\,K at an orbital semimajor axis of 0.015 AU). The effect is equivalent to the temperature inversion due to absorption of solar irradiated energy by ozone in the Earth's atmosphere.

   \section{The effects of microphysical cloud formation}

To study the effects of microphysical cloud formation in substellar atmospheres, the MSG code couples \texttt{MARCS} and \texttt{StaticWeather} to one another  in a self-consistent manner (see section \ref{sec:3codes}). This implies the cloud radiative feedback is accounted for in the radiative transfer scheme. The cloud radiative effect is added to the radiative transfer by considering the cloud's opacity contribution, as well as the gas-phase elemental depletion caused by the cloud formation process. Given the cloud is composed of a mix of materials, we calculate its opacity with spherical particle Mie theory \citep{Mie1908} combined with effective medium theory.

	\begin{figure}[h!]
               \vspace{+0.4cm}
               \hspace{0.5cm}
               \includegraphics[width=7.7cm,angle=-0.]{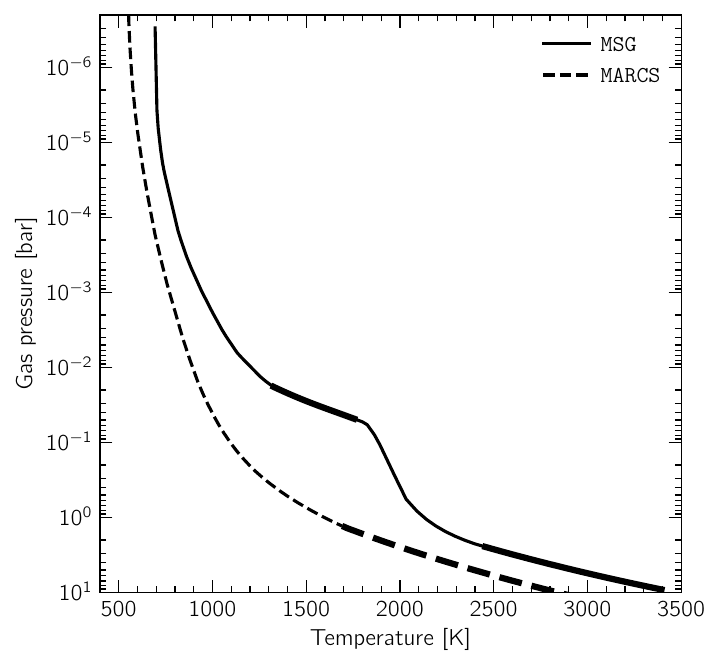}

               \hspace{0.5cm}
               \includegraphics[width=7.7cm,angle=-0.]{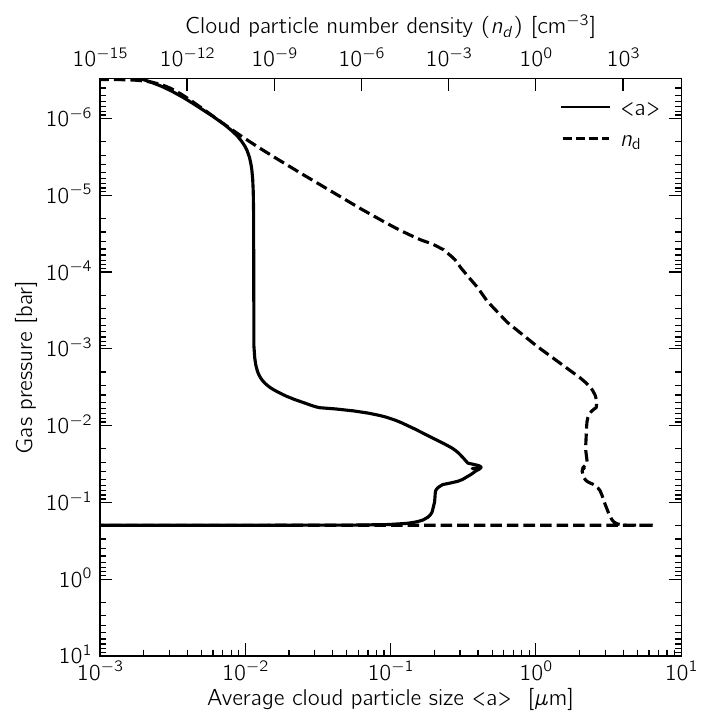}

 \caption{\textbf{Upper panel}: Pressure-temperature profiles for MSG models at \teff=1500K and log(g)=4.0 respectively with clouds (labeled MSG) and without clouds (labeled MARCS). Note that the convective part of the atmosphere is represented with a thicker line. \textbf{Lower panel}: Average cloud particle size (solid curve) and cloud particle number density (dashed curve) for the MSG model shown in the upper panel. We note that the spike in the number density at the bottom of the cloud is numerical and due to the rapid decrease of the average cloud particle size.} 
 \label{fig:cloudy_pt}
\end{figure}

	\begin{figure}[h!]
               \vspace{-0.0cm}
               \hspace{0.5cm}
               \includegraphics[width=7.4cm,angle=-0.]{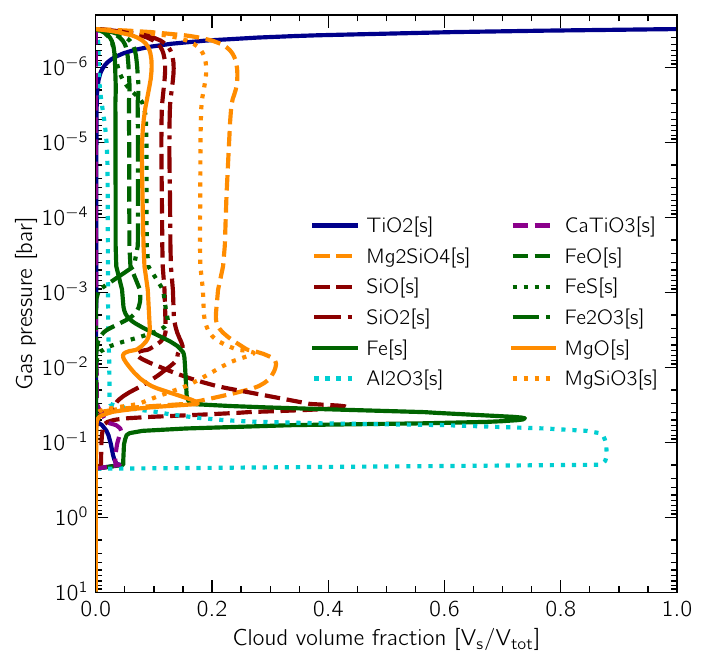}

               \hspace{0.5cm}
               \includegraphics[width=7.4cm,angle=-0.]{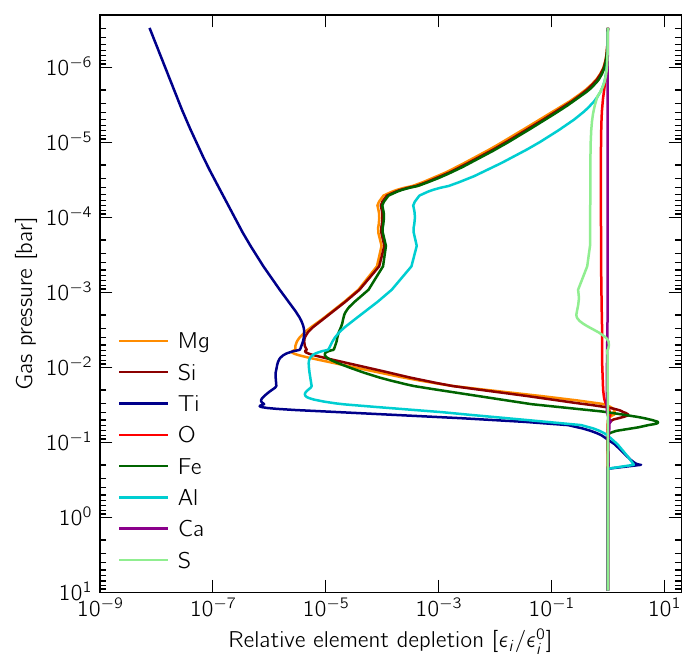}
 \caption{Cloudy MSG model at an effective temperature of 1500\,K and log($g$)=4.0. \textbf{Upper panel}: Cloud volume fraction of the 12 cloud species present in the atmosphere. \textbf{Lower panel}: Relative elemental depletion in the gas phase of elements affected by the cloud formation.}
 \label{fig:cloud_comp}
\end{figure}

	\begin{figure}[h!]
               \vspace{-0.0cm}
               \hspace{0.5cm}
               \includegraphics[width=7.4cm,angle=-0.]{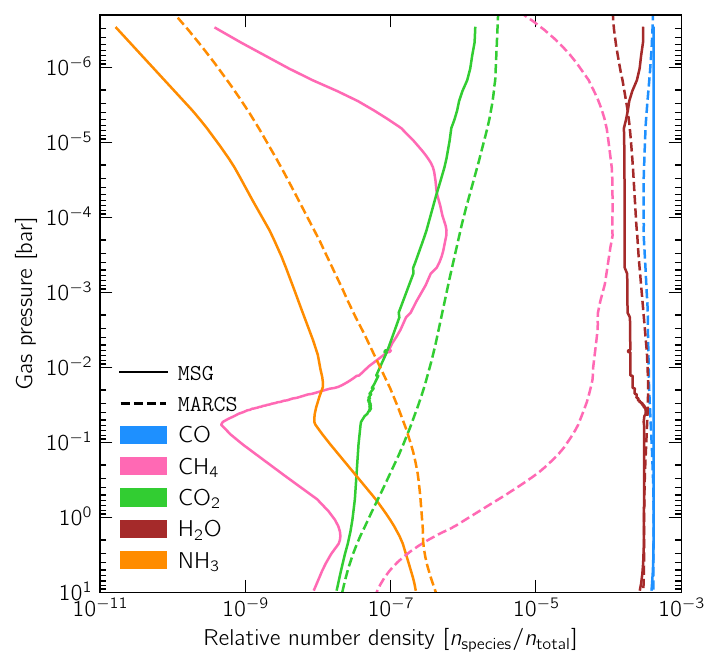}

               \hspace{0.5cm}
              \includegraphics[width=7.4cm,angle=-0.]{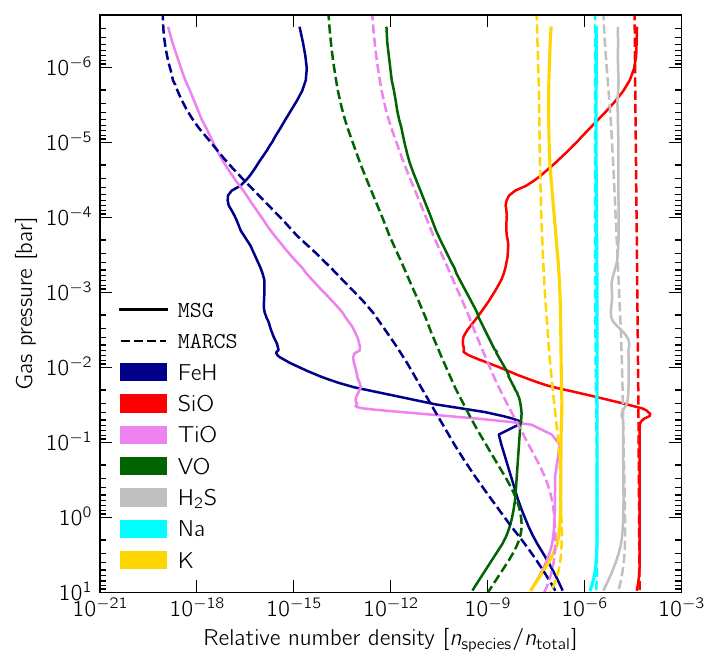}

 \caption{Relative number density of important molecular and atomic gas species in the cloudy {MSG} (solid curves) and cloud-free {MSG} (dashed curves labeled MARCS) models at 1500\,K and log(g)=4.0 (shown in Figure~\ref{fig:cloudy_pt}).}
 \label{fig:cloud_elabund}
\end{figure}

\begin{figure*}
\centering
\begin{minipage}[b]{0.7\linewidth}
\includegraphics[width=\linewidth]{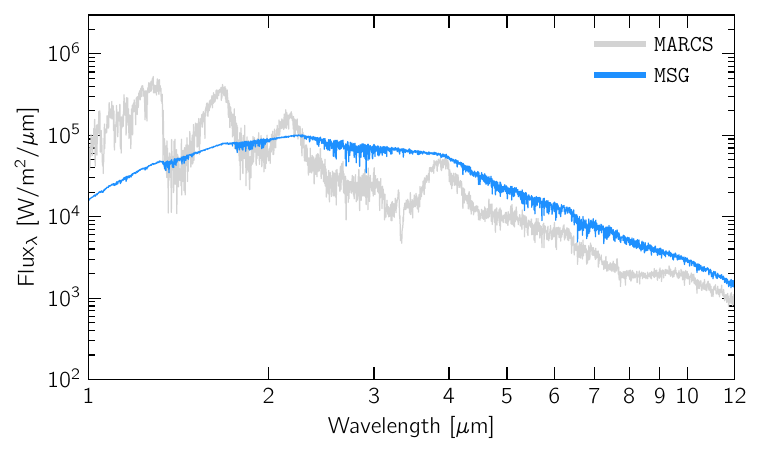}

\end{minipage}
 \caption{Cloudy (blue) and cloud-free (grey) synthetic spectra (binned to a rsolution of 1000) of a self-luminous body with an effective temperature of 1500K and log(g)=4.0, with solar metallicity and solar C/O ratio. 
 } 
 \label{fig:cloud_spectra}
\end{figure*}

The effective optical constants for the material mixtures are obtained with effective medium theory. We generally use the Bruggeman method \citep{Bruggeman1935} with the exception of rare cases where we find non-convergence and therefore use the analytic Landau-Lifshitz-Looyenga [LLL] method \citep{Looyenga1965} (see section 2.4.1. of \citet{Lee2016} for more details). The clouds extinction and scattering coefficients are computed with Mie theory using the routine developed by \citet{Wolf2004}, which is based on the widely used \citet{Bohren1983} routine. However, the \citet{Wolf2004} routine allows for the consideration of larger dust grain sizes. 

Here we consider 12 condensates: TiO$_2$[s], Mg$_2$SiO$_4$[s], MgSiO$_3$[s], Al$_2$O$_3$[s], Fe[s], SiO[s], SiO$_2$[s], FeO[s], FeS[s], Fe$_2$O$_3$[s], MgO[s] and CaTiO$_3$[s]. We use TiO$_2$[s] as the nucleation species. For the dust optical constants, we use the data tables compiled by \citet{Kitzmann2018} (see Table 1 in \citet{Kitzmann2018} for references). We note we use the amorphous (sol-gel) data for Mg$_2$SiO$_4$[s] and MgSiO$_3$[s], and the amorphous data for SiO$_2$[s]. The most relevant data are given in Tab.\,2 below,
and more details are given in Campos Estrada (2024, in prep.).

We run \texttt{MARCS} and \texttt{StaticWeather} iteratively until we find a converged solution. Besides the typical convergence criteria in the radiative-transfer scheme, we require both the cloud opacity and the emission spectra between the last two iterations to be converged.

The upper panel of Figure~\ref{fig:cloudy_pt} shows the effect of clouds on the atmospheric temperature versus gas pressure structure of a MSG model of a self-luminous object with $T_{\rm{eff}}=1500K$ and log($g$) = 4.0, at solar abundances and C/O. The radiative feedback of the clouds has a warming effect in the entire atmosphere (solid curve labeled MSG) compared to a cloud-free atmosphere of the same \teff\ and log($g$) (dashed curve labeled MARCS). 
We note in Figure~\ref{fig:cloudy_pt} that the radiative regions are plotted with a thinner line width than the convective regions which are plotted with a thicker line width. The cloudy model presents a detached convective zone between 0.01 and 0.1 bar. This detached convective zone has its origin on the cloud’s back warming effect. We discuss the emergence of detached convective zones in our cloudy models in detail in Campos Estrada et al. (2024, in prep). Such detached convective zones are also observed in other model grids (e.g. \citet{Morley2024}).
The lower panel in Figure~\ref{fig:cloudy_pt} shows the average cloud particle size and cloud particle number density for the model with clouds. We note where the average cloud particle size is larger, there is a drop in the number density as physically expected. 

The composition of the clouds and the gas elemental depletion are shown in Fig.\,\ref{fig:cloud_comp}. 
At the very top of the atmosphere (TOA) we see that TiO$_2$[s] is the major component of the cloud particles as it is the nucleation species in this specific model presented (Fig.\,\ref{fig:cloud_comp} upper panel). This also explains why Ti in its gas-phase is highly depleted at the TOA (Fig.\,\ref{fig:cloud_comp} lower panel). Although it is not as visible, O is also depleted at the TOA. However, O is a lot more abundant than Ti, and therefore this depletion is not as drastic. After TiO$_2$[s] nucleates, other cloud species grow on top of the TiO$_2$[s] seeds. Throughout most of the atmosphere, the bulk of the cloud is composed of Mg$_2$SiO$_4$[s], MgSiO$_3$[s], SiO[s] and SiO$_2$[s]. Deep in the atmosphere where the temperatures become too high for silicate to be thermally stable high-temperature condensates take over the bulk of the cloud (Al$_2$O$_3$[s] and Fe[s]). Towards the bottom of the atmosphere, we can see regions where there is an enrichment of the cloud forming elements in their gas-phase. These enrichments occur when particular cloud species evaporate: for example, we can see that there is an enrichment of Mg and Si when the silicates and magnesium-silicates evaporate. This is because these elements were trapped in the cloud particles and transported down until those cloud species evaporates, at which point we see an enrichment of the elements involved in that region of the atmosphere. We note that the gas-phase equilibrium chemistry is always computed after we have computed the gas-phase element depletions from cloud formation.

Figure~\ref{fig:cloud_elabund} shows the relative number density of some important molecules and atoms in substellar atmospheres, including H$_2$O, CO and CH$_4$,  along the atmosphere for the {MSG} and {MARCS} models shown in Figure~\ref{fig:cloudy_pt}. The gas-phase chemistry is of course different in each model, not only because of the effect of the cloud formation surface reactions but also because of the distinct temperature gas pressure structures (see CH$_4$ for example).

Clouds present a challenge to atmospheric characterisation as they can often hide spectral features of the gaseous components of atmospheres. Ignoring the presence of clouds can lead to a wrongful determination of various atmospheric parameters, such as metallicity. 

Figure~\ref{fig:cloud_spectra} shows the comparison between the emission spectra of the cloud-free and cloudy models shown in Fig.\,\ref{fig:cloudy_pt}. The spectrum of the cloudy model shows that a number of the molecular features present in the cloud-free spectrum are diminished by the presence of clouds. This is due to a combination of the decrease of molecular abundances due to the higher atmospheric temperatures of the cloudy model, as well as the contribution of the cloud opacity, which is significant in the observable atmosphere. 

In the final version of the cloudy MSG models, to be added to the present grid, we will present and analyse atmospheric
models of a larger range in \teff\  (Campos Estrada et al.\,2024, in preparation), as well as discuss the importance of the cloud radiative feedback, the cloud composition, the nucleation, the mixing timescales, and other intricacies of the microphysics processes.

\begin{table}[hbt!]            
\label{tab:dust_opac}      

\centering
\caption{Dust input data}
\begin{tabular}{l l}    
\noalign{\smallskip}
\hline\hline
\noalign{\smallskip}
Condensate & Reference \\    
\hline
\noalign{\smallskip}
Al$_2$O$_3$[s] & \citet{Begemann1997}\\
CaTiO$_3$[s] & \citet{Posch2003}\\
Fe[s] & \citet{Palik1991}\\
FeO[s] & \citet{Henning1995}\\
FeS[s] & \citet{Henning1997}\\
Mg$_2$SiO$_4$[s] & \citet{Jager2003}\\
MgSiO$_3$[s] & \citet{Jager2003}\\
MgO[s] & \citet{Palik1991}\\
SiO[s] & \citet{Palik1985}\\
SiO$_2$[s] & \citet{Henning1997}\\
TiO$_2$[s] & \citet{Zeidler2011}\\
\hline                              
\end{tabular}
\end{table}

	\section{The effect of non-equilibrium chemistry}

A standard adopted assumption in classical (stellar) atmospheric computations, including most versions of the MARCS code, is the local thermo-dynamical equilibrium, $LTE$, approximation, where the relative abundances of the chemical species is assumed to be in equilibrium with the radiation field "locally", i.e.\ layer by layer the kinetic and radiative temperature are assumed to be identical to one another. In optically very thin atmospheres, for example giant or supergiant stars, the radiation field in a given layer is often a combination of local Planck-like radiation augmented with a non-negligible radiation field from neighbouring layers of different temperatures, giving rise to what in astrophysics classically is termed NLTE (Non-Local-Thermo-dynamical Equilibrium). 
In a broader sense the non-local radiation field could even be a layer of a neighbouring object, such as we know it from the Earth's ozone-layer that is affected by absorption of UV-radiation from the Sun because the uppermost layers are transparent to solar-temperature UV radiation. 

In exoplanetary model atmospheres the chemical equilibrium in the higher layers are often assumed to be a combination of LTE and a gas-component from a wind-system that moves the gas fast enough upward to add an LTE component from an atmospheric layer of a non-local temperature (and density), which in its simplest form is often approximated by an almost free (in the case of exoplanets with unknown topography and wind-systems) parameter, expressed by the so-called eddy diffusion coefficients.

\begin{figure}[h!]
               \vspace{-0.3cm}
               \hspace{+0.3cm}
               \includegraphics[width=12.3cm,angle=-90.]{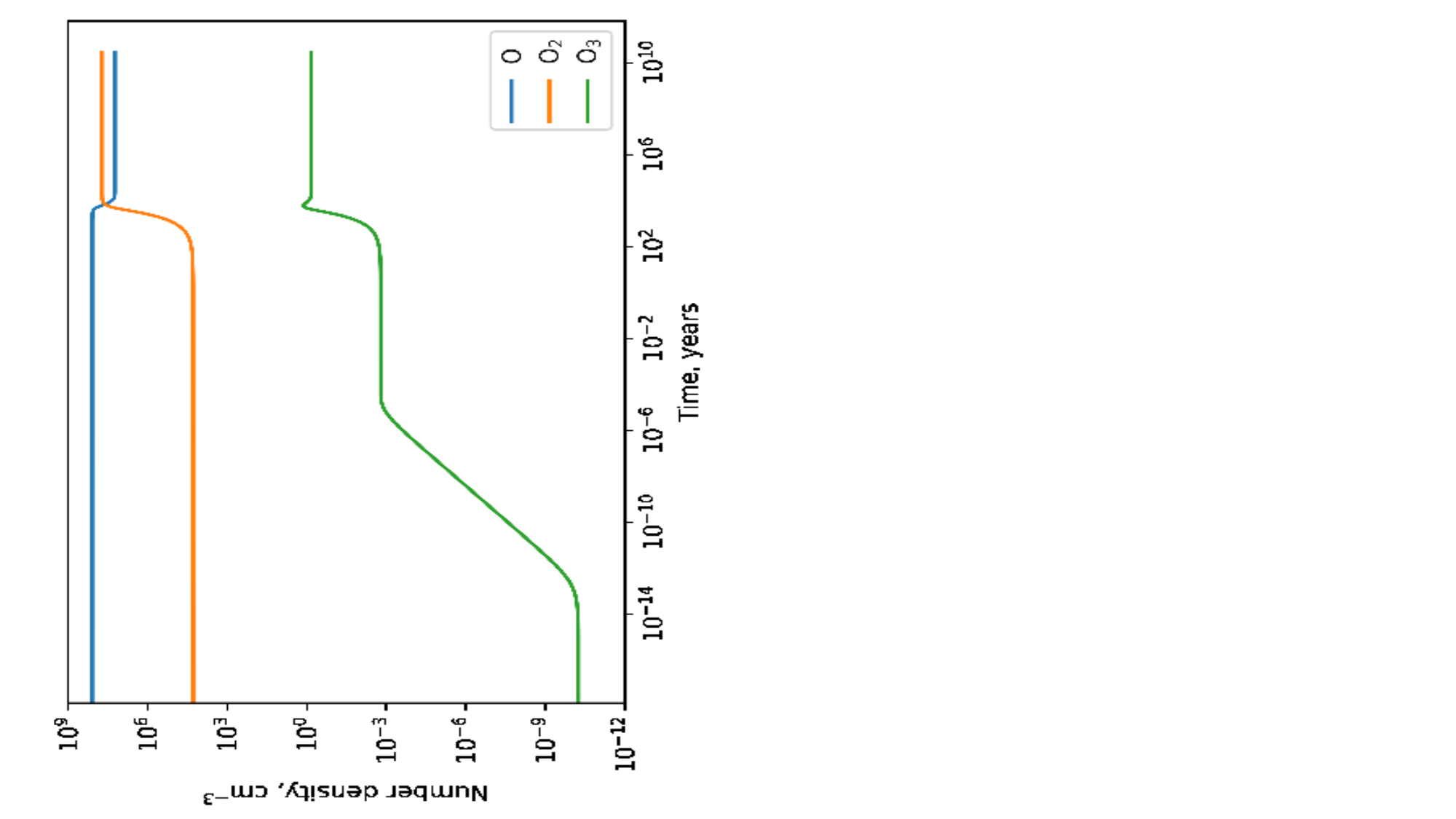}
                \vspace{-6.2cm}
		\caption{Example of the temporal evolution of the O, O$_2$, and O$_3$ number densities from their equilibrium abundances (left) to steady-state (right) for a 4-reaction Chapman mechanism chemical network.
      }
		\label{FigPhotochemistry}

	\end{figure}

\begin{figure}[h!]
               \vspace{-0.0cm}
               \hspace{+0.3cm}
               \includegraphics[width=13.7cm,angle=-90.]{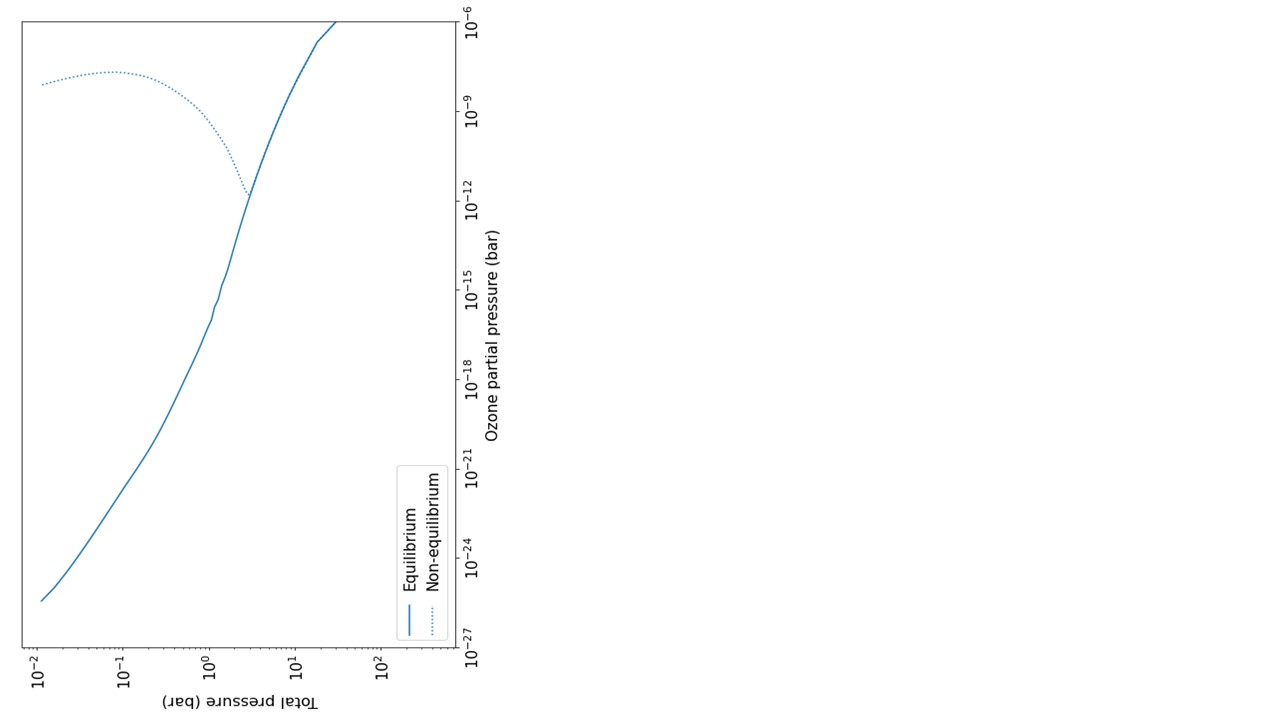}
                \vspace{-8.0cm}
		\caption{The LTE and the non-equilibrium steady state partial pressures of O$_3$ as function of total gas pressure 
(shown as full-drawn and dotted lines, respectively), for a surface-less exoplanet with Earth-like atmospheric 
elemental abundances, log(g) = 3.0 and \teff\ = 1000 K, illuminated by a Solar-type star 
(of log($g$) = 4.4, \teff\ = 5772 K, and solar radius) at 1\,AU from the planet.
      }
		\label{FigOzoneLayer}

	\end{figure}

\begin{figure}[h!]
               \vspace{-0.0cm}
               \hspace{-0.1cm}
               \includegraphics[width=14.4cm,angle=-90.]{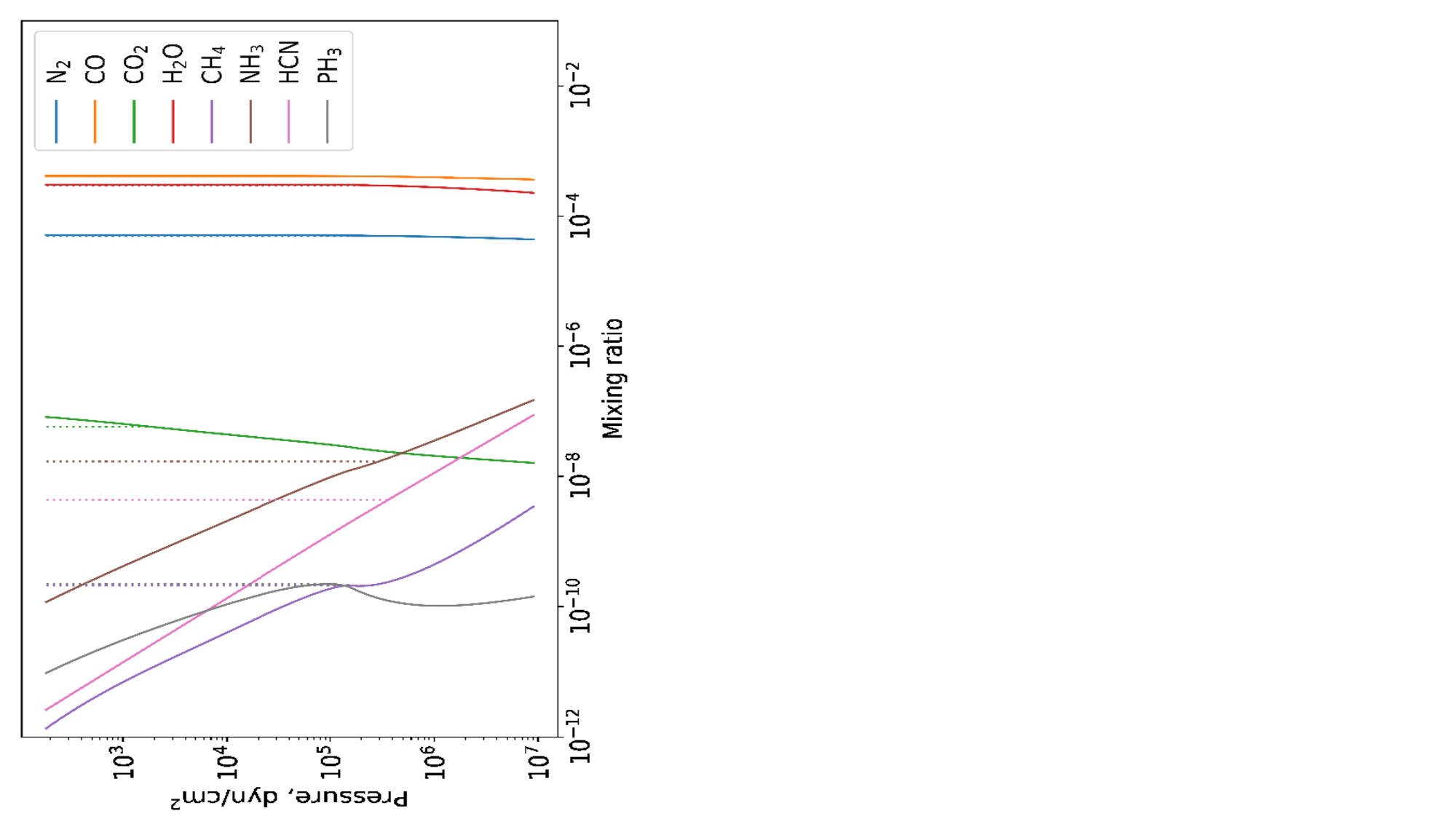}
                \vspace{-7.9cm}
		\caption{Atmospheric mixing ratios of the molecules for which quenching has been implemented into MSG. Solid lines show the equilibrium values, while the dotted lines show the mixing ratios with a fixed eddy diffusion coefficient of K$_{zz}$ = 10$^{10}$\,cm$^2$/s. The model has \teff\ = 2500\,K and log($g$) = 4.5.
      }
		\label{FigQuenching}

	\end{figure}

 In a more detailed description of the deviation from LTE one would like to include a full network of forward and backward reaction rates (to obtain a local steady state) and a realistic dynamical mixing of 
the layers. Only with such a more complex treatment of the gas chemistry would it be possibly to realistically quantify further contributions from potential entropy-reducing entities (including biology), which is the final goal of the additions of more complex MSG models to the basic grid presented here. Earth's atmosphere reflects the existence of all of these non-equilibrium processes, from classical NLTE (in the common astrophysical use of the term) represented most extreme by the 
effect of the solar radiation field on the ozone formation in the upper atmosphere and the photochemistry of the OH radical, 
over the dynamical non-equilibrium estimated from empirically confined eddy mixing, to the introduction of (biologically and non-biologically produced) oxygen and methane on dynamical timescales faster than the corresponding reaction rate timescales. Figures
\ref{FigPhotochemistry}, \ref{FigOzoneLayer}, and \ref{FigQuenching} illustrate preliminary results from our first steps of including steady state computations as well as dynamical mixing into the MSG model grid toward the aim of including self-consistent biological effects in the modelling.

As a first illustration (Fig.\,\ref{FigPhotochemistry}) of the effect of including time-dependent reaction rates into the MSG modelling, we have constructed a limited 4-reaction Chapman mechanism chemical network capable of computing the O, O$_2$ and O$_3$ abundances. The input starting abundances (left side of the figure; time zero) are the LTE abundances from GGchem, and the network is then propagated for long enough in time to assure that steady state has been reached, as shown in the right side of Fig.\,\ref{FigPhotochemistry}. 
The figure shows the result from a single layer of an MSG model atmosphere of solar abundance, \teff\ = 2500\,K and log($g$) = 4.5. The chosen layer had T = 1950\,K and P$_{gas}$ = 5.5\,10$^4$\, dyn/cm$^2$. The photochemistry is driven by irradiation from a solar type star at 1\,AU from the modelled object (for example 
an exoplanet), and the rate coefficients were taken from the NIST database.
We note how the stellar irradiation in a timescale of a few thousand years increases the relative ozone abundance with a factor of $\sim$10$^{10}$ compared to an LTE modelling.

As a next example (Fig.\,\ref{FigOzoneLayer}) is shown the O$_3$ partial pressure as function of total gas pressure, calculated by assuming respectively thermo-dynamic equilibrium and non-equilibrium from a 4-reaction Chapman-like mechanism (as in Fig.\,\ref{FigPhotochemistry}), now for a surface-less exoplanet with Earth-like atmospheric elemental abundances, log($g$) = 3.0 and \teff\ = 1000 K, illuminated by a Solar-type star (of \teff\ = 5772 K and solar radius) at 1 AU from the planet.
We note how the ozone abundance starts deviating from the LTE abundances at gas pressures below a few bars, and compared to its LTE abundance reach more than 25 orders of magnitude increased partial pressure in the upper most layers of model.  
Combining the irradiation and non-equilibrium modelling capabilities of the new MSG code with the Earth-like elemental abundances, allows for modelling of the photochemical Chapman-mechanism responsible for formation of the ozone layer in Earth’s atmosphere. The photolytic dissociation of O$_2$ results in formation of oxygen atoms, which combine with O$_2$ to form O$_3$. This results in a local maximum in O$_3$ concentration comparable to the ozone layer in Earth’s atmosphere. At an effective temperature of 1000 K, the effect of the ozone layer is too small to affect the pressure-temperature structure of the model, but it could be visible in the spectrum.

To allow modelling of more general steady-state conditions, we have developed a (beta-)version of the MSG that is 
coupled to the KROME package, designed specifically to include time-dependent chemistry into astrophysical 
models (\citet{Grassi2014}). 
It attempts to solve the chemistry by treating all relevant chemical processes with their respective rate coefficients, as a system of ordinary differential equations. The relevant equation that needs to be solved would therefore be
\begin{equation}
    \frac{dn_i}{dt}=\sum_{j\in F_i}\left(k_{j}\prod_{r\in R_j} n_{r(j)}\right)-\sum_{j\in D_i}\left(k_j \prod_{r\in R_j} n_{r(j)}\right)
\end{equation}
where $n_i$ is the i-th species in the network, and the first sum represent all the processes that would create the i-th species, with their respective rate coefficients $k_{j}$ and the number densities of the reaction partners $n_{r(j)}$, whilst the second sum represents all processes that would destroy the i-th species, with again their respective rate coefficients $k_{j}$ and the number densities of the reaction partners $n_{r(j)}$. For more details on how the system of differential equations is setup and solved, and which processes KROME is able to include, we refer again to the original publication \citet{Grassi2014}.

Given a chemical reaction network in the KROME format, the MSG species included in the reaction network are treated kinetically in each layer starting from the GGchem equilibrium concentrations, as illustrated in Fig.\,\ref{FigPhotochemistry} and \ref{FigOzoneLayer}.
The MSG atmospheric structure can then be iterated until self-consistent convergence is obtained with the non-equilibrium steady state concentrations of the selected species. All remaining species retain their equilibrium abundances. 

As a final example of non-equilibrium, we show in Fig.\,\ref{FigQuenching} effects of the often used quenching approximation in exoplanetary atmosphere computations (for example PICASO-3.0; \citet{Mukherjee2023}), 
which assumes that vertical flow of air under some conditions can bring bubbles of air from deeper layers to higher layers fast enough that the chemical equilibrium 
of the deeper layer is conserved in the higher layers in spite of their different temperatures (and gas pressures). Hence, the assumption is that the convective (quencing) timescales are shorter than the chemical reaction timescales. A typical introduction of quencing could be to let the chemical abundance of the first layer (from bottom and upward) where the vertical convective timescales 
was found to be smaller than certain reaction timescales in that layer, persist throughout the remaining part of the atmosphere (upward). 

In Earth's atmosphere such movements could typically take place when a mountain range forces air flow to move 
quickly upward, expressed by eddy diffusion with the diffusion coefficients fitted to atmospheric observations. Corresponding coefficients for a given exoplanet could be treated as a free parameter that could be adjusted to obtain a desired structure or reproduce an observed spectrum. We tested quenching based in MSG computations as an option for the 8 species CO, CH$_4$, H$_2$O, NH$_3$, N$_2$, CO$_2$, HCN and PH$_3$. The typical chemical timescales for conversion are calculated based on parametrizations (\citet{Zahnle2014} and \citet{Visscher2006}) and compared to the timescale of vertical mixing as approximated by a fixed eddy diffusion coefficient, K$_{ZZ}$. In regions above the lowest layer where the vertical mixing time scale for the chosen K$_{ZZ}$-value in this sense is lower than the estimated timescale for chemical conversions, the mixing ratios of the selected species are then kept fixed. An example is shown in Fig.\,\ref{FigQuenching} for a chosen fixed value of K$_{ZZ} = 10^{10}$\,cm$^2$/s. 
This option will allow testing the quenching approximation against a full self-consistent chemical non-equilibrium layer-by-layer computation, which will be the preferred solution, in particular when no particular topology and wind system is known.
 
	\section{Partial pressures and spectral diagnostics}

The presented MSG grid covers the temperature range of late-M, L, T, and early-Y spectral type objects, 
representing the
temperatures of cool dwarf stars, brown dwarfs, hot gas-giant exoplanets, as well as the 
temperature range of the atmospheres of hot to temperate rocky exoplanets. This diversity of objects cannot be described by a simple sequence of \teff, log($g$), and scaled solar chemical composition.  
The ambition of this first basic MSG grid of these complex objects is therefore also only to present some fundamental 
features, understanding and methods relevant for the sequence, focusing on the basic gas phase chemistry and self-consistent radiative transfer. Ongoing work on inclusion of more complexity into the grid includes the issues illustrated in the three previous sections. 
In these last two short sections we will based on the existing basic grid first touch upon the fundamental reasons behind the transitions between the empirically defined late-type spectral classes (the present section), and some of the expected major changes in the self consistent atmospheric structure when shifting from solar-like to jupiter-like and Earth-like compositions (next section). 

	\begin{figure}[h!]
               \vspace{-0.2cm}
               \hspace{-1.5cm}
              \hspace{1.0cm}
               \includegraphics[width=10.3cm,angle=0.]{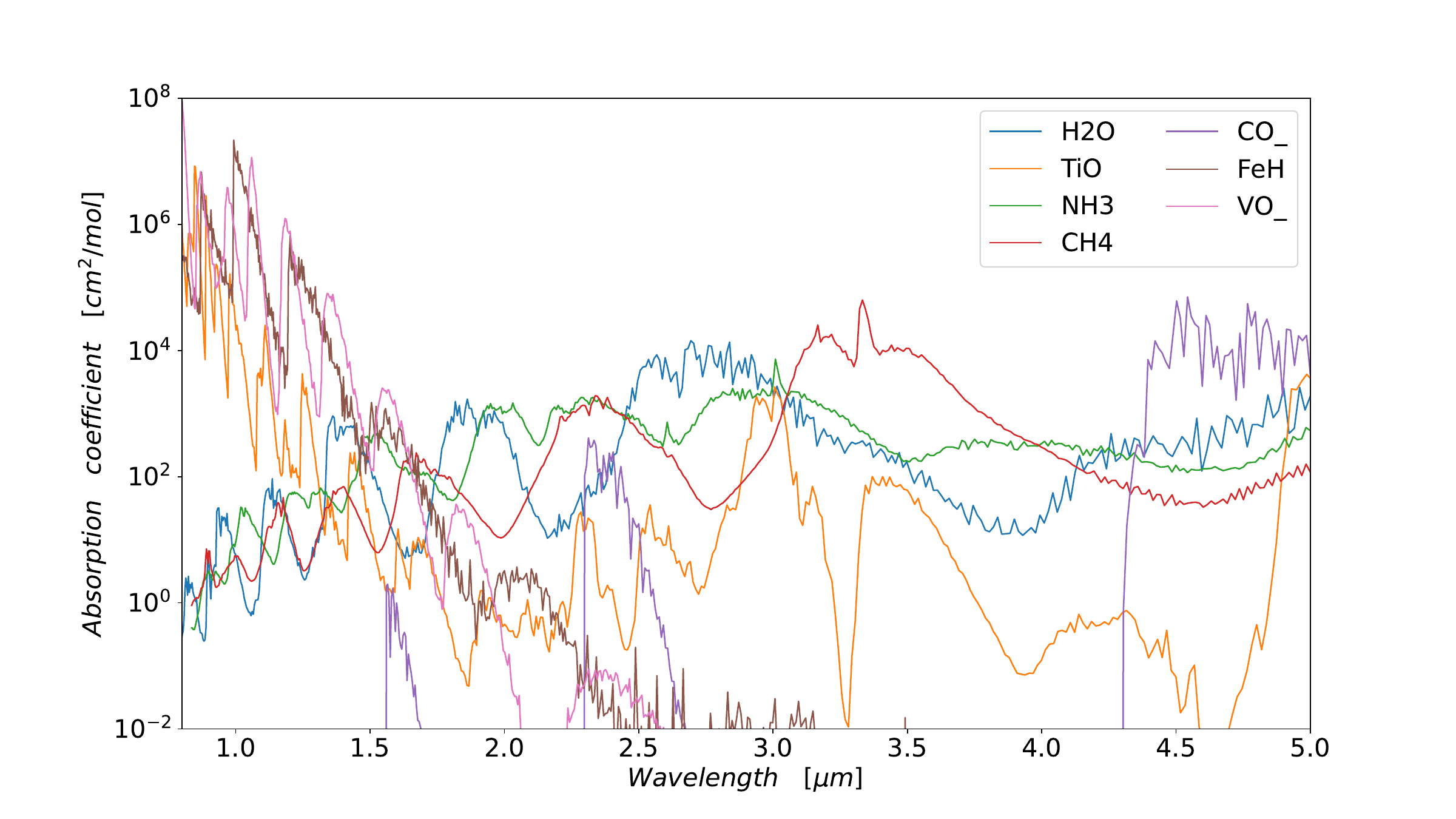}
                \vspace{-0.3cm}
		\caption{The absorption coefficient at T=1500\,K of the most important molecules (listed in the legend) for 
         the structure and understanding of the coolest dwarfs 
		}
		\label{FigOS}
	\end{figure}

\begin{figure}[h!]
               \vspace{0.3cm}
               \hspace{-.0cm}
               \includegraphics[width=11.7cm,angle=0.]{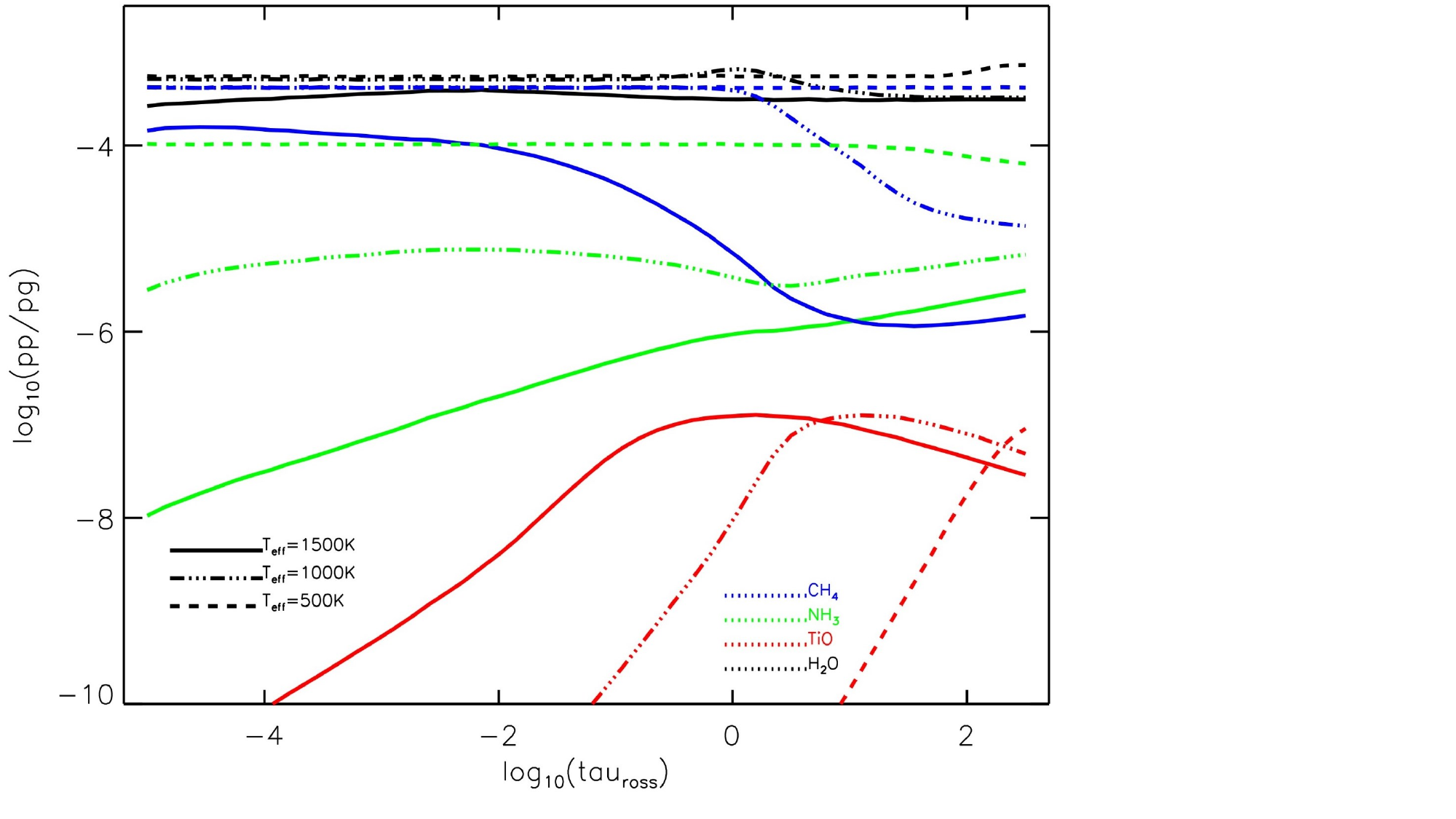}
                \vspace{-0.5cm}
		\caption{The partial pressures of the most important molecules in models of 
                    \teff\ = 1500\,K, 1000\,K, and 500\,K, respectively.
		}
		\label{Figpp}
	\end{figure}
\begin{figure*}
\centering
\begin{minipage}[b]{0.8\linewidth}
\includegraphics[width=\linewidth]{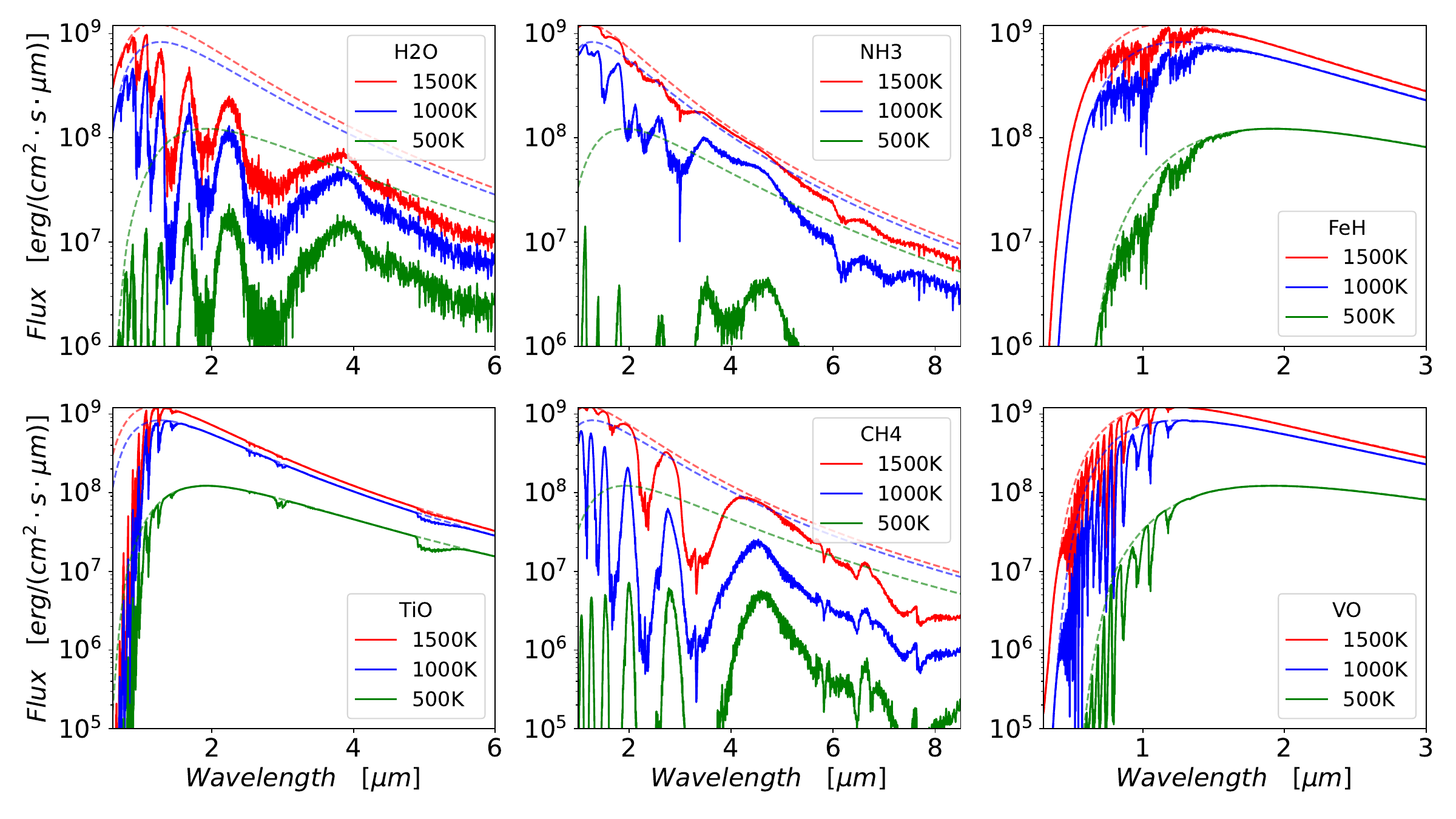}
\end{minipage}
 \caption{The development of the spectrum of H$_2$O and TiO (the two left panels), 
CH$_4$, and NH$_3$ (middle panels), and FeH and VO (right panels), for models of \teff\ = 1500\,K  (red), 1000\,K (blue) and 500\,K
(green). The dashed curves are the true continua (ie.\ the spectrum computed by allowing only the continuum opacities into the spectrum computation, but all opacity sources into the model computation), while the full drawn lines likewise are the spectra of each individual molecule listed in the legend, computed by allowing the opacity from only that molecule in the spectrum computation, but all opacity sources into the model computation.
Remark that for the coolest models (and the plotted resolution) the combined spectra do not at any 
place reach even close to the (true) continuum, which is the continuum that should be the reference in abundance analyses.
 } 
 \label{FigMLTspectra}
\end{figure*}

The radiative properties of self-consistent, cloud-free, static atmospheres are determined by the opacities, which are determined by the line transitions, line-profiles and sampling techniques as described in previous sections, which determine the absorption coefficients (Fig.\,\ref{FigOS}) and hence the partial pressures (Fig.\,\ref{Figpp}) and the radiative transfer leading to the spectrum 
(Fig.\,\ref{FigMLTspectra}).  

Fig.\,\ref{FigOS} shows the monochromatic absorption coefficient for the most important molecules needed for the 
understanding of the M-, L-, T- (and Y-)sequence. Due to its many electronic transitions, TiO stands out from most other species by the enormously strong absorption coefficient per molecule in the visual and photographic infrared region. It also has a relatively strong transition in the region from $10-12\mu m$, that per molecule is comparable to that of water, methane and ammonia. Due to the low abundance of Ti relative to C, N, and O, it is usually only the visual transitions that are discussed in observational spectra, but comparison of the visual and infrared absorption holds a rather unexplored diagnostic potential. VO has an absorption coefficient relatively similar to that of TiO, and also
the absorption coefficient of FeH is similar to that of TiO at least in the photographic infrared. The solar 
abundance of Fe is more than two orders of magnitude higher than that of Ti, while the abundance of V is almost an order of magnitude below than that of Ti.

The integrated absorption coefficients per molecule are rather similar for H$_2$O, NH$_3$ and CH$_4$, 
but due to the difference in the internal configuration of the three molecules, water is the most temperature dependent of the three and CH$_4$ the least. For water, the amount of weak lines increase rapidly with temperature, making the 
absorption coefficient stronger and more "continuum-like" for high temperatures, giving it an important structural role for stellar and sub-stellar objects of a wider range of effective temperatures than the two other abundant polyatomic molecules. The difference in the C-H and C-O stretching (and bending) energies happens to be so that the most pronounced  H$_2$O and CH$_4$ bands alternate with one another throughout large parts of the spectrum (which is not the case for NH$_3$), and hence give rise to a much stronger structural effect than would have been from individual ones of them. This is also the basic reason for why methane in even tiny amounts have a relatively large green-house effect on Earth's atmosphere, and it will make it possibly to quantify even small biological contributions to exoplanetary atmospheres by measuring trace elements in (coming) high-resolution exoplanetary atmospheres if the biology includes methane as it does on Earth.

CO also has a strong absorption per molecule. Being a diatomic molecule it has (as TiO) much fewer vibrational transitions than water, methane and ammonia and hence more discrete infrared bands, so even though the individual bands can be strong, CO has less importance for the structure in cool stellar and substellar environments than the corresponding polyatomic molecules.
Most pronounced is its fundamental vibrational transition around 5 $\mu$m, but also the overtones in the more accessible JHK region are strong. Due to its unusual high binding energy it will often consume most of the less abundant of either C or O, and therefore remain with a relatively constant band intensity in the spectrum independent of the atmospheric C/O ratio. The ratio of the H$_2$O and the CH$_4$ band intensities toward the CO band intensities are therefore good indicators of the atmospheric C/O ratio in cool stellar and substellar environments.

CO$_2$ has a strong near- and mid-infrared absorption coefficient, and is well known from Earth, Venus and Mars to be able to produce a strong absorption around 5\,$\mu$m, even at low abundances. If it does show up in the 
spectrum simultaneously with CO and CH$_4$ it has enormous biological diagnostic potential, and its presence together with methane, water, and ozone in Earth's infrared spectrum is considered evidence of photosynthesis on Earth. At high temperatures the absorption coefficient becomes "continuum-like" resembling that of water a wavelengths beyond 10\,$\mu$m. Its strongest transition overlap with the corresponding CO transition, and depending on temperature, gas pressure and relative oxygen abundance, it should be expected to have smaller structural effects on habitable exoplanets than methane.

While the absorption coefficients are basic for identifying the spectral features, it is the product of the 
absorption coefficient and the partial pressure that determines the opacities and hence the atmospheric structure and the quantitative spectroscopy. The development of the partial 
pressures (normalized to the gas pressures) through the M-, L-, T-sequence is shown in Fig.\,\ref{Figpp} as function of optical depth. Colour coding representing the different molecules is indicated with the legends, and the 
results are shown for models of \teff\ = 1500\,K (full drawn lines), 1000\,K (dash-dot lines), and 500\,K (dashed lines). 

In the observational classification of the M-, L- and T-type (and Y-type) objects, 
the transition between M-type and L-type is seen as 
the gradual disappearance of the TiO bands from their maximum intensity around M7 (\teff\ $\sim$2700\,K) toward a disappearance (in the visual to near-IR) at spectral type L3 to L6 (\teff\ $\sim$1700\,K). VO plays a similar role in the classification, although appearing and disappearing over a more narrow range of spectral classes, usually not seen before in the very latest M-types. Parallel with the disappearance of the small oxides, is observed an intensified appearance of metal hydrides and hydroxides (FeH, CaH, MgH, CaOH and others) reaching maximum
intensities for mid L-types and disappearing again with late L-types. 

Somewhat surprising, atomic lines, that would usually have been associated with much warmer objects, 
show up in the observationally defined  
L-type sequence, and gets broader and broader toward L5 to L8. In spectral type L5 the Na-triplet diverges into
what is interpreted as one broad line almost 1400\,{\AA} wide, and the K-triplet similarly reach a broadness of 1000\,{\AA} in type L8. It is hard to understand this as a temperature effect, and one suggestion has been that the decreasing electron density results in decreasing continuum opacities, which in turn result in a more transparent atmosphere, such that one sees deep into the atmosphere through a large column density of neutral atoms which would increase the line intensities and in particular contribute to broader Van der Walls wings due to the higher pressures in the warmer layers at the bottom of this clear photosphere. This interpretation is not immediately in agreement with our MSG models. Our models do show a particularly marked decreasing electron pressure (by more than 15 orders of magnitude) in what could be interpreted as the late L-type sequence (Fig.\,\ref{FigPe-tau}), but our continuum is not very dependent on the contributions from electrons already at higher values of \teff\ (Fig.\,\ref{FigMLTspectra}). The interpretation is open for more detailed analysis by use of the present grid. The understanding obviously also raises the question of the general assumption of more dust-clouds influence on the cooler atmospheres, and it all underlines the complexity of the processes compared to warmer objects. At what is interpreted as even lower effective temperatures, the observed metal lines disappear from spectra again.

At lower effective temperatures than the L-type sequence, methane (CH$_4$) appears in the spectrum, and the traditional classification feature of the beginning of the T-type sequence is the appearance of the CH$_4$ 
overtone bands at 1.6 and 2.2 microns, but the fundamental band at 3.3 micron is of course even stronger, and we 
see it in our cloud free models already in what is traditionally classified as mid-L type objects at \teff $\sim$1500\,K 
(Fig.\,\ref{FigMLTspectra}). The T-type objects are sometimes called the methane brown dwarfs, and the whole T-type sequence belongs to sub-stellar objects. For even cooler temperatures, ammonia (NH$_3$) becomes abundant, although these objects (some times called Y-type objects) usually are so dim that the spectra are open for several interpretations. Our cloud-free solar metallicity models predicts the NH$_3$ bands to be very strong in the \teff\ = 500\,K models, but overlapping strongly with the CH$_4$ bands and probably best identifiable around 6\,$\mu$m
(Fig.\,\ref{FigMLTspectra}).

Just as for the changing depth distribution of TiO in the atmosphere (Fig.\,\ref{FigPP2500} and \ref{FigPP1000}), 
we have seen in Fig.\,\ref{Figpp} and \ref{FigMLTspectra} that in the coolest MSG models presented 
also NH$_3$ changes depth distribution with changing values of \teff\ rather than changing total atmospheric abundance, but in the opposite depth direction that TiO (and neutral atoms).
Depending on the height of cloud formation, NH$_3$ may therefore form above a potential cloud layer 
in the coolest layers (just like in the much cooler atmosphere of Uranus), and the relative intensity of at one hand
the atomic lines and the strongest TiO bands and on the other hand ammonia bands may 
therefore be a valuable detailed diagnostics of the level of cloud formation in self consistent modelling.
The relative weakening of the TiO and NH$_3$ bands will depend on how high in the atmosphere the clouds
form and the T to Y transition is therefore not solely a measure of 
a temperature sequence, but depend among many factors also on the height in the atmosphere that the clouds form. High level clouds will weaken or erase the NH$_3$ features, while low lying clouds weakens the TiO bands but not the NH$_3$ bands.
In our preliminary cloud models (\cite{Juncher2017}) the first thin clouds can form already in the upper layers of models of \teff$\approx$2500\,K, and will therefore move the M-L (and L-T) transition to higher values of \teff\ than in the cloud-free models.

	\section{Solar versus Jupiter-like and Earth-like chemical compositions}
\begin{figure}[h!]
               \vspace{-0.1cm}
               \hspace{-0.5cm}
               \includegraphics[width=10.0cm,angle=-180.]{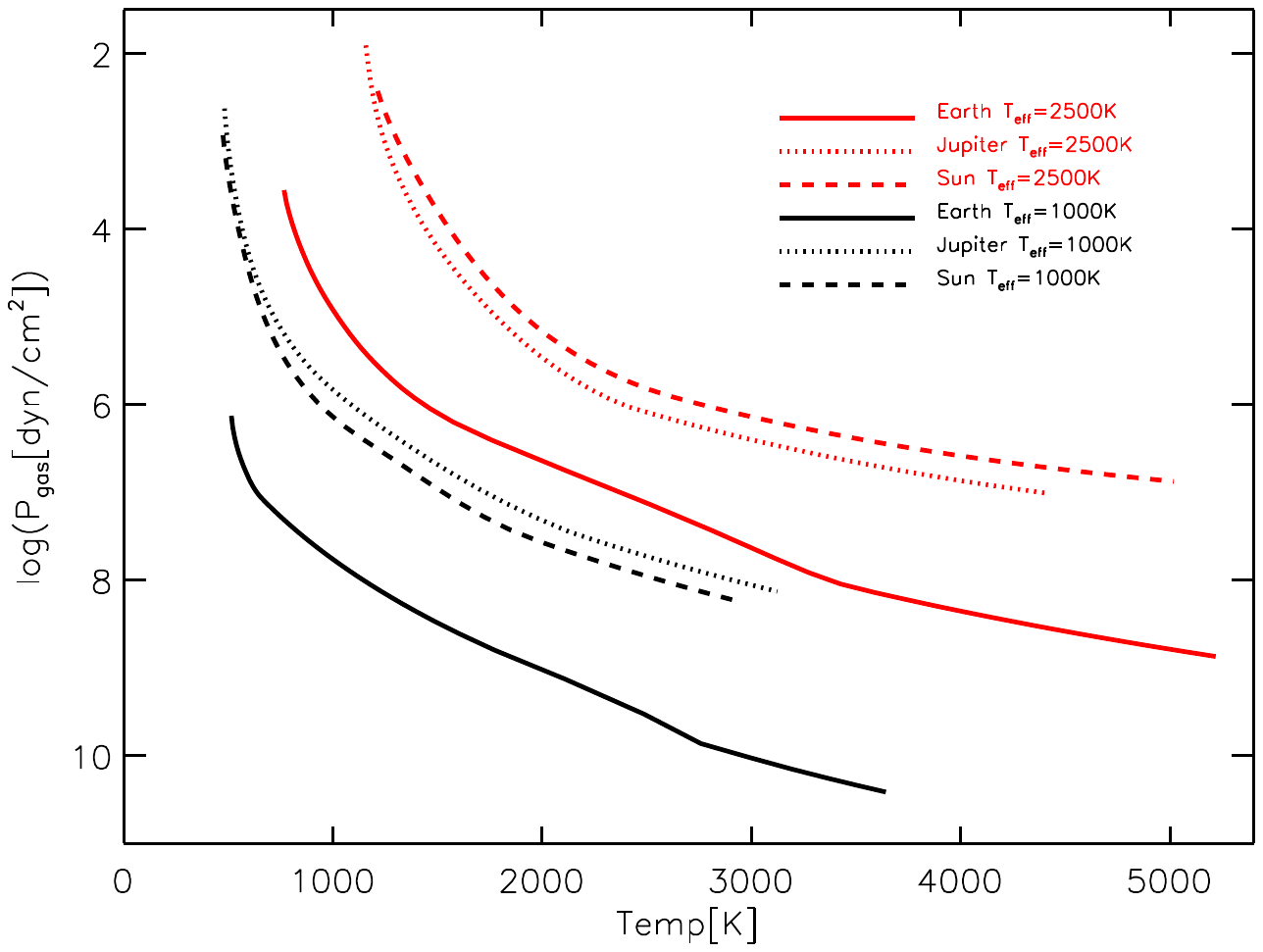}
                \vspace{-0.4cm}
		\caption{The $T-P_g$ structure of models at \teff\ = 1000\,K (black) and 2500\,K (red), with respectively 
                   solar, Jupiter-like and Earth-like chemical atmospheric compositions. All models are for log($g$) = 4.5.
		} 
      \vspace{0.0cm}
		\label{FigEarthSpec2}
	\end{figure}

Primordial planetary atmospheres of all kinds, as well as present day gas-giant exoplanetary atmospheres, may well be close to (scaled) solar composition for many elements. Jupiter’s atmosphere has close to the solar relative abundance of C, N, S (and the noble gasses Ar, Kr, and Xe), although they all are a factor of approximately 3 increased relative hydrogen, compared to their solar ratios. He and P are almost solar (relative to hydrogen) while Ne and O are of the order a factor 10 down compared to their solar values. Expressed in the standard $\epsilon$-scale (ie.\ as the logarithm of the number abundance of an element $A$ per 10$^{12}$ hydrogen atoms, 
$\epsilon$($A$) = log($A$/H)+12), then Jupiter's abundances \citep{Atreya2003}
would translate into epsilon values of $\epsilon$(He, C, N, O, P, S, Ne, Ar, Kr, Xe) = (10.896, 9.021, 8.593, 7.930, 5.486, 7.607, 7.090, 6.957, 3.638, 2.640). Since these abundances are not widely different from the solar values, 
$\epsilon$(He, C, N, O) =  (10.93, 8.39, 7.78, 8.66), the corresponding T-Pg structures are not widely different either, as illustrated in Fig.\,\ref{FigEarthSpec2} for models of \teff\ = 1000\,K and 2500\,K. One sees that the solar and the Jupiter composition models span approximately the same temperature and gas pressure ranges, with the (hot, cloud-free) Jupiter composition models having only slightly higher/lower gas pressures for given temperature in the 
warmest/coldest of the models presented in  Fig.\,\ref{FigEarthSpec2}, caused by the only slightly different (and temperature dependent) opacities from the two different compositions. Jupiter itself has log($g$) close to 3.5 rather than the solar value of (close to) log($g$) = 4.5 that is illustrated in Fig.\,\ref{FigEarthSpec2}. The figure is meant to illustrate the effect of chemical composition by isolating the elemental parameter differences. A main difference between the temperature response of an object with a Jupiter-like (C/O$>$1) and a solar-like (C/O$<$1) elemental abundance pattern is hidden in the different temperature dependence of dominating carbon molecules and oxygen molecules, as was illustrated for TiO, H$_2$O, and CH$_4$ in Fig.\,\ref{Figpp}. In Fig.\,\ref{FigEarthSpec2} this effect is reflected in which of the two (Jupiter-like or solar-like) models is the warmest at the two values of \teff\ presented.

The epsilon values of Earth’s atmosphere, on the other hand, are widely different from the solar composition, mainly because of the presence of biology, and can be expressed as $\epsilon$(He, C, N, O, Ne, Ar, Kr) = (8.418, 10.319, 13.892, 13.219, 8.958, 11.670, 7.756) as converted from the Earth’s molecular abundances given in the so-called NASA standard Earth (dry) atmosphere \citep{Williams2024}. The T-Pg structure is therefore also widely different from that of corresponding solar or Jupiter composition models. Fig.\,\ref{FigEarthSpec2} therefore also include an illustration of the physical effect of such a chemical composition. The main difference of the Earth-composition models
from the Jupiter or solar composition models, is a substantially higher gas pressure for given physical temperature. 
This is a direct effect of the high nitrogen and oxygen abundances which give rise to Earth-composition models being dominated by dipole in-active homonuclear molecules, making the dry atmosphere almost completely transparent. Together with the very low hydrogen and carbon abundances in Earth-composition models, this also makes the temperature gradient as function of height very sensitive to small changes in the carbon and hydrogen abundances, which is well known as the Earth's extreme greenhouse sensitivity to tiny changes in carbon and/or hydrogen abundances. The 
very much higher temperature in the bottom layers of Earth composition models seen in Fig.\,\ref{FigEarthSpec2} is not a greenhouse effect, but just an artefact of the way Rosseland optical depth is defined, the physical important difference between the models being the changing gas pressure values at given temperature, which defines the changing transparency (spectrum) of the models.

The gravity $g$ of the Sun in units of cm/s$^2$ is close to log($g$)=4.5, Jupiter has log($g$)$\approx$3.5, 
while Saturn and Earth both have log($g$)$\approx$3.0. 
A change of one order of magnitude in the value of $g$ will give rise 
to a change in $P_g$ of a similar amount as the differences seen in Fig.\,\ref{FigEarthSpec2} due to the abundance difference between Earth-like and Jupiter/solar-like compositions. 
However, while the changing abundances will give rise to substantial changes in the resulting spectrum, the changing value of log($g$) will in most cases hardly be visible in the spectrum, as long as the opacity carrying species remain unchanged.    This is a well known and well described effect for cool stars, because it makes it difficult to estimate the mass of cool stars based on their spectra (as opposed to hotter stars). 
The reason being that in a self-consistent energy balance, the amount of energy absorbed in a column of atmospheric material (the "integrated opacity") depends
on the flux (the effective temperature) and not the gravity, such that a change in the gravity will adjust the height structure until the absorbed energy reflects the flux, and hence to a first approximation will reach the same spectrum of the opacity dominated species at the top of the atmosphere independent of log($g$).
As already discussed above, the spectra of the trace elements will, however, be widely different in two such self-consistent models of different values of log($g$), and therefore immediately reveal for example nitrogen or oxygen biological features even if oxygen itself would not be visible, provided the energy balance in the analysing model atmosphere is computed self-consistently. 
In exoplanetary spectra we will probably have to await large enough instruments, such as the ELT, to be able to identify the spectral features from trace molecules in the atmosphere.

	\section{Outlook and conclusions}

We spent a considerable effort throughout the paper to investigate the quality of assumptions 
that were used in previous versions of the MARCS code, by comparing results from computations where these assumptions were included with corresponding MSG models where such assumptions have now been 
possible to relax. The logic behind these comparisons were to secure consistency with the previous MARCS grids and to study the effects of various limitations that 
are no longer needed (for example due to the existence of more advanced input data, or due to new computational 
methods), but also to give advice to the huge community of users of previous versions of the MARCS code (judged
from the more than 3000 citations to the code in the literature) as to whether some results should need revisions.

\begin{figure}    [h!]
               \hspace{-0.45cm}
               \includegraphics[width=10.0cm,angle=-180.]{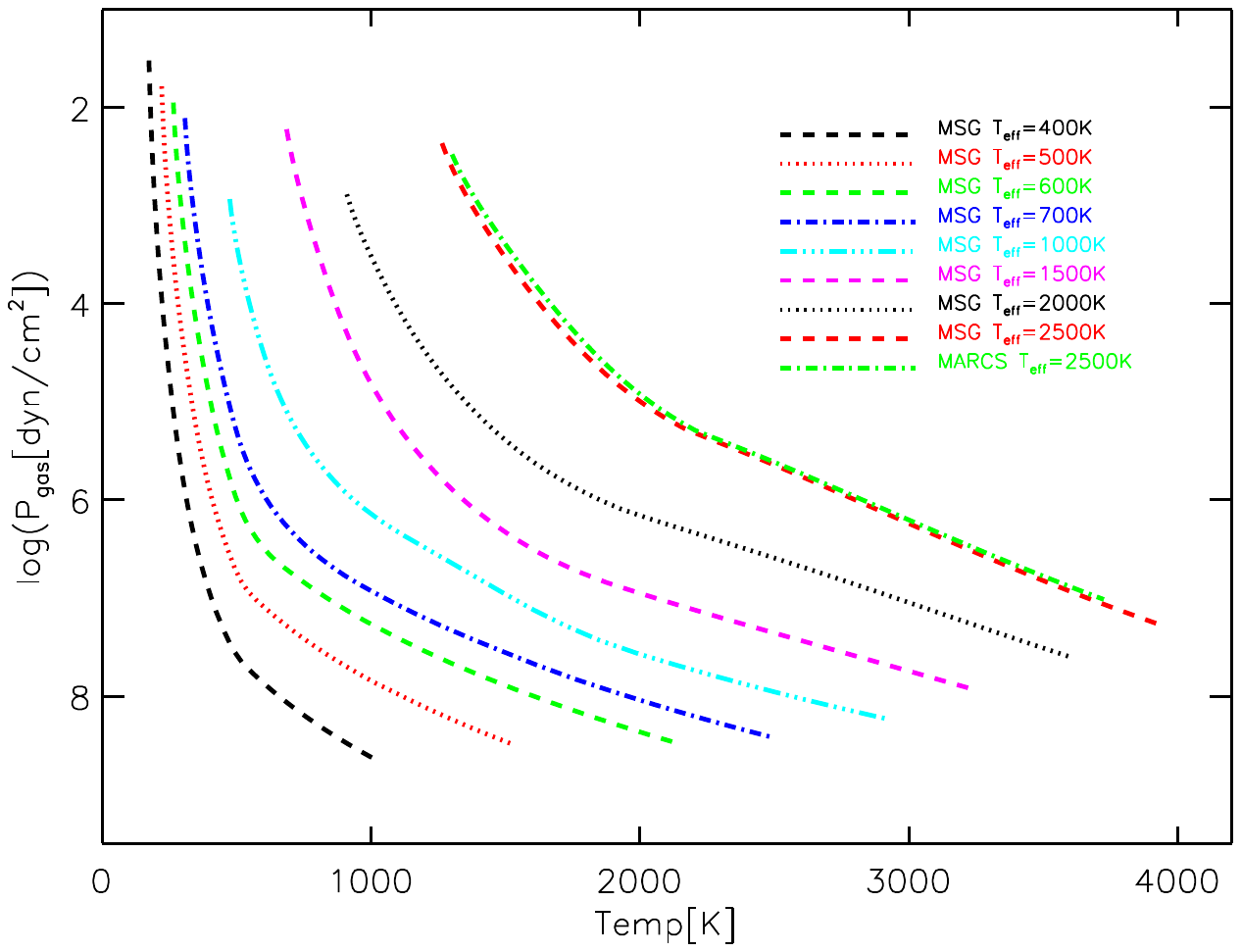}
                \vspace{-1.0cm}
		\caption{The $T-P_g$ structure of basic MSG models from \teff\ = 400\,K to 2500\,K (plus the coolest classical 
        MARCS model for comparison as well).} 
      \vspace{0.0cm}
		\label{FigBasicGrid}
	\end{figure}

By comparing models based on opacities from the limited number of molecules that were available in 2008 with
models based on a more extensive number of molecules now available in the ExoMol database, we concluded that the limitation to including only 16 molecules in the classical
2008-grid (and 22 in the 2017 DRIFT-MARCS computations) did not affect the accuracy of the models, even for the coolest models of the 2008-grid of \teff\ = 2500\,K (Fig.\,\ref{FigTPg2500K} and Fig.\,\ref{FigBasicGrid}).

We also conclude that it has been ok to exclude atomic line opacities in any of the applications 
(that we know of) where it was "just intuitively" assumed to be ok (ie.\ typically for models 
of \teff\ $\approx$ 3500\,K and cooler).
This conclusion could of course be challenged for compositions deviating substantially from solar composition and not 
included in the tests performed in the present study, so in general all MSG models do include atomic line opacities (taken from the DACE database), even when they are expected to be without importance for the structure.

The intuitive assumption in all MARCS models that it was sufficient to use Gaussian line profiles in the computation 
of the molecular opacities was, however, challenged. One might have assumed that the uncertainty in the model
 structure of older models due to lacking molecular line profile data could be simulated by 
introducing models with $\xi$ = 2 km/s and 5 km/s in the Gauss profile, which might however underestimate the true uncertainty. We found that the difference between models based on line opacities
from Gaussian profiles of $\xi$ = 3 km/s and based on extensive Voigt profile datasets now available, was larger that the difference  between models based on the $\xi$ = 2 km/s and $\xi$ = 5 km/s in the molecular Gaussian line profiles. 
However, the presently available Voigt profile based opacity data for molecules in the ExoMol data base that we
tested here are computed only for $\xi$ = 0, so more work is needed on this issue in order to reach a final conclusion. This uncertainty goes of course not only for MARCS and MSG models, but for all other self consistent exoplanetary models in the literature as well.

We have focused in the present paper on a description of the basic principles that have made it possible to 
expand the previous grids of well cited MARCS models to lower effective temperatures, to a greater variety of 
chemical compositions, to irradiated models with cloud formation, and to models of time dependent chemistry that
are necessary for a self consistent description of the effect of biological activity on exoplanetary atmospheres and spectra. In the present paper we have given examples of these models, and we illustrate in Fig.\,\ref{FigBasicGrid} 
the rough changes in $T-P_g$ structure through the new grid of the basic models.
In additional MSG models, to be added to the present grid, we will present and analyse atmospheric
models of a larger range in \teff, log($g$), chemical elemental composition, stellar irradiation, surface structures, clouds based on additional nucleation and dust species, etc.

The code in its present form is able to self consistently compute the host star irradiation (and atmospheric structure) of a vide range of stars and substellar objects together with the atmosphere and spectrum of their system of planets from Earth-like temperatures to even the hottest clear and cloudy gas-giant exoplanets, to include spectral effects of biological non-equilibrium activity, and in future versions rocky planets of various surface compositions as well to allow for example studies of outgassing from the planetary surface that can give insights into the formation processes, surface composition, and biological activity patterns of early versus late-type Earth-like exoplanets. \\


	\begin{acknowledgements}
   We acknowledge funding from the European Union H2020-MSCA-ITN-2019 under Grant no. 860470 (CHAMELEON)
   and from the Novo Nordisk Foundation Interdisciplinary Synergy Programme grant no. NNF19OC0057374.
	\end{acknowledgements}

    \end{document}